\documentclass[english,useAMS, usenatbib]{mnras}
\usepackage[T1]{fontenc}
\usepackage[latin9]{inputenc}
\usepackage{units}
\usepackage{rotating}
\usepackage{url}
\usepackage{amsmath}
\usepackage{commath}
\usepackage{amssymb}
\usepackage{graphicx}
\usepackage[caption=false]{subfig}
\usepackage{pdfpages}
\usepackage{esint}
\usepackage[authoryear]{natbib}
\usepackage{listings}
\usepackage{textcomp}
\usepackage[export]{adjustbox}
\usepackage[para,online,flushleft]{threeparttable}

\newcommand{\be}{\begin{equation}}    
\newcommand{\ee}{\end{equation}}

\newcommand{\Enstatite}{{\rm MgSiO$_3$}}
\newcommand{\Forsterite}{{\rm Mg$_2$SiO$_4$}}
\newcommand{\Magnetite}{{\rm Fe$_3$O$_4$}}
\newcommand{\Silica}{{\rm SiO$_2$}}
\newcommand{\Alumina}{{\rm Al$_2$O$_3$}}

\usepackage{etoolbox}
\makeatletter
\patchcmd\@combinedblfloats{\box\@outputbox}{\unvbox\@outputbox}{}{%
   \errmessage{\noexpand\@combinedblfloats could not be patched}%
}%
\makeatother

\title[SN dust yields: metallicity, rotation, and fallback]{Supernova dust yields: the role of metallicity, rotation, and fallback}
\author[Marassi et al.]{S. Marassi$^{1}$\thanks{E-mail:stefania.marassi@oa-roma.inaf.it}, R. Schneider$^{2,1,3}$, 
M. Limongi$^{1,3}$, A. Chieffi$^{4,5}$, L. Graziani$^{1}$ and S. Bianchi$^6$ \\
$^{1}$INAF/Osservatorio Astronomico di Roma, Via di Frascati 33, 00040 Monte Porzio Catone, Italy \\
$^{2}$Dipartimento di Fisica, ``Sapienza'' Universit$\grave{a}$ di Roma, Piazzale Aldo Moro 5, 00185 Roma, Italy \\
$^{3}$Kavli Institute for the Physics and Mathematics of the Universe, Todai Institutes for Advanced Study, the University of Tokyo, 
Kashiwa, \\ 
Japan 277-8583 (Kavli IPMU, WPI)\\
$^{4}$INAF/IAPS, Via Fosso del Cavaliere 100, 00133 Roma, Italy \\
$^{5}$Monash Centre for Astrophysics (MoCA), School of Mathematical Sciences, Monash University, Victoria 3800, Australia\\
$^{6}$INAF/Osservatorio Astrofisico di Arcetri, Largo E. Fermi 5, 50125 Firenze, Italy}

\begin{document}

\date{28 September 2018}

\pagerange{\pageref{firstpage}--\pageref{lastpage}} \pubyear{2017}

\maketitle
\label{firstpage}

\begin{abstract}
Supernovae (SNe) are considered to have a major role in dust enrichment of high redshift galaxies and, 
due to the short lifetimes of interstellar grains, in dust replenishment of local galaxies.
 Here we explore how SN dust yields depend on 
the mass, metallicity, and rotation rate of the progenitor stars, and on the properties of the explosion. 
To this aim, assuming uniform mixing inside the ejecta, we quantify the dust mass produced by a sample of SN models with 
progenitor masses $13~M_{\odot} \leq M  \leq 120~M_{\odot}$, metallicity $\rm -3 \leq [Fe/H] \leq 0$, 
rotation rate $\rm v_{\rm rot} = 0$ and $300$~km/s, that explode with a fixed energy of $1.2 \times 10^{51}$~erg
(FE models) or with explosion properties calibrated to reproduce the $\rm ^{56}Ni$ - $M$ relation inferred
from SN observations (CE models). We find that rotation favours more efficient dust production, particularly for more
massive, low metallicity stars, but that metallicity and explosion properties have the largest effects on the dust mass and its
composition. In FE models, SNe with $M \leq 20 - 25 ~M_{\odot}$ are more efficient
at forming dust: between 0.1 and 1 $M_\odot$ is formed in a single explosion, with a composition dominated by
silicates, carbon and magnetite grains when $\rm [Fe/H] = 0$, and by  carbon and magnetite grains when $\rm [Fe/H] < 0$. 
In CE models, the ejecta are massive and metal-rich and dust production is more efficient. 
The dust mass increases with $M$ and it is dominated by silicates, at all [Fe/H]. 

\end{abstract}


\section{Introduction}
Dust in astrophysical environments has an important role
as it regulates the physical and chemical conditions in the 
interstellar medium (ISM). Expanding ejecta of core-collapse supernovae (SNe)
are possible sites of dust formation.
Knowledge of the dust mass condensed in SN explosions and 
injected in the ISM is of primary importance for the understanding 
of early dust enrichment in galaxies. 

Infrared (IR) and submillimeter (submm) data obtained using different 
space and ground-based telescopes ({\it Spitzer}, {\it Herschel}, SOFIA, AKARI and
ALMA) have provided strong evidence of dust formation in 
the ejecta of SN remnants in the Milky Way and the Large Magellanic Cloud
\citep{2009MNRAS.394.1307D,2010A&A...518L.138B,2010A&A...518L.139O,
2011Sci...333.1258M,2012ApJ...760...96G,2017MNRAS.465.3309D,2015ApJ...799..158T,2017ApJ...836..129T}. 
Newly formed dust masses in SNe in more distant galaxies have been recently 
inferred through the modeling of the blue-red asymmetries of late-time optical and near-IR line profiles 
(SN1980K and SN1993J, \citealt{2017MNRAS.465.4044B}). 
The masses of cold dust inferred from far-IR and submm observations span a large range of values, from $0.1 \, M_{\odot}$ 
of cool dust in Cas A (\citealt{2010A&A...518L.138B}, but see the recent up-ward revision by \citealt{2017MNRAS.465.3309D} 
who estimate $0.3 - 0.6 \, M_\odot$ of silicate/carbon grains) to $(0.4 - 0.7)\,M_{\sun}$ 
in SN1987A \citep{2011Sci...333.1258M,2015ApJ...800...50M,2014ApJ...782L...2I,2016MNRAS.456.1269B} and the minimum
estimated dust mass of $\gtrsim 0.3 \, M_\odot$ for SNR G54.1+0.3 \citep{2017ApJ...836..129T,2017arXiv170708230R}.
For a discussion on the dependence of dust formation on the progenitor mass and supernova type we refer to
the review by \citet{2011A&ARv..19...43G}.

Theoretical models have attempted to predict the amount of freshly formed dust in SN ejecta 
adopting nucleation theory
\citep{1991A&A...249..474K,2001MNRAS.325..726T,2003ApJ...598..785N,2010ApJ...713..356N,
2004MNRAS.351.1379S,2007MNRAS.378..973B,Marassi2015,2016ApJ...817..134L} or a
kinetic approach \citep{2009ApJ...703..642C,2010ApJ...713....1C}, including a description of grain 
growth \citep{2013ApJ...776..107S,2015A&A...575A..95S} and coagulation \citep{2018MNRAS.480.5580S}.
We refer the interested reader to \citet{Marassi2015} for a 
discussion of the differences between the two approches.

Classical nucleation theory (CNT) has proven to be suitable 
to follow dust condensation in SN ejecta \citep{2013ApJ...776...24N,2015ApJ...811L..39N}, 
despite its dependence on parameters, such as the sticking coefficient or the minimum 
number of monomers forming the first seed nuclei \citep{2007MNRAS.378..973B}.
Grids of SN dust yields have been computed using CNT starting from different
set of progenitors and SN explosion models. \citet{2001MNRAS.325..726T} and
\citet{2007MNRAS.378..973B} have considered SN progenitors with masses
in the range $[12 - 35]\,M_{\sun}$ and metallicity values in the range $[0 - 1]~Z_\odot$
using \citet{1995STIN...9622970W} SN models. \citet{2003ApJ...598..785N} built a grid
of dust yields for Pop~III core-collapse SNe with metal-free progenitors in the mass range 
$[13 - 30]\, M_{\sun}$ and for Pop~III pair instability SNe (PISNe) with stellar progenitors
masses of 170 and 200 $M_\odot$. For all these cases they adopted the SN explosion
and nucleosynthesis calculations of \citet{2002ApJ...565..385U}. \citet{2004MNRAS.351.1379S} 
have considered the nucleation of dust in the ejecta of Pop~III PISNe with
progenitor masses in the range $[140 - 260]~M_\odot$ using the grid of PISN models
by \citet{2002ApJ...567..532H}. More recently, \citet{Marassi2015} have used an
improved version of the \citet{2007MNRAS.378..973B} model to estimate the dust yields
of {\it standard} and {\it faint} Pop III SNe, starting from a homogeneous set of 
pre-supernova models with progenitor masses in the range $[13 - 80]~M_{\sun}$ \citep{2012ApJS..199...38L} 
and varying the degree of mixing and fallback during the explosion.

The above studies consistently show that the composition, size distribution and total mass of dust
formed in the ejecta depend on the physical properties of the stellar progenitors (mass and metallicity),
on the explosion energy, and on the ejecta temperature and density profiles 
\citep{2001MNRAS.325..726T,2003ApJ...598..785N,2004MNRAS.351.1379S,2007MNRAS.378..973B,
2009ASPC..414...43K,2009ApJ...703..642C,2010ApJ...713....1C,2010ApJ...713..356N,
2013ApJ...776..107S,2015A&A...575A..95S,Marassi2015}. 
However, none of these studies have explored the effects of stellar rotation
on the dust mass formed in the ejecta. In fact, many stellar evolutionary processes are
affected by rotation and this is reflected in the properties of the star at the pre-SN
stage (see e.g. \citealt{2017ApJ...836...79C,Limongi2017}). 
Finally, the physical properties of the ejecta depend on the supernova type \citep{2010ApJ...713..356N}.
The most commonly observed dusty SNe are core-collapse Type II-P, but there 
are evidence of dust formation also in SN type IIb and Ib \citep{2011A&ARv..19...43G}. 
Conversely, dust grains formed in the ejecta of SNe Ia are almost completely destroyed in the shocked gas before 
being injected into the ISM \citep{2011ApJ...736...45N}. 

The composition and size distribution of grains formed in SN ejecta
a few hundreds days after the explosions are critical information to estimate the dust 
mass that survives the subsequent passage of the reverse shock on timescales
of  $10^3$ and $10^5$~yr and effectively enrich the 
ISM \citep{2007MNRAS.378..973B,2007ApJ...666..955N,2010ApJ...715.1575S,2012ApJ...748...12S,
2014ApJ...794..100M,2014A&A...564A..25B,2016A&A...589A.132B,2016A&A...587A.157B,2016A&A...590A..65M}. 
With the exception of SN 1987A, which is too young for the reverse shock to have affected the dust mass, 
theoretical models suggest that the dust mass currently observed in SN remnants 
is only a fraction (ranging between 60 to 90\%) of the initial dust mass formed in the explosion \citep{2016A&A...587A.157B}, 
in agreement with the observational evidence for ongoing dust destruction in Cas A \citep{2010A&A...518L.138B}. 

The goal of this study is to provide a tabulated set of dust masses 
that takes into account the great diversity of SN events, spanning 
a large range of progenitor masses and metallicity. 
We study the dependence 
of the mass of dust on metallicity, rotation and fallback, to assess 
the relative importance of these processes
on dust formation. Although mass loss can be significant, especially at solar
metallicity, we do not consider dust formation in stellar winds and we restrict
our analysis to dust formation in SN ejecta.
Having this goal in mind, we investigate two 
samples of SN models: a {\itshape fixed} energy sample (hereafter FE models) 
and a {\itshape calibrated} explosion sample (hereafter CE models). The 
first sample is made by SN progenitor masses in the range 
$[13 - 120]~\,M_{\odot}$ that explode with a fixed energy of
$1.2\times 10^{51}$ erg \citep{Limongi2018}. 
This is divided in two sub-data set that differ for the adopted 
rotation degree: non-rotating (NR) and rotating (ROT) progenitors 
models with initial equatorial rotational velocities of $\rm v=0$  and $\rm v=300$ km~s$^{-1}$, 
respectively. These values are meant to provide a mimum and maximum 
estimate of the impact of rotation as they bracket the bulk of the observed rotation velocities
in the Milky Way \citep{2006A&A...457..265D}, and in 
SMC/LMC \citep{2008A&A...479..541H, 2017A&A...600A..81R}.
In turn, each sub-data set contains four different 
classes of progenitor metallicity: class 0 ([Fe/H]=0), 
-1 ([Fe/H]=-1), -2 ([Fe/H]=-2), and -3 ([Fe/H]=-3). 
In the second sample, CE models 
have the same structure of FE models but their explosion energy is not fixed 
a priori. Rather, its value is calibrated 
requiring that the exploding SNe eject the entire stellar mantle \citep{Limongi2017}.

\begin{figure*}
\includegraphics[width=9.55cm]{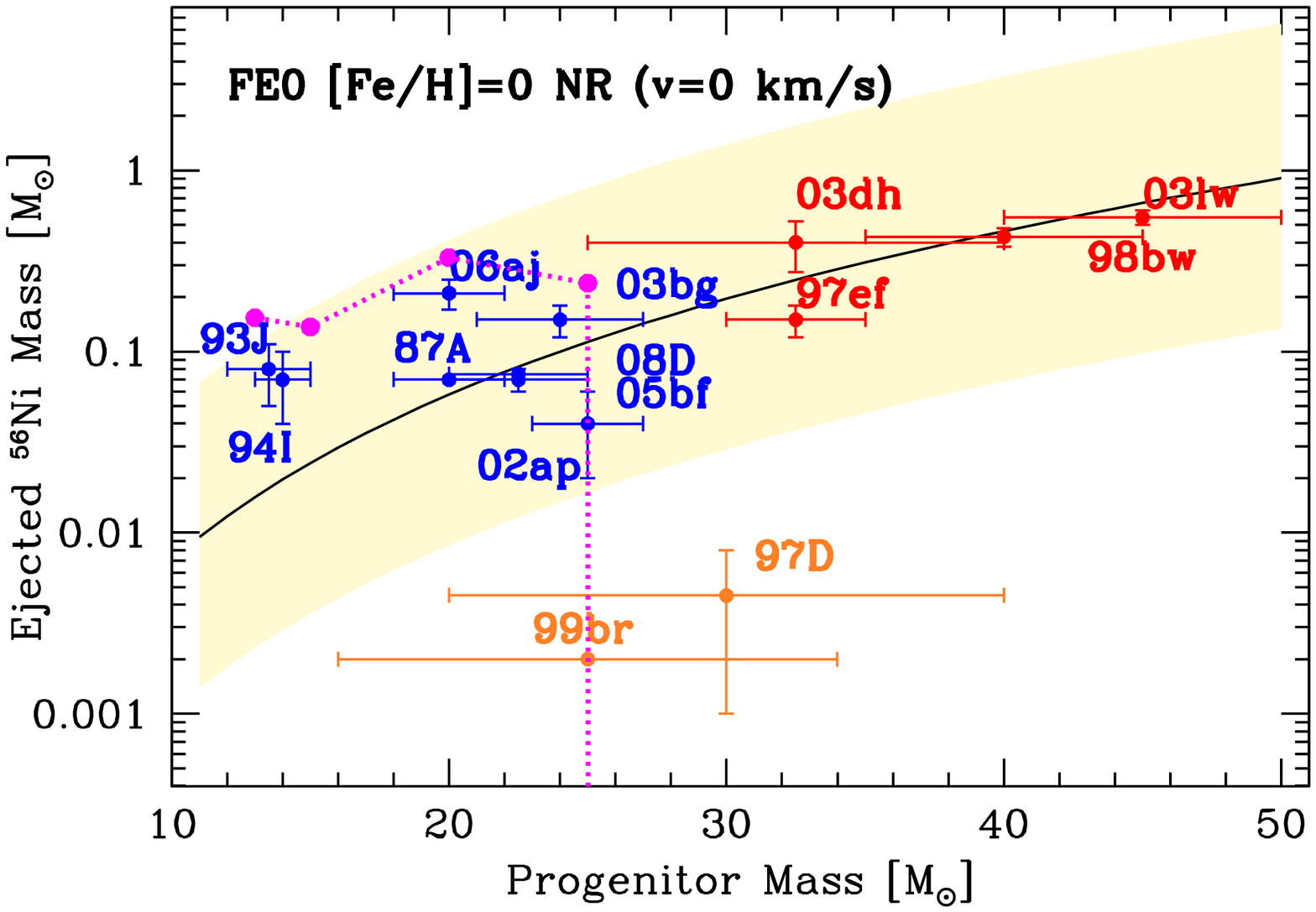}
\hspace{-1.6cm}
\includegraphics[width=9.55cm]{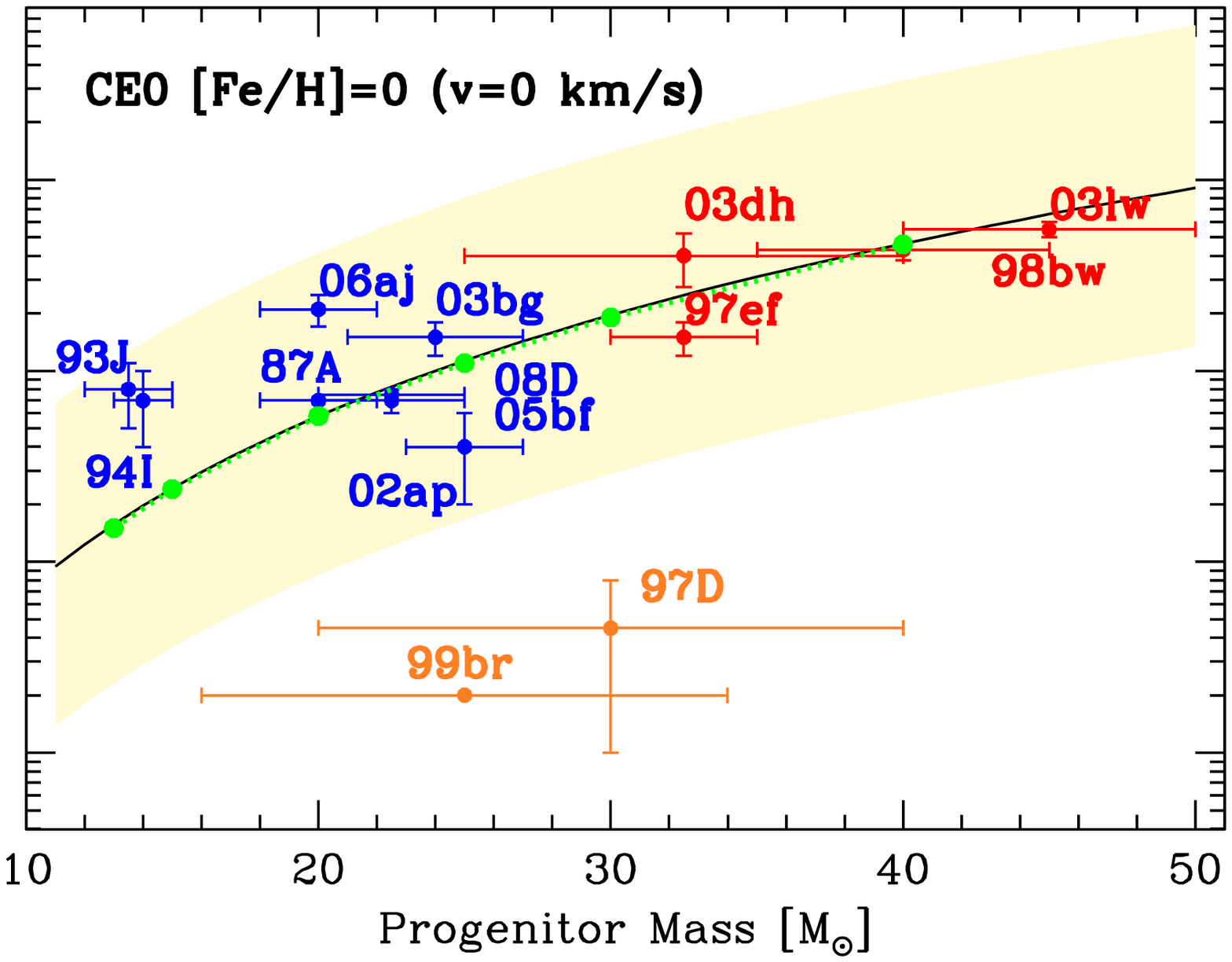}
\caption{Ejected $\rm ^{56}$Ni mass as a function of the 
main-sequence mass of the progenitors for several supernovae (normal SNe in blue, hypernovae in red and faint SNe in orange).
Data are from Nomoto et al.~(2013). The black curve is the best fit to the
data. {\it Left panel}: the dashed magenta line with points shows the prediction
of FE, solar metallicity, non-rotating models. 
{\it Right panel}: the green solid line with points show the prediction
of CE, solar metallicity, non-rotating models after the calibration procedure.}
\label{Nomoto}
\end{figure*}
With this choice, we can perform a {\it parametric} 
study to investigate how metallicity, rotation, and fallback impact ({\it i})
the nucleosynthetic output of the explosion, and ({\it ii}) the total mass, size and composition of 
dust formed in the ejecta.  All the SN dust yields
are available upon request.

The paper is organized as follows. In Section 2 we describe the two samples 
of FE and CE SN models. For each of these, in Section 3 we discuss the
most important properties, such as the masses of the stellar remnants
and the ejecta metal composition, as a function of rotation and metallicity. 
In Section 4 we briefly recall the main features of the adopted dust formation 
model \citep{2007MNRAS.378..973B,Marassi2015}. In Section 5 we discuss the dust 
masses obtained for the FE and CE samples. Finally, in Section 6 we draw our main conclusions. 
\section{Calibration of SN models}
\begin{figure*}
\vspace{\baselineskip}
\hspace{-1.0cm}
\includegraphics[width=5.90cm]{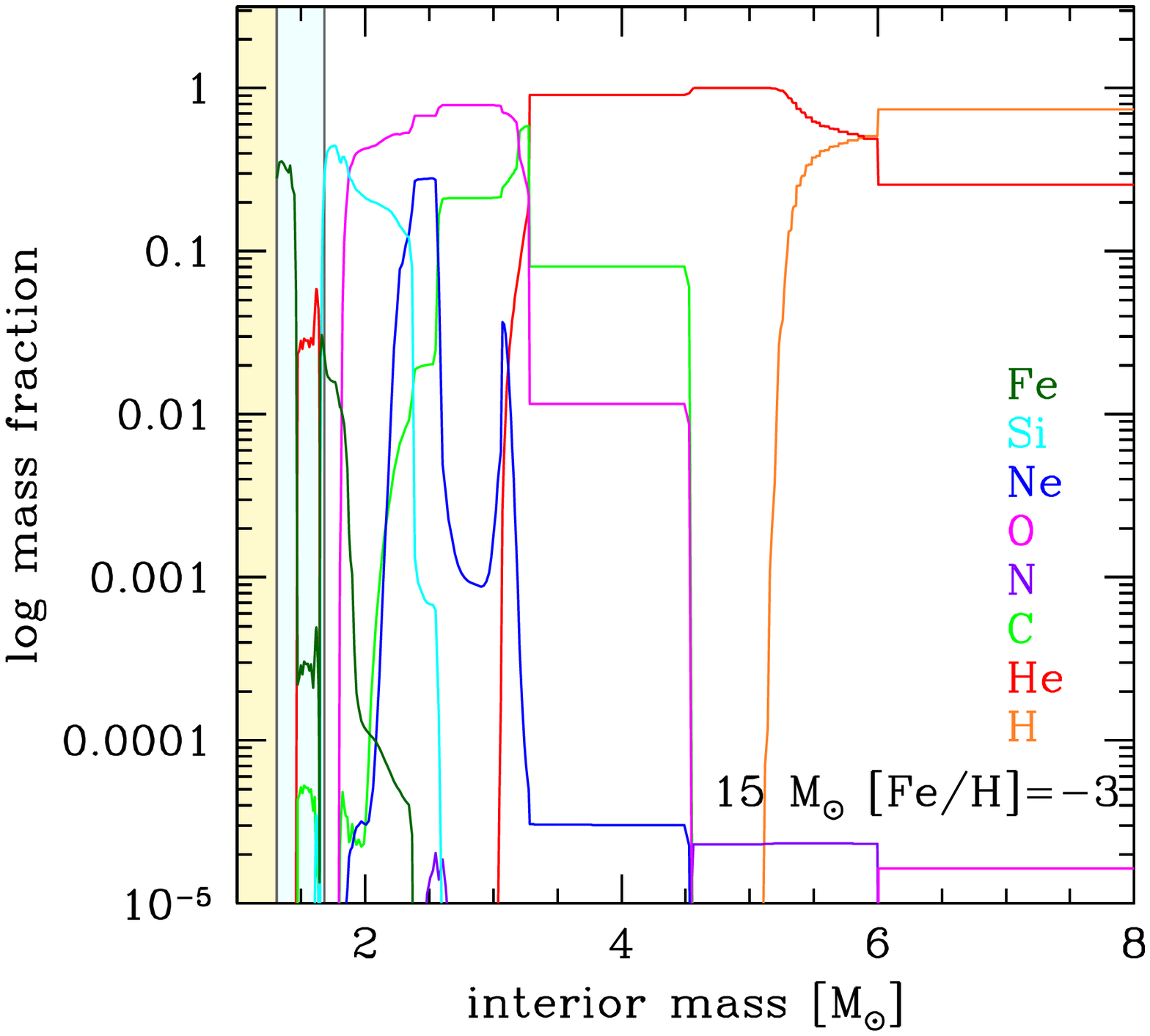}
\hspace{-1.85cm}
\includegraphics[width=5.90cm]{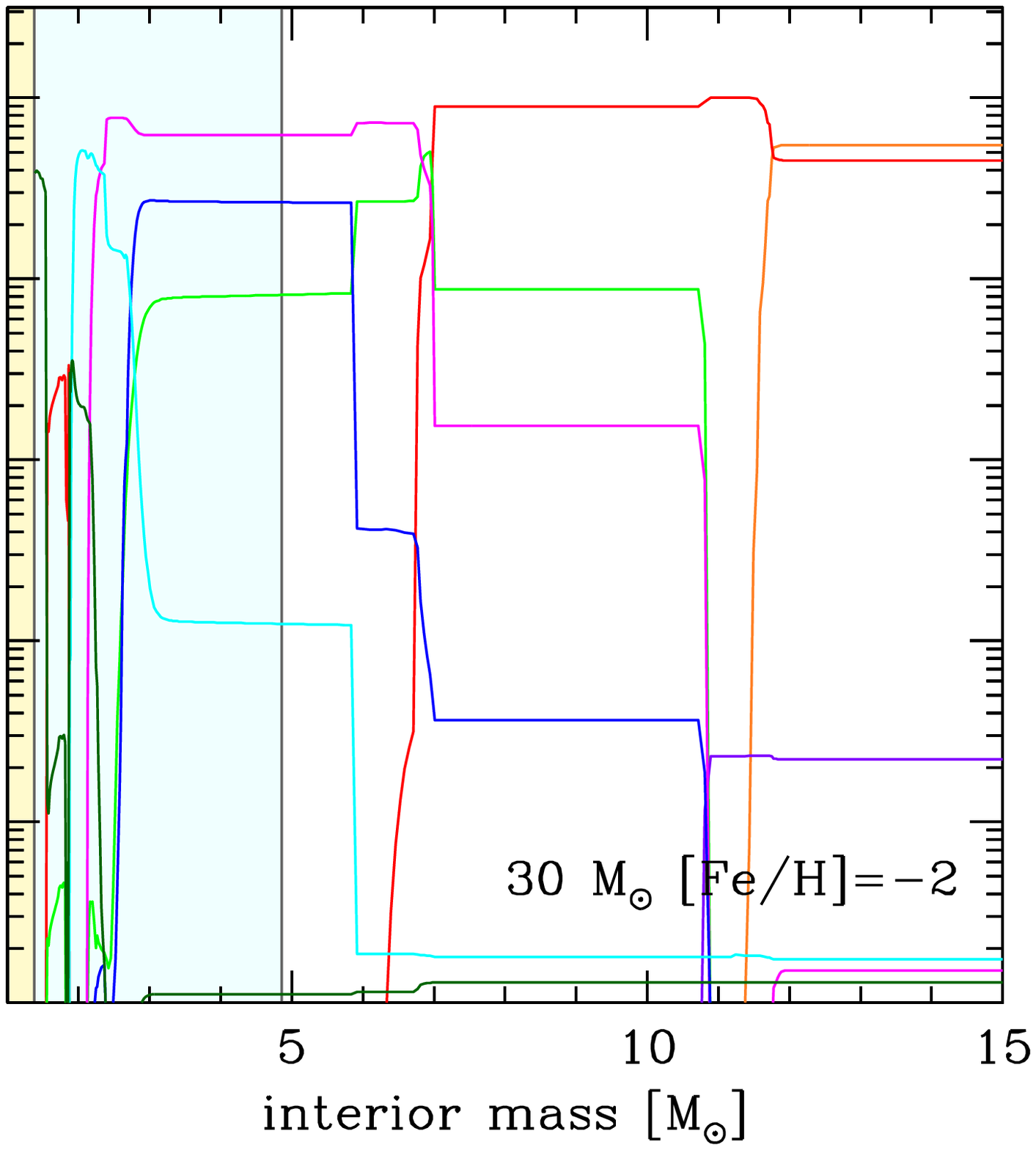}
\hspace{-1.85cm}
\includegraphics[width=5.90cm]{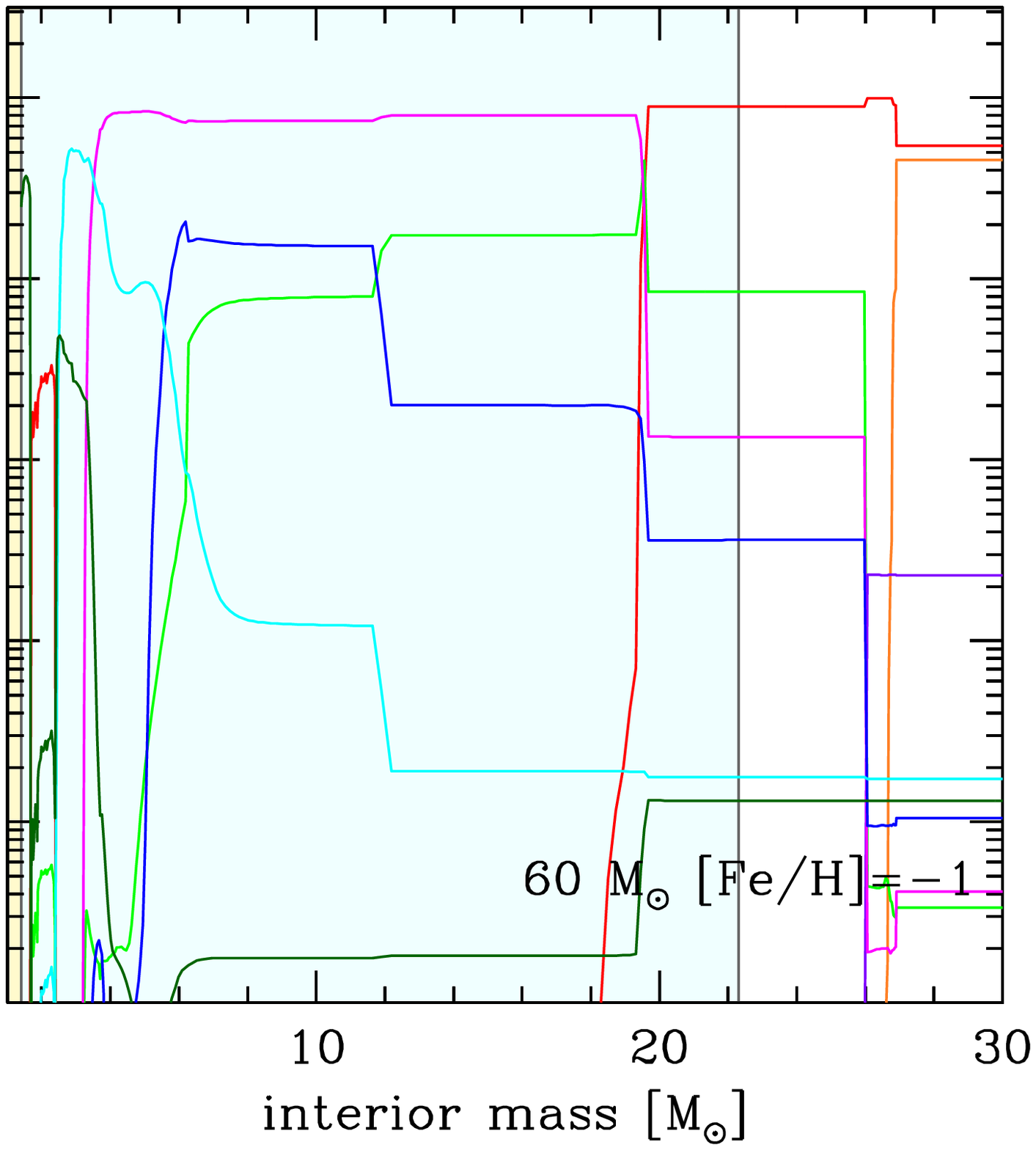}
\hspace{-1.85cm}
\includegraphics[width=5.90cm]{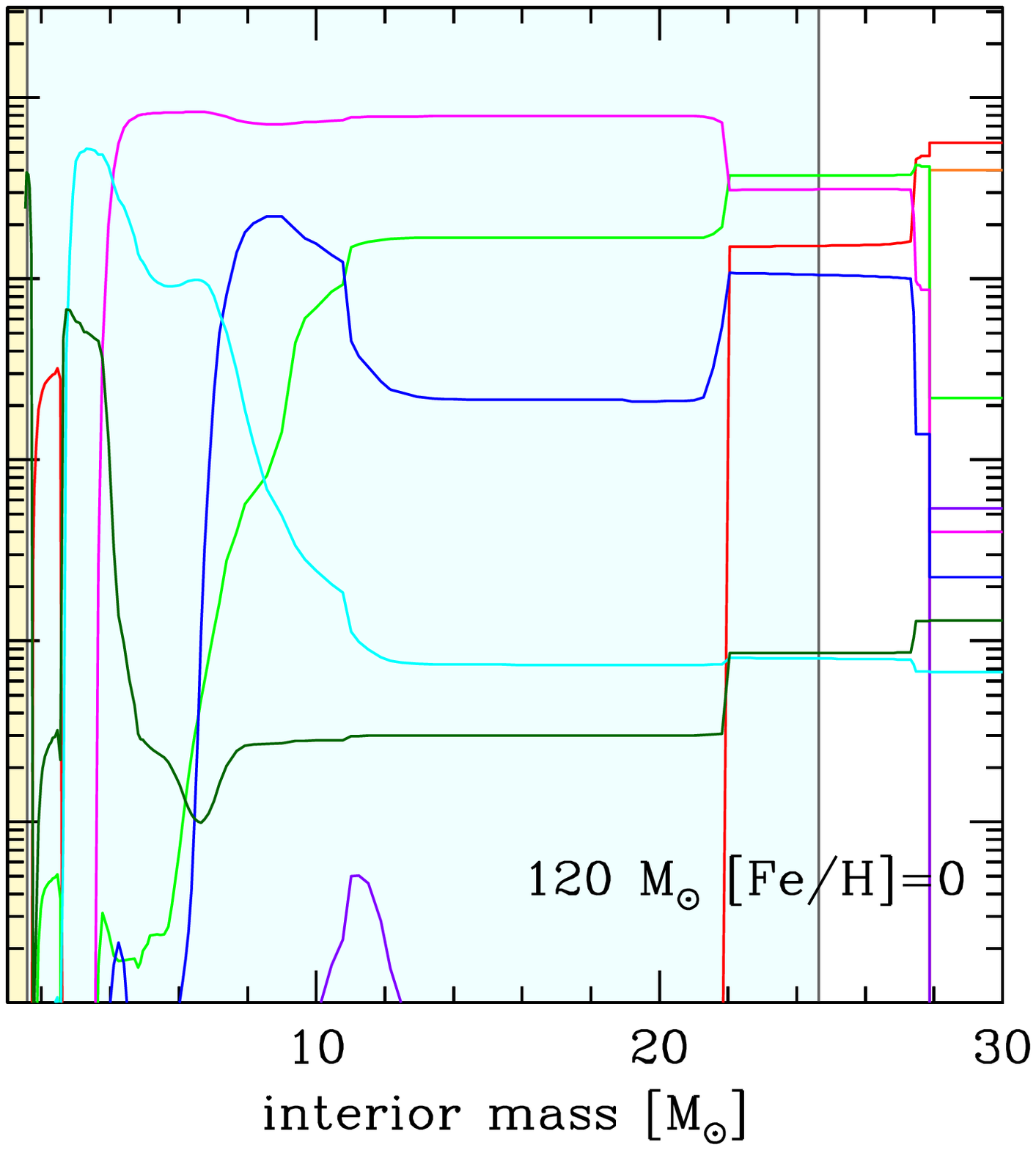}
\caption{Pre-supernova chemical structure for four selected non rotating models. From left to right:  $\rm 15 \, M_{\odot}$  with $\rm [Fe/H] = -3$, $\rm 30 \, M_{\odot}$ with $\rm [Fe/H]=-2$, $\rm 60 \, M_{\odot}$ with $\rm [Fe/H]=-1$, and $\rm 120\, M_{\odot}$ with $\rm [Fe/H]=0$.  Different colored shaded regions
indicate the remnant mass for CE  (yellow) and FE (azure) models.
A colour version of the figure is available online.}
\label{prog_cut}
\end{figure*}
SNe show large differences in physical properties, and their  final outcome depends on
the mass, metallicity, rotation and mass-loss rate of their progenitors 
(see e.g. \citealt{2009ARA&A..47...63S,Limongi2017} and references therein).
One possible choice to construct a reference sample is to assign
the explosion energy and vary the progenitor mass and metallicity, as done
by \citet{1995STIN...9622970W}. On the other hand, the observed SNe exhibit a range of
explosion energies. Fig.~\ref{Nomoto} shows a collection of data
where the main-sequence progenitor mass and the $\rm ^{56}Ni$ mass 
have been estimated comparing theoretical and observed light curves 
\citep{2013ARA&A..51..457N}. The data span a large variety 
of SNe, ranging from hypernovae (red data points) to faint supernovae (orange 
data points). As an attempt to account for this diversity of objects 
we decided to adopt two different SN reference samples.

The first sample comprises SN progenitors with masses in the range 
$[13 - 120]~M_{\odot}$ which explode with a fixed energy of $1.2\times 10^{51}$ ergs
(FE models). This sample is divided in two additional sub-set: 
non-rotating (NR) models ($\rm v=0$) and rotating (ROT) models ($\rm v= 300 $ km s$^{-1}$).
Depending on the SN progenitor metallicity, each sub-set contains four different classes:
class FE0 (FE, [Fe/H] = 0), FE-1 (FE, [Fe/H ]= -1), FE-2 (FE, [Fe/H] = -2),  and FE-3 (FE, [Fe/H] = -3). 
The FE sample allows us to understand how metallicity and rotation affect both the
fallback (hence the mass of the stellar remnant) and the physical properties of the ejecta, 
such as its mass and chemical composition. 

In the left panel of Fig.~\ref{Nomoto} we show the $\rm ^{56}Ni$ 
mass ejected by solar FE non-rotating models (FE0-NR SNe, dashed magenta line with points): in these 
SN models, the ejection of $\rm ^{56}Ni$ occurs only in low mass progenitors. 
These models are not able to reproduce the larger $\rm ^{56}Ni$ masses ejected by
hypernovae with $\rm  > 30~M_\odot$ progenitors.
Rotation does not alter this conclusion because in solar FE rotating 
models (FE0-ROT SNe) only the $\rm 13~M_{\odot}$ and $\rm 15~M_{\odot}$ progenitor models eject a non zero 
$\rm ^{56}Ni$ mass (see Table 5). Besides, observations show that 
the same SN types do not eject the same amount of $\rm ^{56}Ni$ mass 
\citep{2003ApJ...582..905H,2013ARA&A..51..457N}. As an example, the blue data points
encompass a factor of ten in nichel masses (shaded yellow region in Fig.~\ref{Nomoto}, 
\citealt{2003ApJ...582..905H}). 

For these reasons, we considered a second sample of SN models  
which spans the same range of progenitor masses of FE models, $[13 - 120]~M_{\odot}$, but the
properties of the explosions are calibrated to reproduce the amount 
of $\rm ^{56}Ni$ obtained from the best fit to the observations, as shown by 
the black line in the right panel of Fig.~\ref{Nomoto} (CE models). 
1D simulations, in the framework of the kinetic bomb, require that the initial velocity is tuned in some way to get a succesful explosion, e.g. to obtain a given value for the final kinetic energy.
In CE progenitor models, on the other hand, the initial velocity is taken as the minimum 
initial velocity which provides the ejection of the whole mantle above the 
iron core (see \citealt{2013ApJ...764...21CL} for further details). As explained 
in \citet{2013ApJ...764...21CL}, this approach allows to choose
the mass cut (i.e., the mass coordinate which separates the SN ejecta from the 
compact remnant) a-posteriori and to calibrate the model 
by requiring the ejection of a specific amount of $\rm ^{56}Ni$.
This procedure has been tested and does not alter significantly the final yields 
\citep{2003PASA...20..324C}.
The structure of the CE sample is the same as that of FE models: it is divided in two additional sub-set, 
non-rotating (NR)/rotating (ROT) models, and each sub-set contains four different 
classes, depending on the progenitor metallicity: set CE0 (CE, [Fe/H]=0), set 
CE-1 (CE, [Fe/H]=-1), set CE-2 (CE, [Fe/H]=-2), set CE-3 (CE, [Fe/H]=-3). 
The same observed best fit $^{56}$Ni - progenitor mass relation is adopted to calibrate
models with different metallicity values and rotation rates, due to the limitations of available
observations.

It should be noted that the $\rm ^{56}Ni$ mass and the progenitor mass reported in 
Fig.~\ref{Nomoto} for the observed SN sample have been derived using
theoretical models \citep{2013ARA&A..51..457N}. As an example, in type II-P SNe 
(as e.g SN1987A in Fig.~\ref{Nomoto}), the $\rm ^{56}Ni$ mass is derived from 
the brightness of the SN exponential tails, assuming that all the gamma 
rays due to decay of $\rm ^{56}Co \rightarrow \, ^{56}Fe$ are fully thermalized 
\citep{2003ApJ...582..905H}. The uncertainties/approximations of this approch 
are quantified by the errors on the estimated $\rm ^{56}Ni$ and stellar progenitor masses.

\section{Properties of SN models}
In this section, we discuss some of the most important physical properties of the SN models. 
Both FE and CE samples are simulated with the latest version of the FRANEC code which takes into account 
the effects of rotation and metallicity on the evolution of the star (for more details see  \citealt{2013ApJ...764...21CL}
and \citealt{Limongi2018}).

\citet{2012MNRAS.422...70H} classify core-collapse SN models depending on H and He 
envelope masses in the pre-supernova phase. In particular, they set (i) the minimum H mass for a type II-P 
SN to be $M({\rm H})_{\rm min,SNIIP} \simeq 0.3 \, M_{\odot}$;  (ii) the H mass for a type IIb SN to be
$0.1 M_\odot \lesssim M({\rm H})_{\rm SNIIb} \lesssim 0.3 \, M_{\odot}$; (iii) when the H mass is 
$\lesssim 0.1 \, M_{\odot}$, the SN is classified as a type Ib if $M({\rm He}) \gtrsim 0.1 \,M_{\odot}$ or
as a type Ic if $M({\rm He}) \lesssim 0.1 \,M_{\odot}$. We adopt the same classification here 
(see also  \citealt{2013ApJ...764...21CL} and \citealt{Limongi2017}) and, for each model, the 
inferred SN type is shown in Tables 1 - 12. We find that none of the SN models in our CE and FE 
samples can be classified as a SNIc.This is a consequence of the rather high He mass present 
in the envelope at the time of the explosion (\citealt{Limongi2017}).
The lack of SNIc is irrelevant for dust production as \citet{2009A&A...508..371H} found no 
evidence of dust condensation in the ejecta of the prototypical Ic SN 2007gr and this 
class of SNe is not considered to be an important source of dust
\citep{2011A&ARv..19...43G}. 

The mass of dust formed in the ejecta depends on the metal abundances and 
on their distribution. Fig.~\ref{prog_cut} illustrates the variety of chemical compositions
that characterize the pre-supernova models. The mass fraction profiles for the most abundant
atomic elements are shown for four selected models at different metallicity: 
$15~M_{\odot}$ with $\rm [Fe/H]=-3$, $30~M_{\odot}$ with 
$\rm [Fe/H]=-2$, $60~M_{\odot}$ with $\rm [Fe/H ]=-1$ and $120~M_{\odot}$ 
with $\rm [Fe/H]=0$. In addition, in each panel we also show 
the range of pre-SN mass that collapses
and forms the remnant for FE and CE SN models (azure and yellow shaded regions, respectively). 
For the $15~M_{\odot}$ pre-SN, the remnant mass is always $1.3~M_{\odot}$ 
(azure and yellow regions are superimposed in the right panel).
When the progenitor mass increases, fallback and remnant mass increase in FE 
models, resulting in smaller and more metal-poor ejecta compared 
to the corresponding CE models\footnote{We assume uniform mixing to take place during the earliest phases of ejecta propagation, hence beyond the mass cut, outside the shaded regions.}. As a result, FE and CE models 
place a lower and upper limit on the dust mass that forms in our grid of SN models.
In the following subsections, we discuss the effects 
of metallicity and rotation on FE and CE SN models.

\subsection{FE models}
\begin{figure*}
\vspace{\baselineskip}
\includegraphics[width=8.50cm]{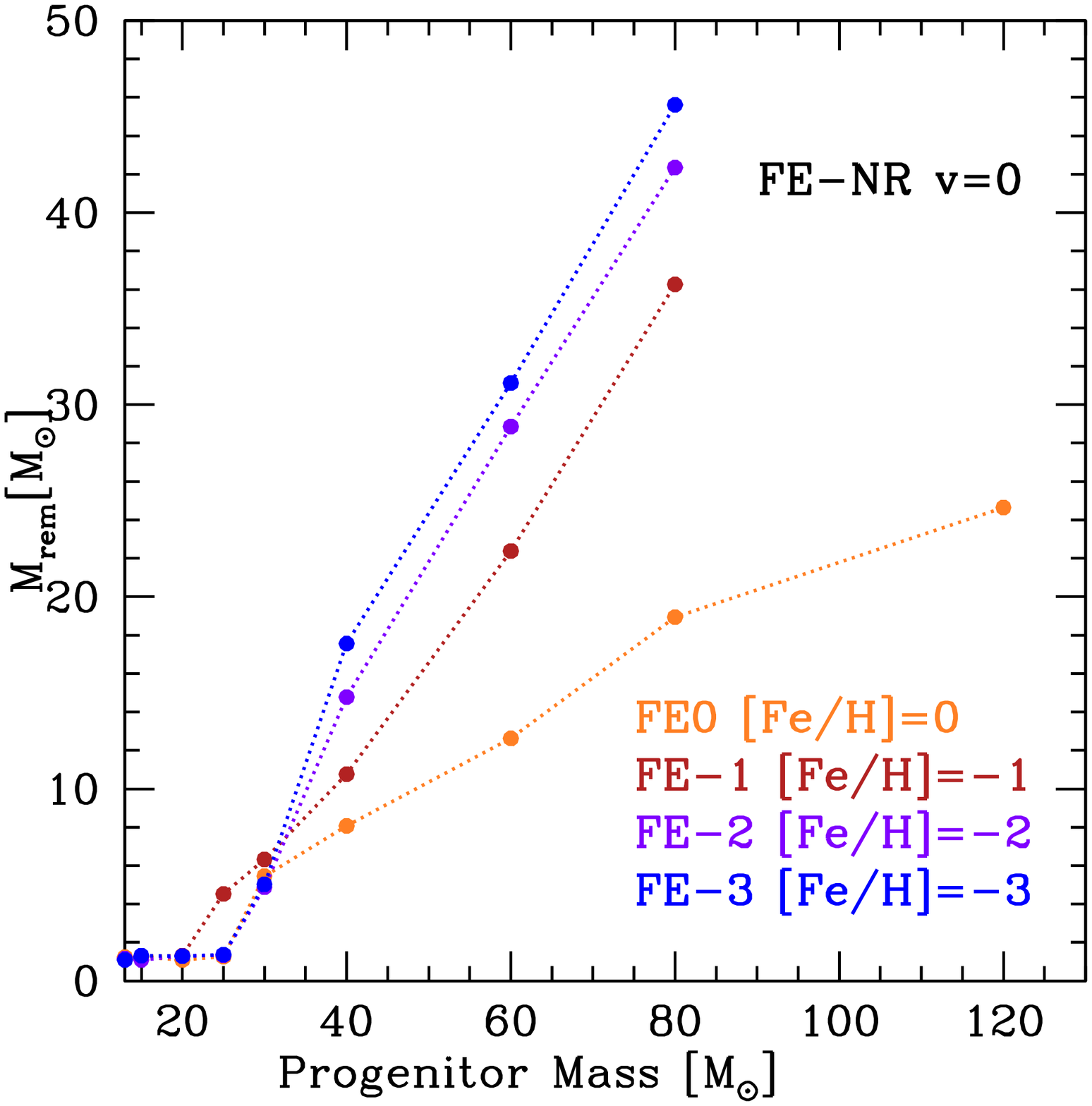}
\includegraphics[width=8.50cm]{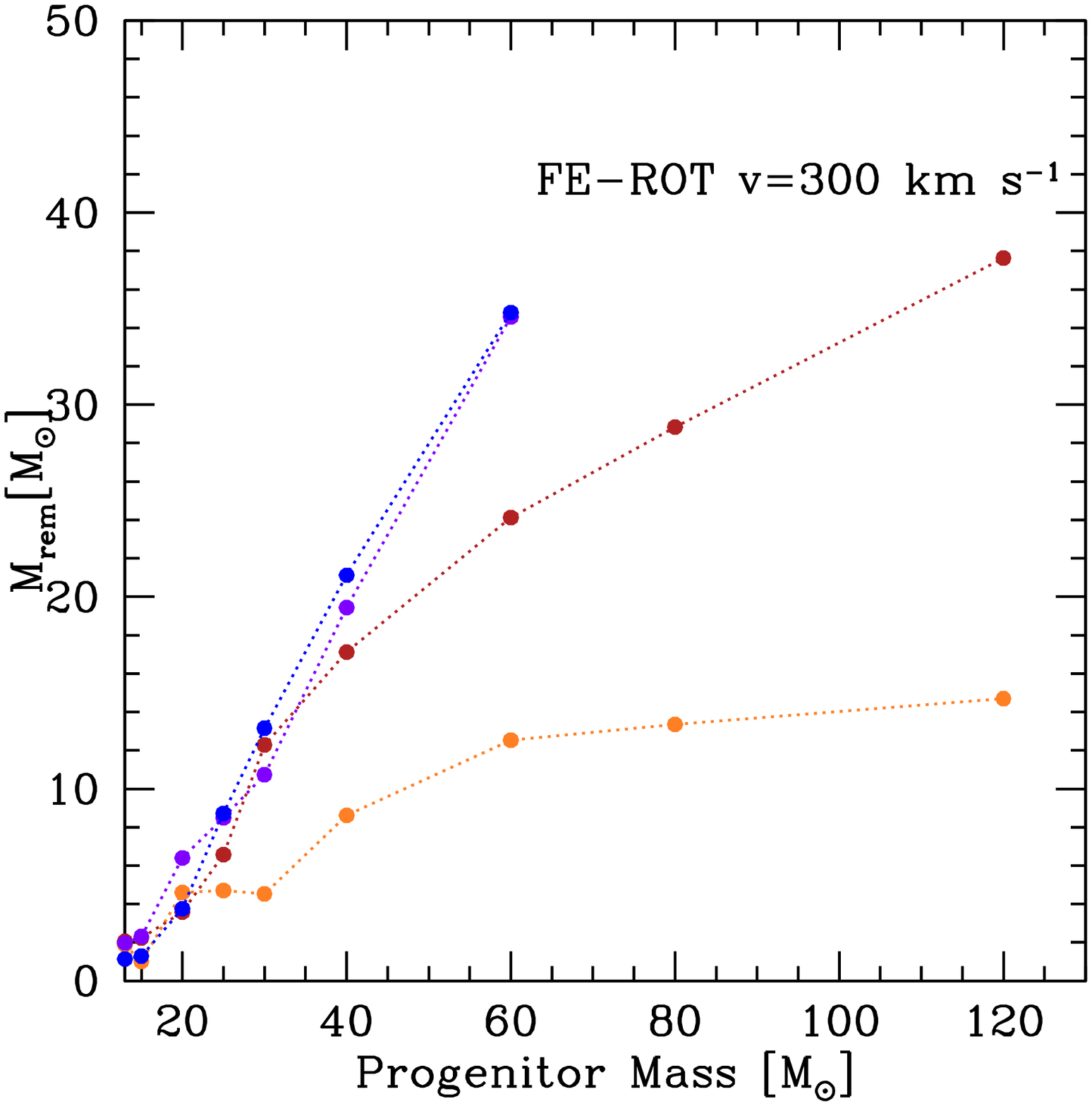}
\includegraphics[width=8.50cm]{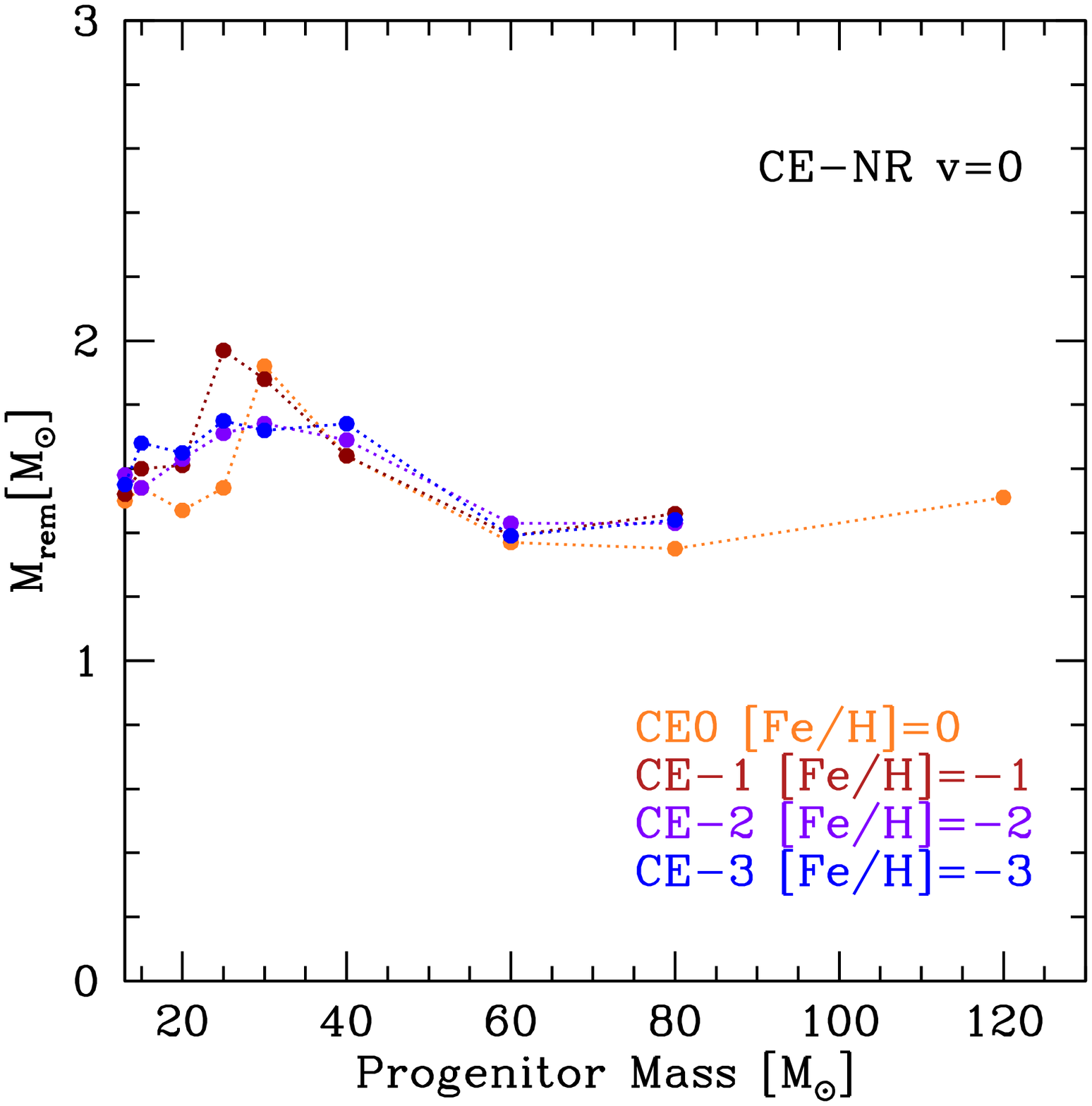}
\includegraphics[width=8.50cm]{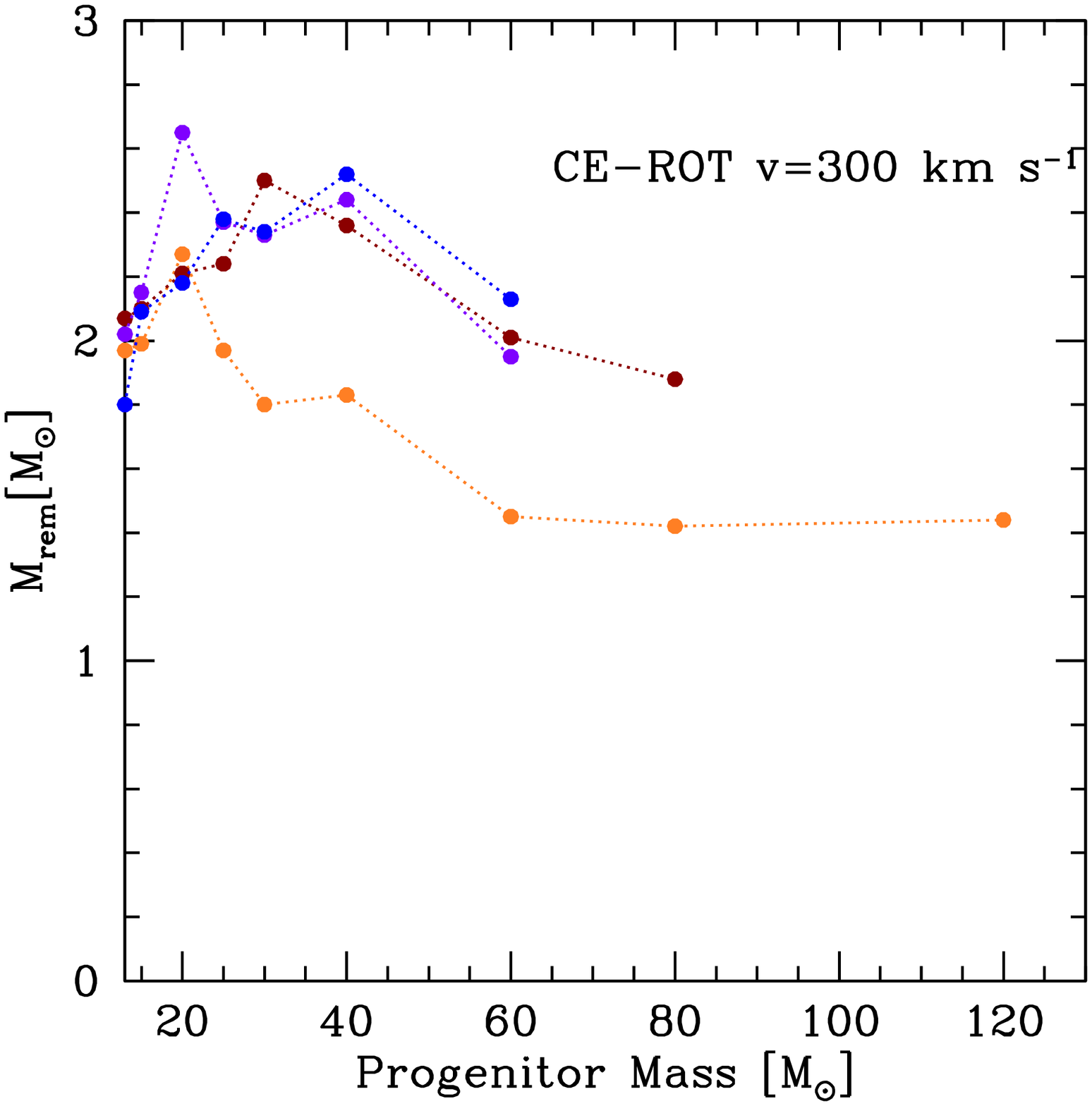}
\caption{The mass of the remnant as a function of the initial stellar progenitor
mass for different metallicity (see the legenda). Upper panels show FE models, 
lower panels show CE models, and left and right panels show non rotating
and rotating models, respectively.}
\label{fig:mass_nr_rot_efix}
\end{figure*}
\begin{figure*}
\vspace{\baselineskip}
\includegraphics[width=8.50cm]{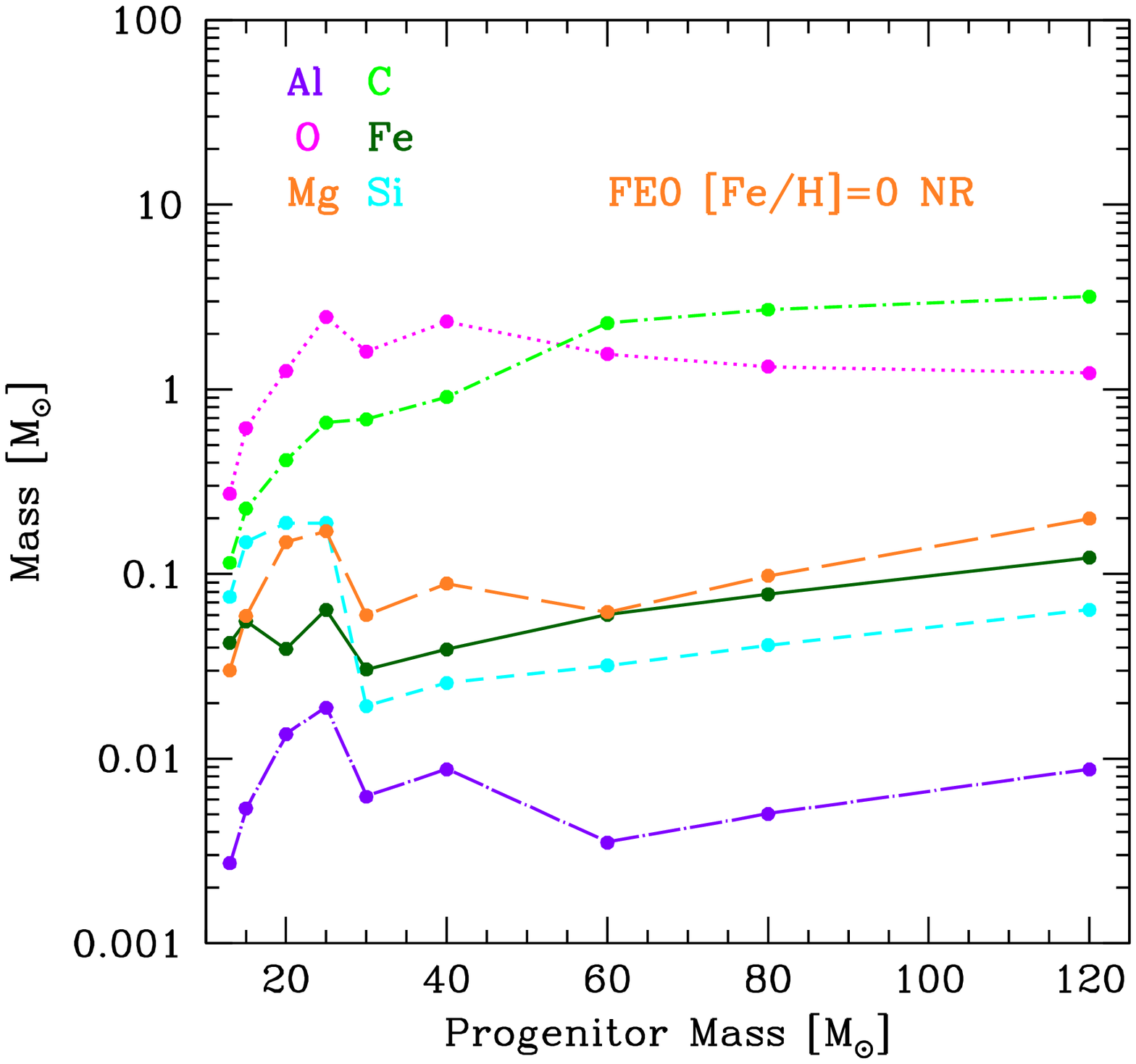}
\includegraphics[width=8.50cm]{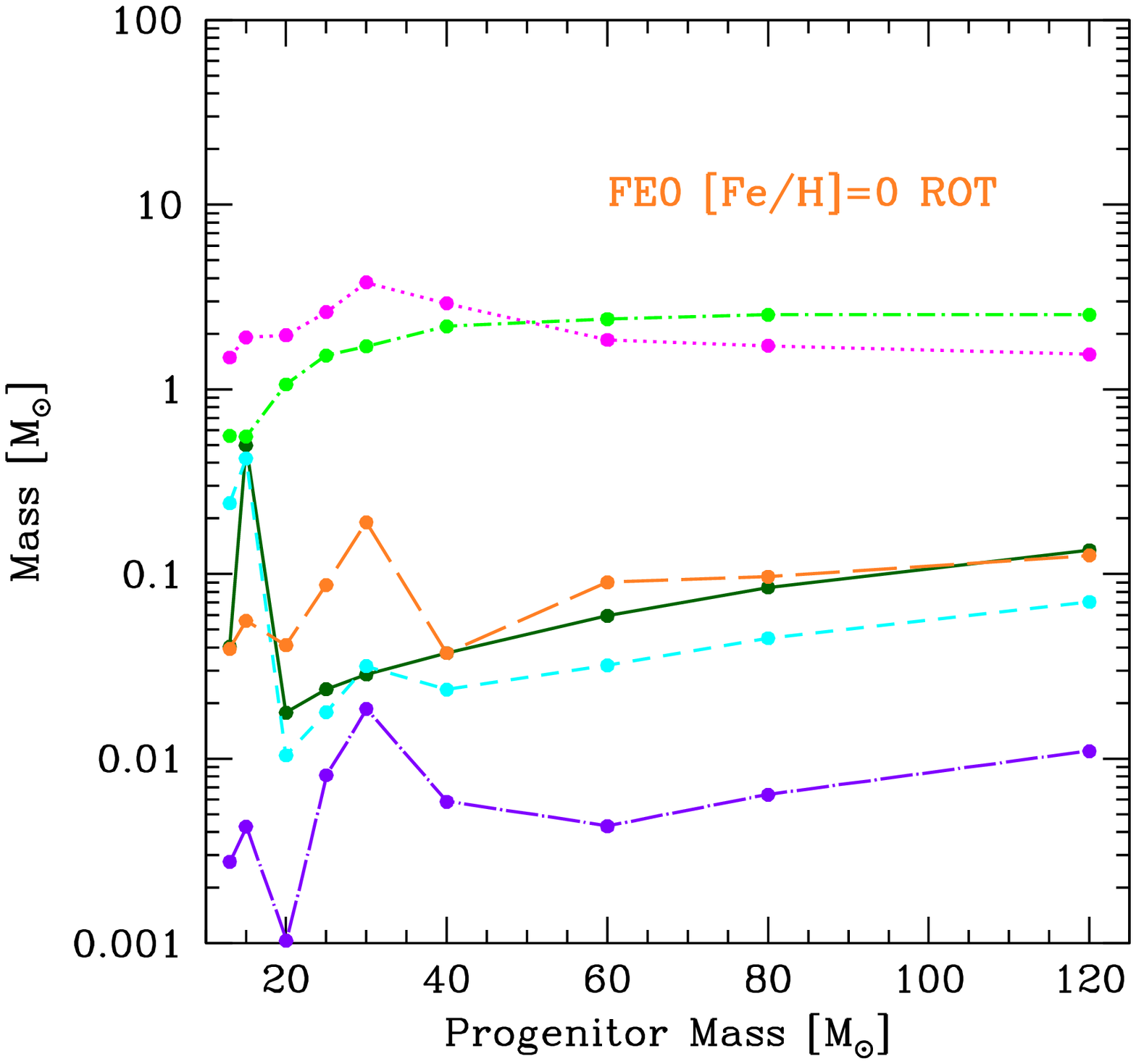}
\includegraphics[width=8.50cm]{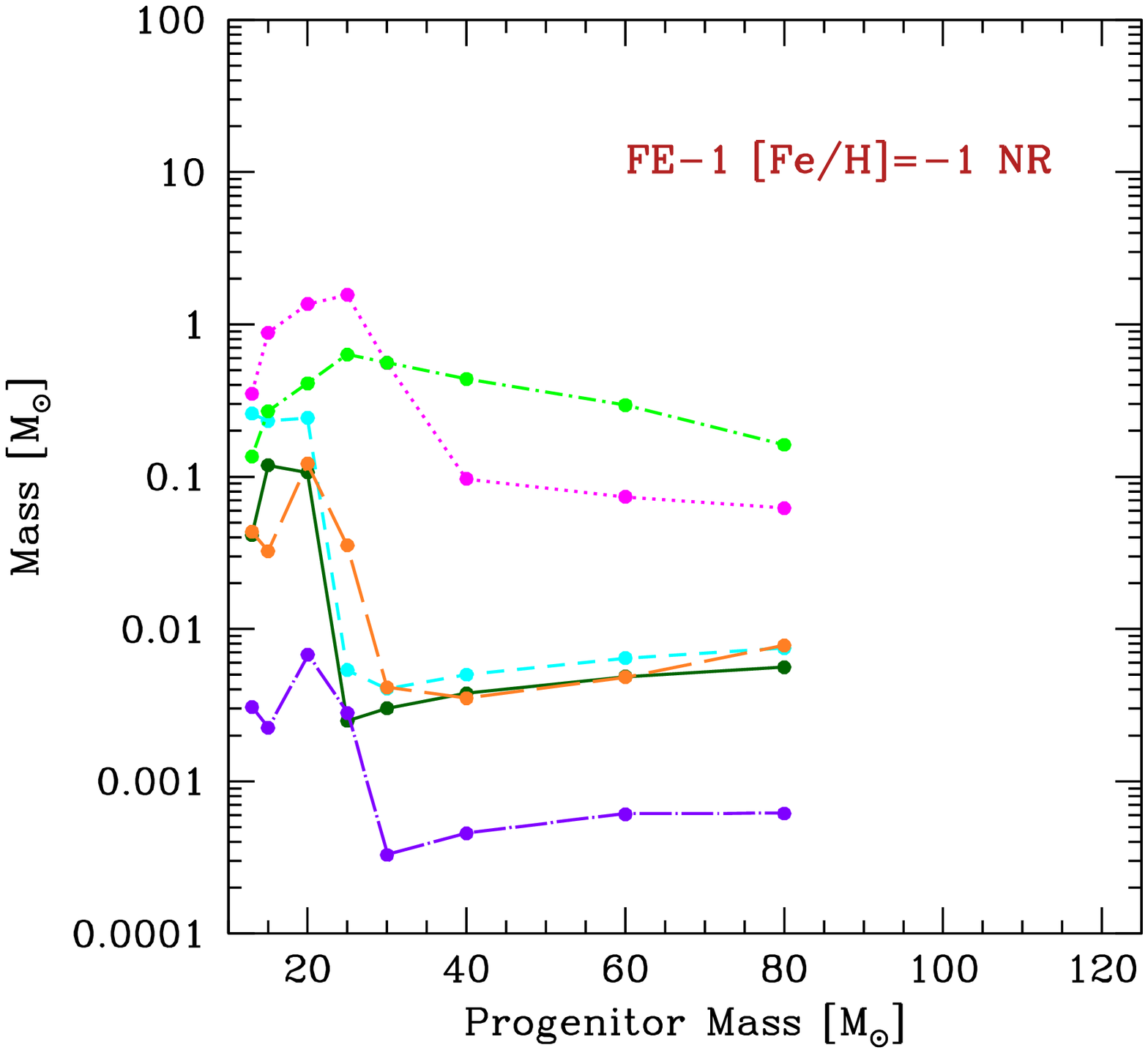}
\includegraphics[width=8.50cm]{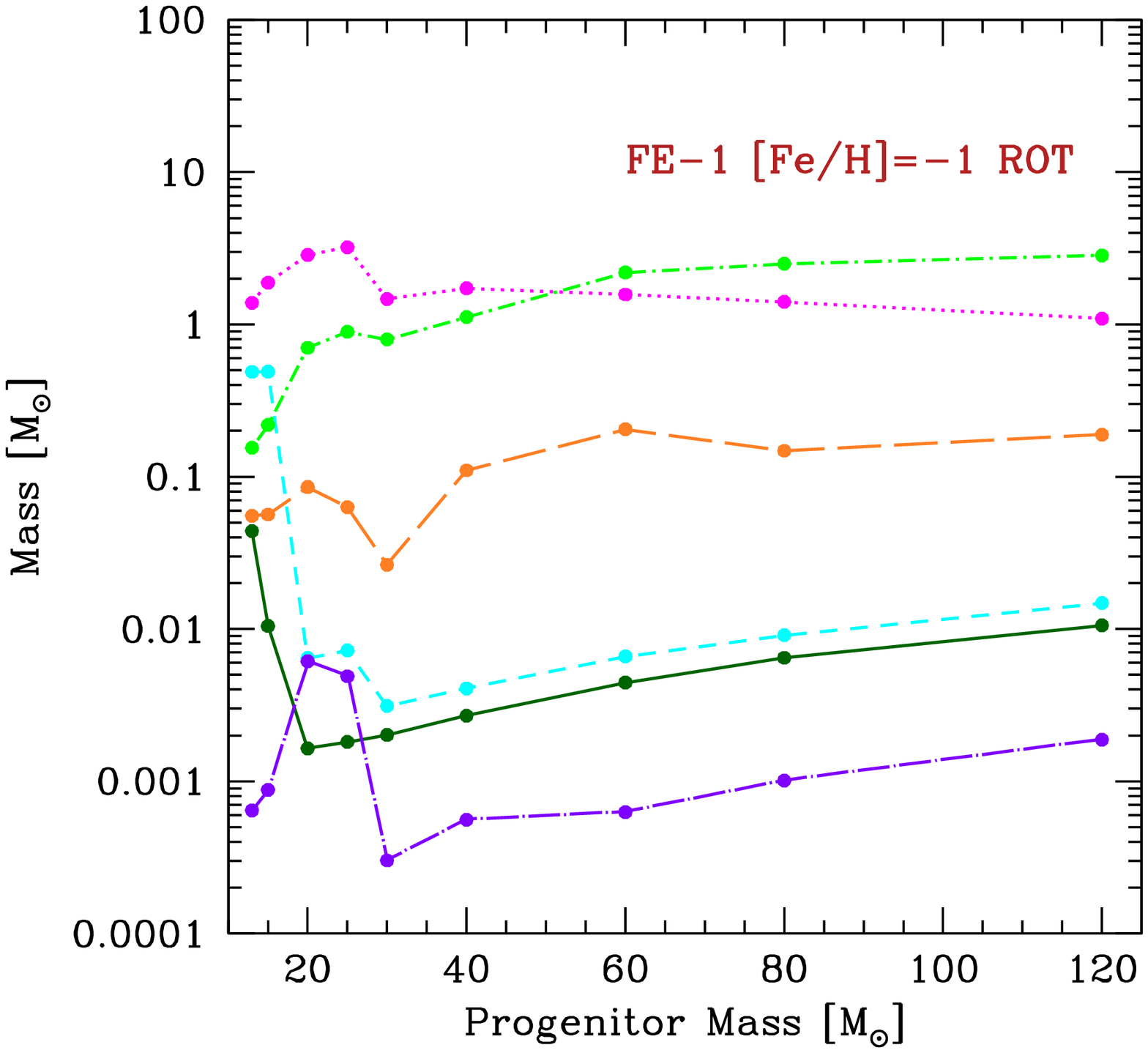}
\caption{Initial C, O, Mg, Si, Al and Fe elemental abundances in the ejecta  (see the legenda)
as a function of the progenitor mass for FE-SN models.
Each pair of panels shows a different initial value of metallicity: set FE0 (top left),
set FE-1 (bottom left).In each pair, left and right panels show non rotating and rotating models.}
\label{Input_metal_efixAB}
\end{figure*}
\begin{figure*}
\vspace{\baselineskip}
\includegraphics[width=8.50cm]{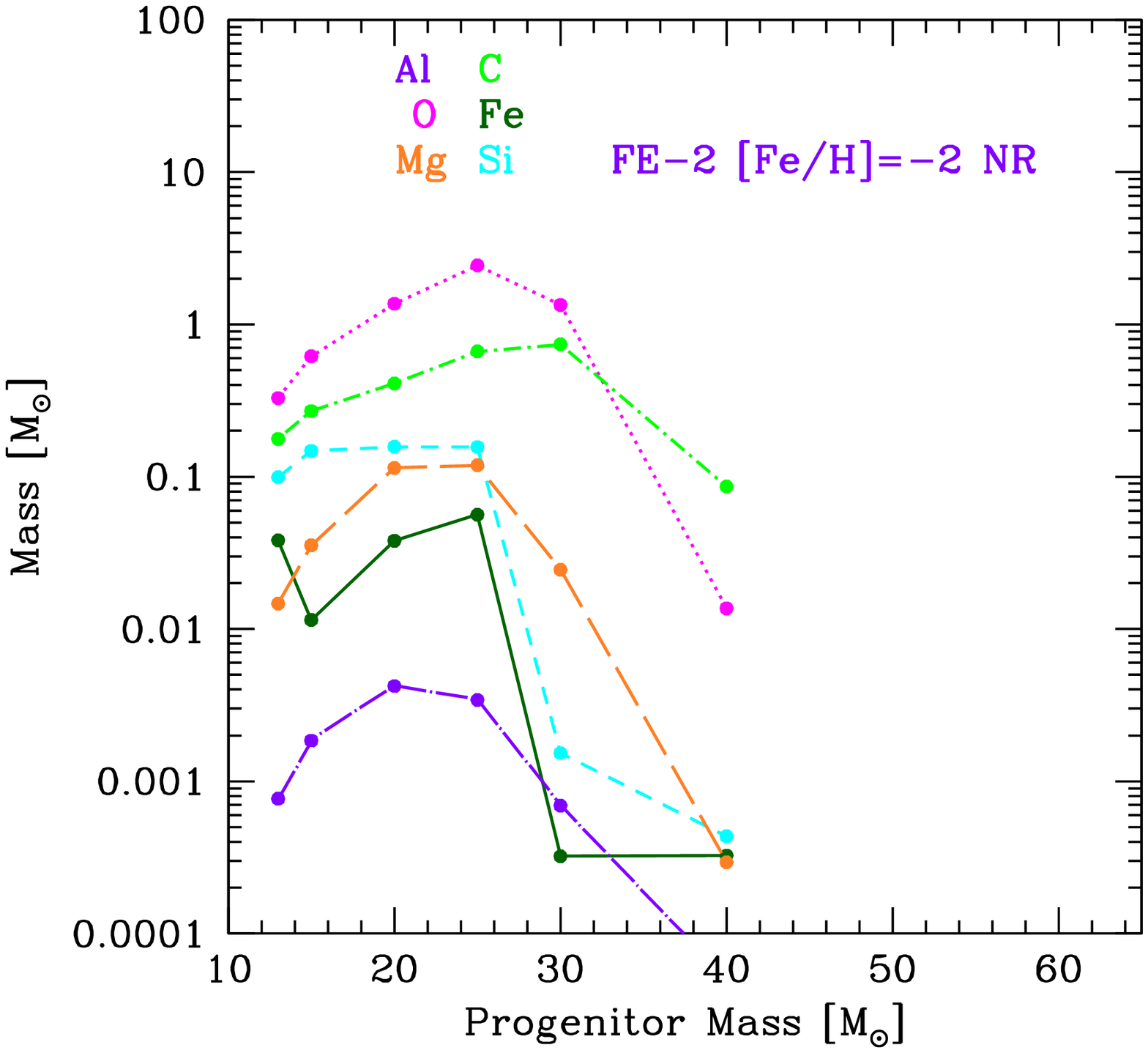}
\includegraphics[width=8.50cm]{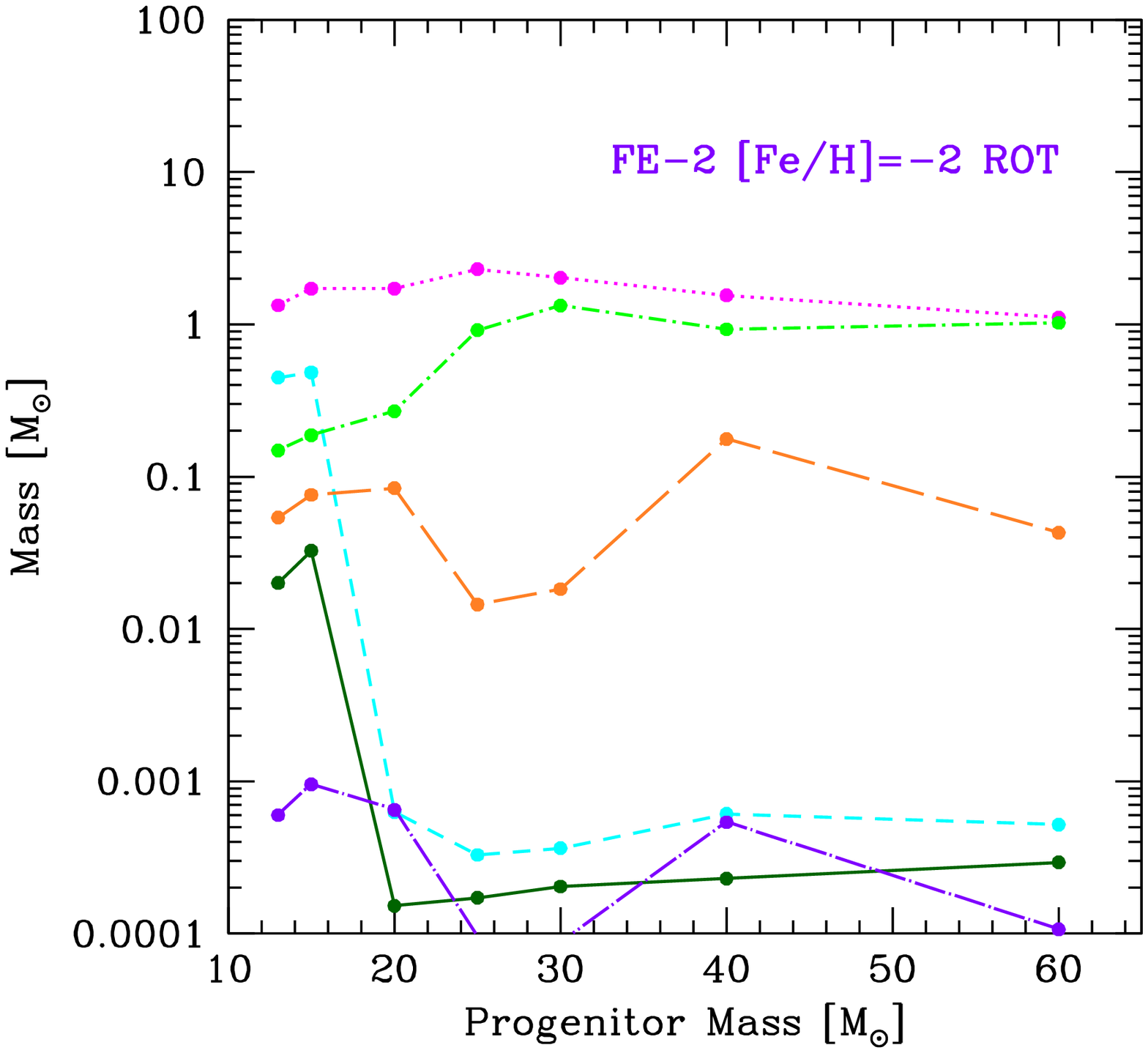}
\includegraphics[width=8.50cm]{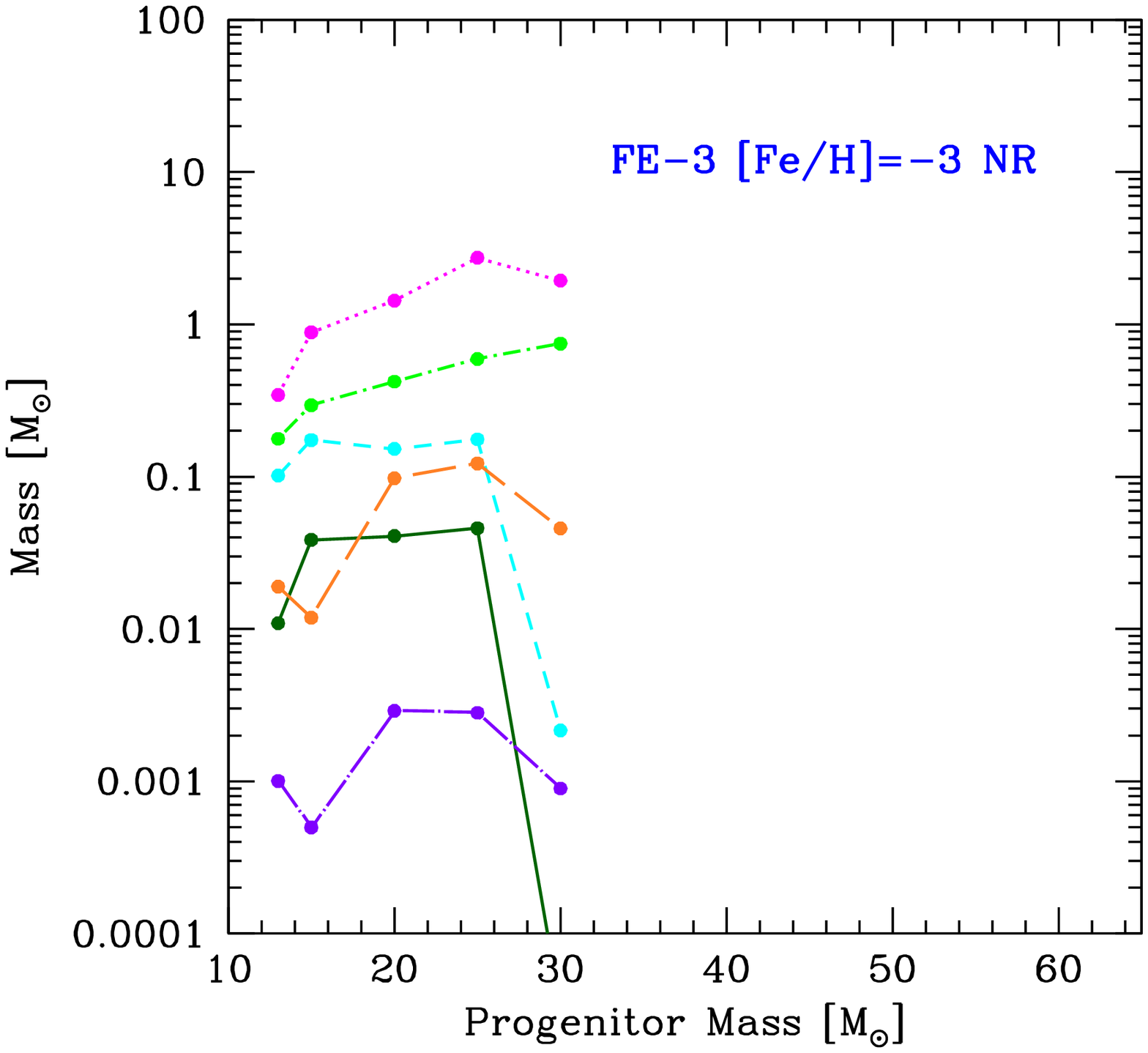}
\includegraphics[width=8.50cm]{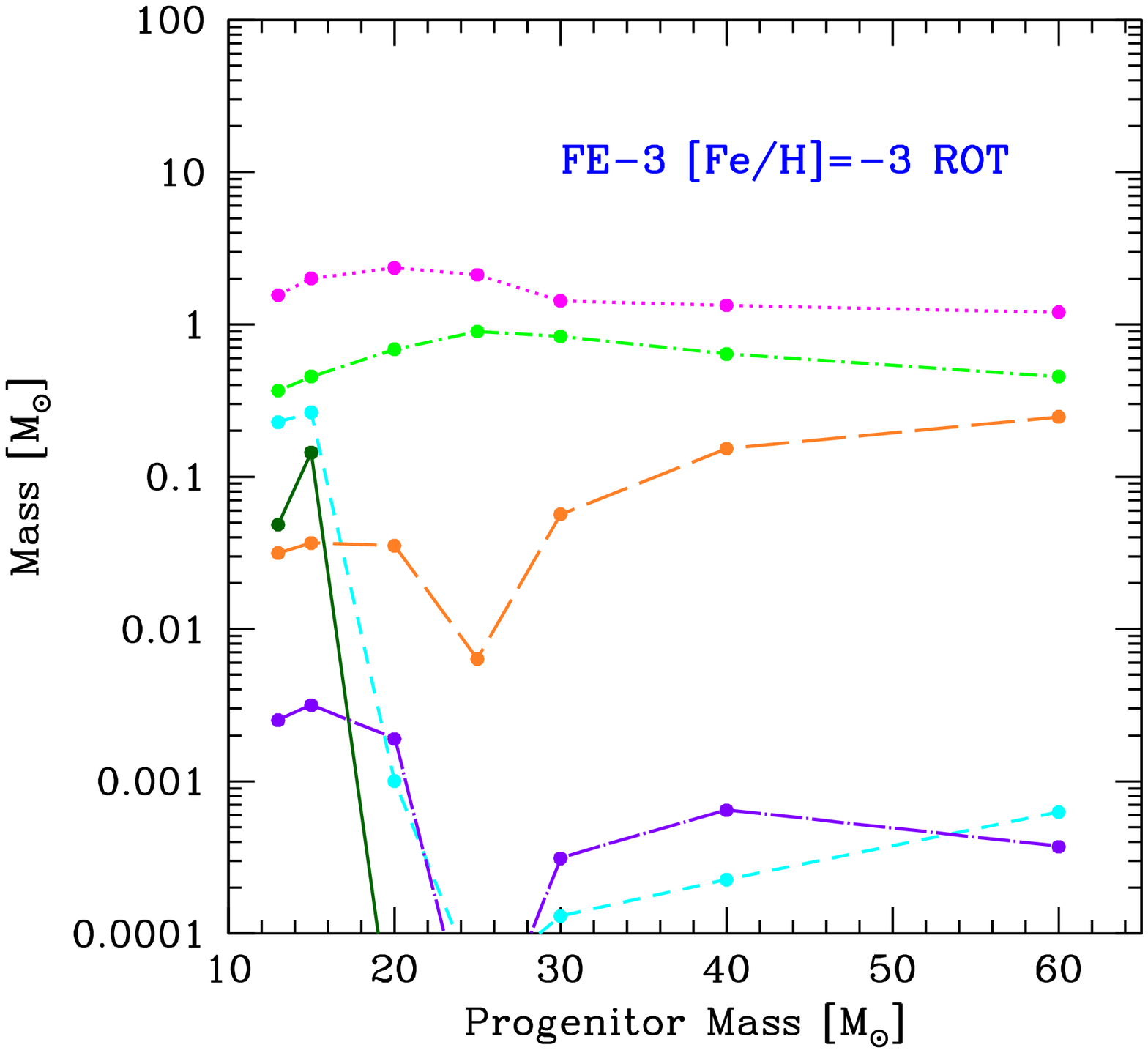}
\caption{Initial C, O, Mg, Si, Al and Fe elemental abundances in the ejecta  (see the legenda)
as a function of the progenitor mass for FE-SN models.
Each pair of panels shows a different initial value of metallicity: set FE-2 (top left),
set FE-3 (bottom left). In each pair, left and right panels show non rotating and rotating models.}
\label{Input_metal_efixCD}
\end{figure*}

We first analize the effects of metallicity on the non-rotating 
FE models. The leftmost panel of Fig.~\ref{fig:mass_nr_rot_efix} shows 
that the remnant mass increases with progenitor mass and with decreasing metallicity.
This is due to the increasing compactness and smaller mass loss
experienced by stars at lower metallicity. As a consequence, not all these models
lead to a successful explosion: stars with $M  > 40~M_{\odot}$ in set FE-2 and 
with $M > 30~M_{\odot}$ in set FE-3 undergo a huge fallback and end their life as a 
failed SN event, forming a black hole. Since these models loose only a small fraction of their 
H-envelope before collapsing to a black hole, they do not contribute to metal and dust enrichment
(we only report their pre-SN and final remnant masses in Tables \ref{FE_C} and \ref{FE_D}).
In addition, when [Fe/H] $\leq -1$ the most massive progenitors, with $M > 80 M_\odot$, enter the 
pulsation pair instability regime \citep{2002ApJ...567..532H} and their final fate cannot be computed 
with precision in the present framework. For this reason, these models are not shown in the figure
and in the corresponding Tables. 

The effect of rotation on the evolution of massive stars is twofold: {\it (i)} rotation driven mixing
leads to more massive CO cores, and {\it (ii)} rotation favors a more efficient mass loss that in turn, in the most
extreme cases, may induce a reduction of the CO core masses\footnote{The CO core is never directly eroded by mass loss.
Rather, mass loss can be strong enough to reduce the He core mass and the star  develops a smaller CO core.}. The interplay between these two different effects
determines the final remnant mass, because (for a fixed explosion energy) this quantity directly depends on the CO core mass at the time of the explosion.

At solar metallicity, the increase of the CO core mass is the dominant effect for stars with $M < 40~M_\odot$.
At larger stellar masses, the more efficient mass loss is the dominant effect.
As a consequence, when compared to non rotating models, the remnant mass of rotating models with [Fe/H] = 0  
increases for stars with $M < 40 M_\odot$ and decreases for stars with $M > 40 M_\odot$.
A similar behaviour is found for models with [Fe/H] = -1.
At lower metallicity, due to the strong reduction of mass loss, rotation always increases the CO core mass. 
Therefore, all rotating models with [Fe/H] $< -1$ lead to more massive remnants than their non rotating counterparts. 
Because of the more massive CO cores, the minimum stellar mass that enters the pulsation pair instability reduces if rotation is taken into account. A comparison between the left and the right panel of Fig. 3 clearly shows such a 
general behavior (see also Tables 1-4).

\begin{figure}
\hspace{-0.80cm}
\includegraphics[width=10.0cm]{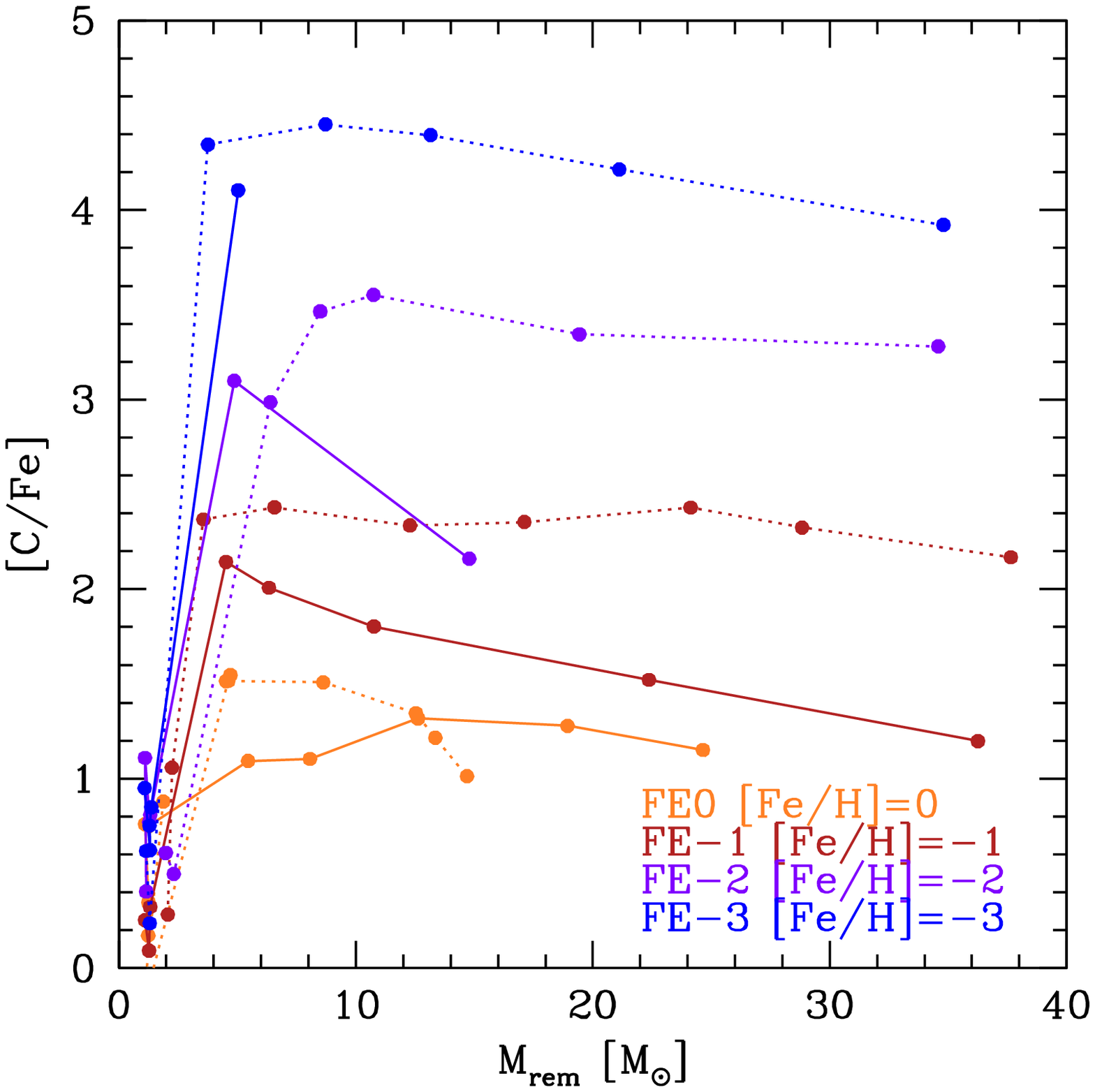}
\caption{Initial [C/Fe] as a function of the remnant mass for non rotating 
(solid lines) and rotating (dotted lines) FE models with different initial metallicity (see the
legenda).} 
\label{CsuFe_efix}
\end{figure}
\begin{figure*}
\vspace{\baselineskip}
\includegraphics[width=8.5cm]{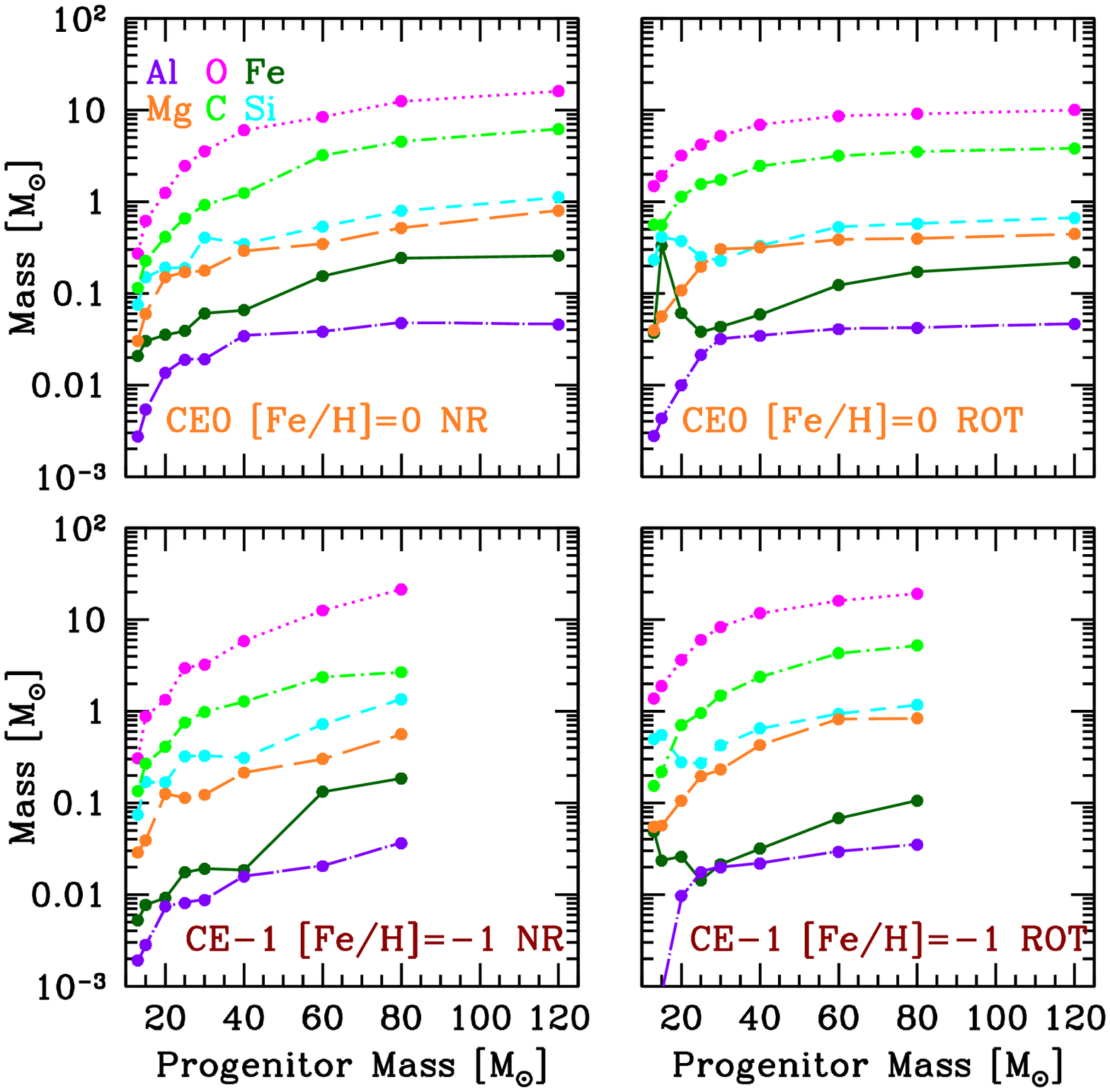}
\includegraphics[width=8.5cm]{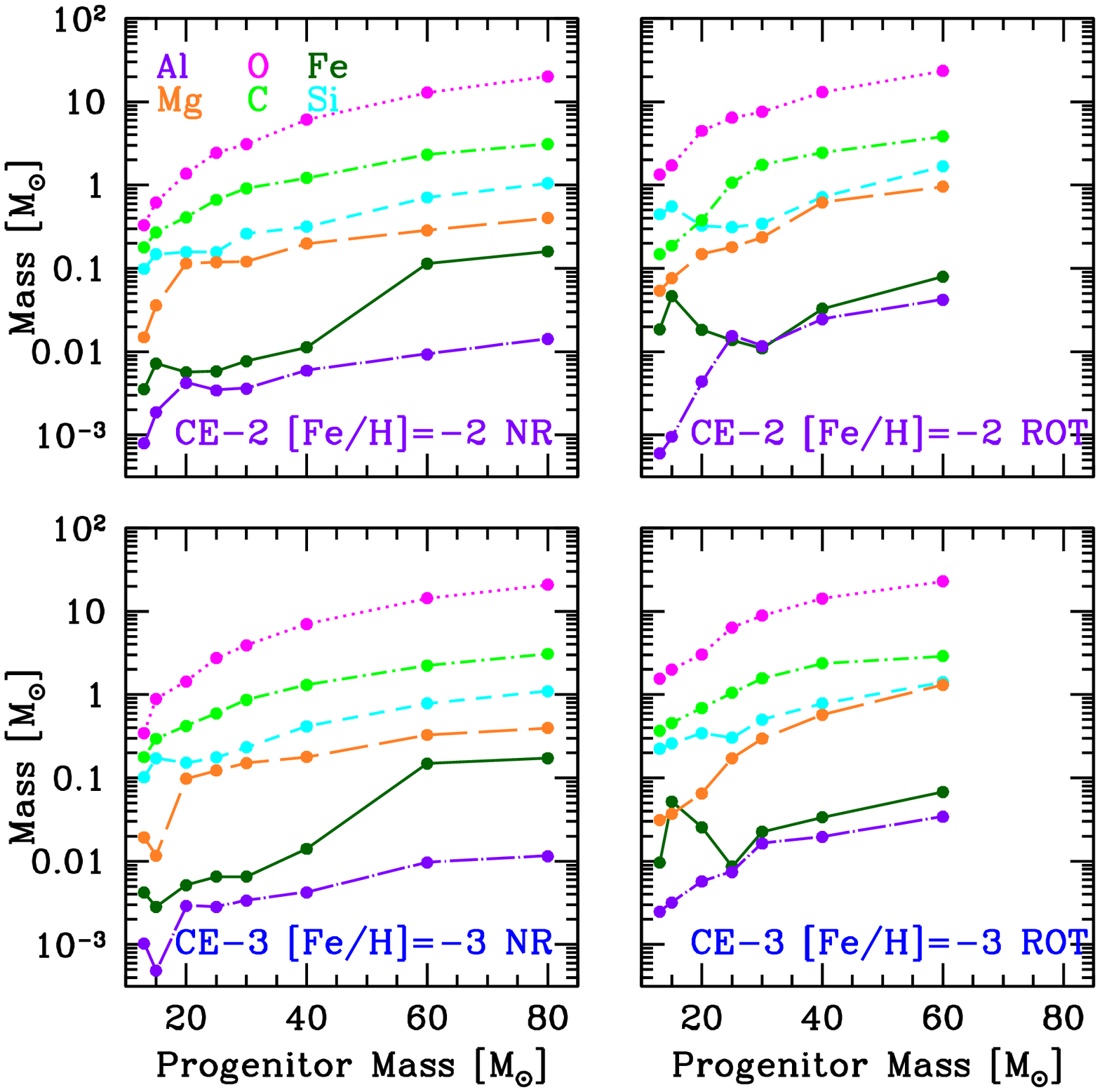}
\caption{Initial C, O, Mg, Si, Al and Fe elemental abundances in the ejecta  (see the legenda)
as a function of the progenitor mass for CE-SN models.
Each pair of panels shows a different initial value of metallicity: set CE0 (top left),
set CE-1 (bottom left), set CE-2 (top right) and set CE-3 (bottom right). In each pair, left and right panels show non rotating and rotating models.}
\label{Input_metal_LC}
\end{figure*}
The metallicity and elemental composition of the ejecta 
are very important for dust formation. Figs.~\ref{Input_metal_efixAB} and
\ref{Input_metal_efixCD} show
the initial metal abundances in the ejecta (prior to dust production) 
as a function of the progenitor stellar mass for the four metallicity data set, 
comparing rotating and non rotating FE models. 

The most abundant metals in the ejecta are O and C. In general,
for smaller progenitor masses, $\rm O > C $. When [Fe/H] $\geq -1$
(set FE0 and FE-1), there is a progenitor mass above which $\rm O < C$. At lower metallicity
(with the exception of FE-3-NR), oxygen remains the most abundant element,
even for very large progenitor masses. 

The abundance of heavier and more internal elements  is
very sensitive to the degree of fallback and rotational mixing. As a result,
the mass of Mg, Si and Fe does not show a simple monotonic trend with
progenitor mass (for masses $\rm \leq 25 - 30~M_{\odot}$), independently of metallicity and rotation.
When [Fe/H] $ \geq -1$ (set FE0 and FE-1), more massive progenitors have 
Mg, Si and Fe abundances that slowly increase with progenitor mass.
Due to their massive remnants, the ejecta of massive rotating models at lower metallicity 
are largely dominated by O, C and Mg.

An interesting consequence of the dependence of light and heavy element
abundances on stellar mass, metallicity and rotation is the different 
degree of [C/Fe] in the ejecta. Using the initial abundances of C ad Fe represented by
the  light and dark green lines in Figs.~\ref{Input_metal_efixAB} and \ref{Input_metal_efixCD},  
we compute [C/Fe] for all FE models and we show it in Fig.~\ref{CsuFe_efix} 
as a function of the remnant mass.
As expected, [C/Fe] is a strong function of metallicity: for [Fe/H] $\geq -1$ 
(set FE0 and FE-1),  [C/Fe] ranges from 0 to $\sim 2.5$, whereas for smaller 
metallicity (set FE-2 and FE-3), [C/Fe] increases up to $\sim 3.5 - 4.5$. 
FE SN models that leave low-mass remnants, with $M_{\rm rem} \lesssim 5~M_\odot$,
are characterized by [C/Fe] $\lesssim 1$, independently of metallicity and rotation.
However, the ejecta of FE SN models that leave more massive remnants are all carbon-enhanced,
particularly if the progenitor stars have low initial metallicity and rotate. 
Indeed, massive, metal-free faint SNe (with mixing and fallback) have been 
invoked to explain the observed [C/Fe] in carbon-enhanced extremely metal-poor stars,
the so-called CEMP-no stars, with [C/Fe]$ > 0.7$ and no traces of rapid or slow neutron capture
elements \citep{2005ARA&A..43..531B}. Indeed, their peculiar properties have been interpreted as 
due to the inprints of metal-free supernovae \citep{2014MNRAS.445.3039D,2015MNRAS.451.2108D} 
that evolve as faint SNe \citep{2003Natur.422..834B, 2003ApJ...594L.123L,2005Sci...309..451I,2014ApJ...794..100M}, 
or of primordial "spinstars'', which experienced mixing and mass-loss because of their
high rotational velocities \citep{2006A&A...447..623M,2015A&A...580A..32M}.
The results shown in Fig.~\ref{CsuFe_efix} suggest a dependence on metallicity 
that may have interesting observable implications.

\subsection{CE models} 

The two lower panels in Fig.~\ref{fig:mass_nr_rot_efix} show the remnant 
mass  as a function of the progenitor mass for CE models.
As a consequence of the calibration procedure described in Section~2, the 
effects of fallback are substantially reduced in CE SN models. 
The resulting remnants are formed with very similar masses (particularly for 
non rotating models) and are much smaller than in FE models, with 
$1~M_\odot \lesssim M_{\rm rem} \lesssim 2~M_\odot$.
In CE models, the ejecta composition is sensitive to the mass cut, that is chosen to
obtain a $\rm ^{56}Ni$ mass in accordance to the fit described in Section~2. 
The resulting C, O, Mg, Si, Al and Fe abundances as a function of the progenitor mass are shown 
in Fig. \ref{Input_metal_LC} for rotating and non rotating models with different initial metallicity.
The ejecta composition is almost indipendent of metallicity and rotation, 
because the abundances of metals in the He core are very similar for 
progenitors with similar mass. The main differences are due to the progenitor mass 
and to rotation, which induces efficient mixing and affect the abundances of heavier
elements (Mg, Si, Al, and Fe), particularly for low mass progenitors.

\section{Dust formation model}
The calculation of dust formation in the ejecta is based on CNT. A similar
approach has been adopted to investigate dust formation in core-collapse 
\citep{2001MNRAS.325..726T,2007MNRAS.378..973B}, pair-instability \citep{2004MNRAS.351.1379S}, 
Pop III core collapse \citep{Marassi2015}, and faint SNe \citep{2014ApJ...794..100M}.
The current version of the model is described in \citet{Marassi2015} and it is 
an improved version of the model adopted by \citet{2007MNRAS.378..973B}. Here
we briefly summarize the main features of the model and we refer the interested reader
to the original papers for more details.

We assume that dust seed clusters are made of ${\cal N} \geq 2$ monomers and that the sticking coefficient 
(defined as the probability that an atom colliding with a grain will stick to it) 
is equal to 1 for all grain species. The onset of grain condensation is controlled 
by the temperature and density in the ejecta, whereas the grain composition depends 
on the chemical composition, which, in turn, depends on the nature of the SN progenitor 
(mass, metallicity, rotation, explosion energy). While the ejecta expands, we follow
the formation and destruction rates of CO, SiO, C$_{2}$, O$_{2}$ molecules (which 
play an important role in subtracting gas-phase elements) and the condensation of 
seven different grain species: amorphous carbon (AC), iron (Fe), corundum (\Alumina), 
magnetite (\Magnetite), enstatite (\Enstatite), forsterite (\Forsterite) and quarz (\Silica). 
Following \citet{Marassi2015}, we investigate dust formation in FE/CE SNe adopting 
the thermal, dynamical and chemical evolution of the ejecta predicted by  
the output of 1D SN explosion simulations. The initial time 
for the calculation ($t_{\rm ini}$)  is fixed by requiring that the gas temperature at the 
radius of the He core, $\rm R_{\rm He_{\rm core}}$ reaches a temperature of $T_{0} = 10^{4}$ K. 
At $t \geq t_{\rm ini}$, the ejecta follow an adiabatic expansion with a temperature 
evolution,  
\begin{equation}
T = T_{0} \left[1+\frac{v_{\rm eje}}{R_{0}}(t-t_{\rm ini})\right]^{3(1-\gamma)},
\label{thermo}
\end{equation}  
where $\rm \gamma=1.41$ is the adiabatic index, and $T_{0}$, $R_0$, and $v_{\rm eje}$ are the temperature
and radius of the He core and ejecta velocity at $t=t_{\rm ini}$, respectively.

\begin{figure*}
\vspace{\baselineskip}
\includegraphics[width=8.5cm]{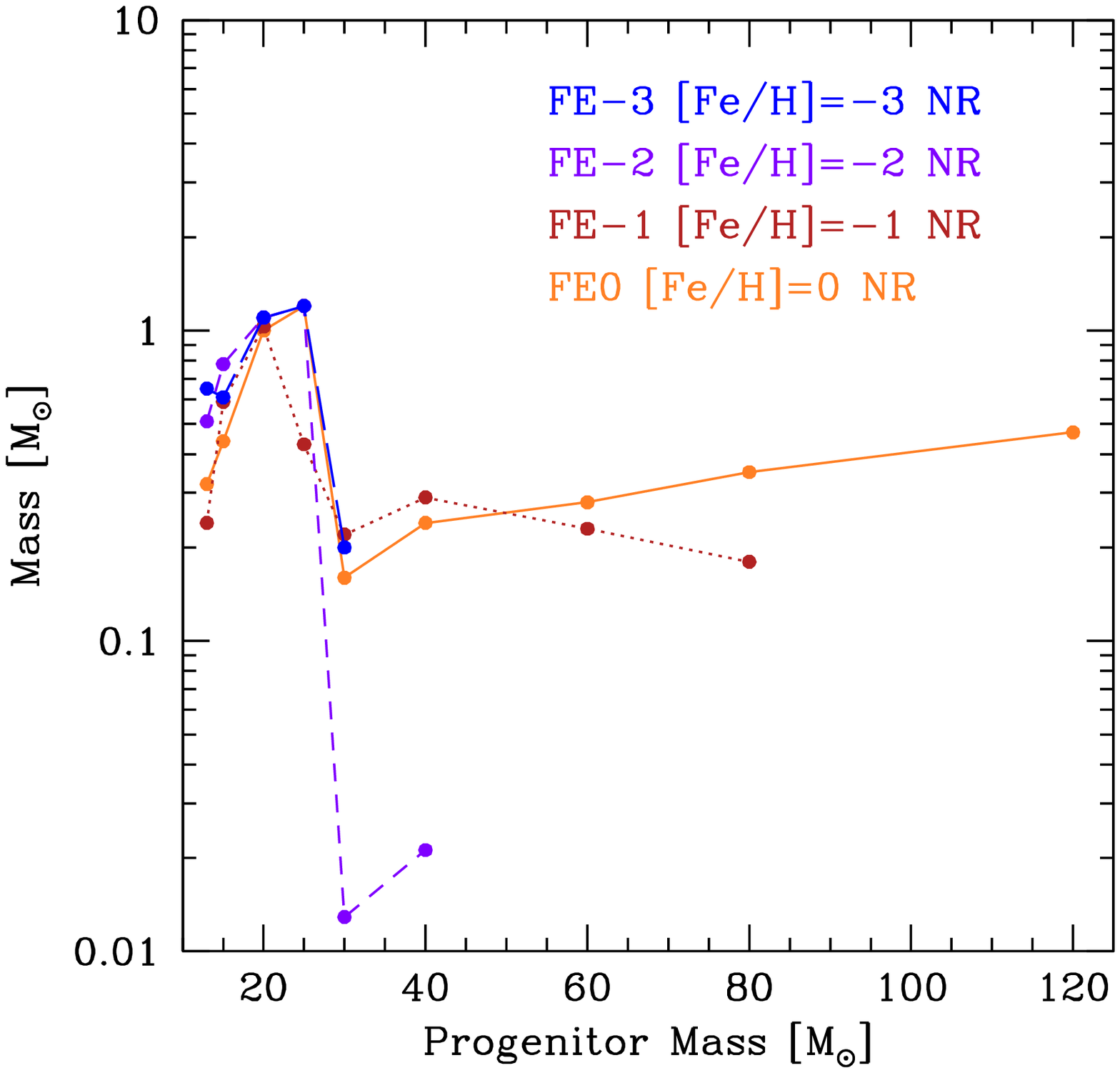}
\includegraphics[width=8.5cm]{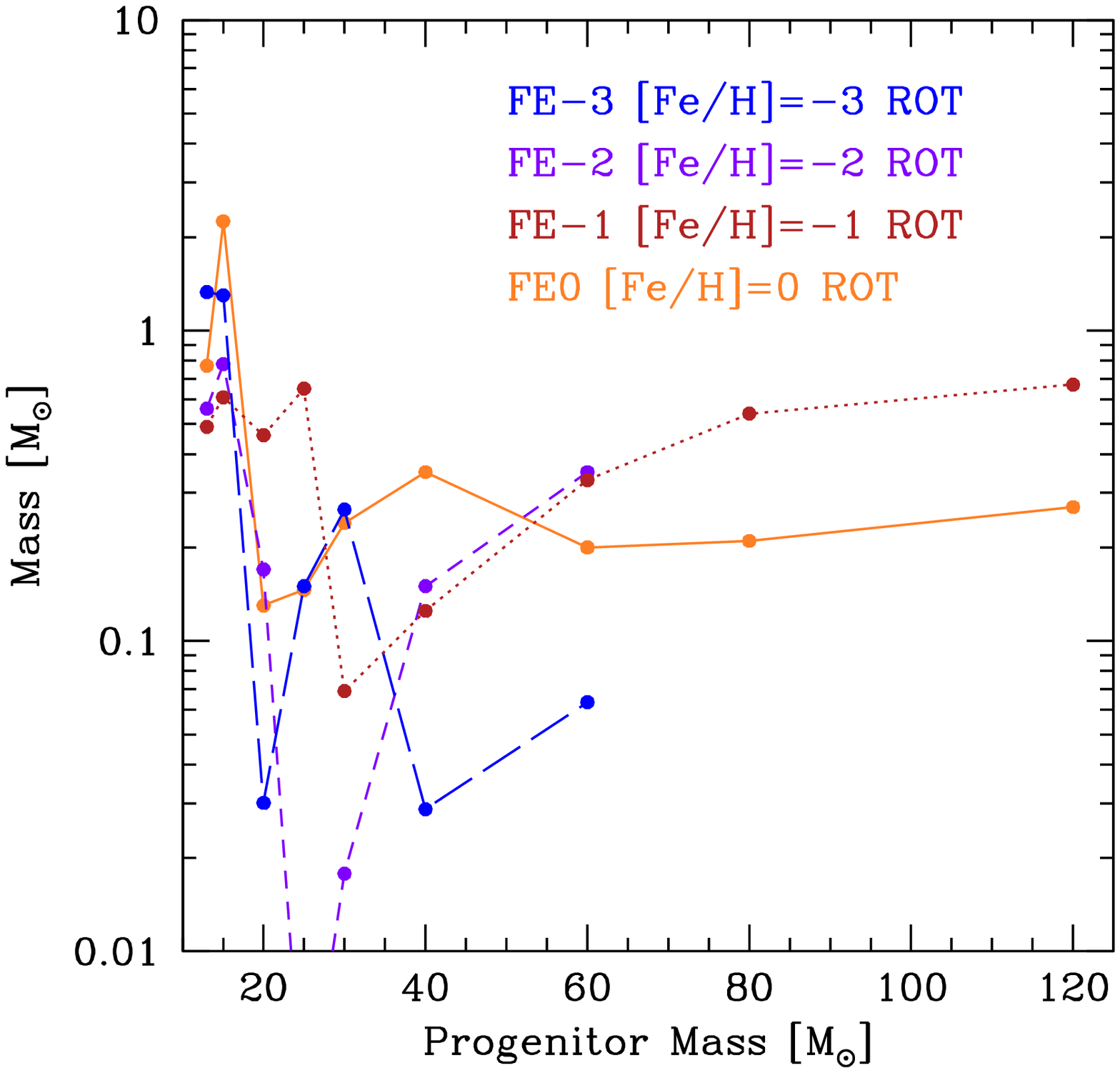}
\caption{Mass of dust formed in the ejecta of FE SN models as 
a function of the mass of the progenitor star, for different values of metallicity. 
Left and right panels compare the results for non rotating and rotating models.}
\label{FE_dust}
\end{figure*}
\begin{figure*}
\vspace{\baselineskip}
\includegraphics[width=8.5cm]{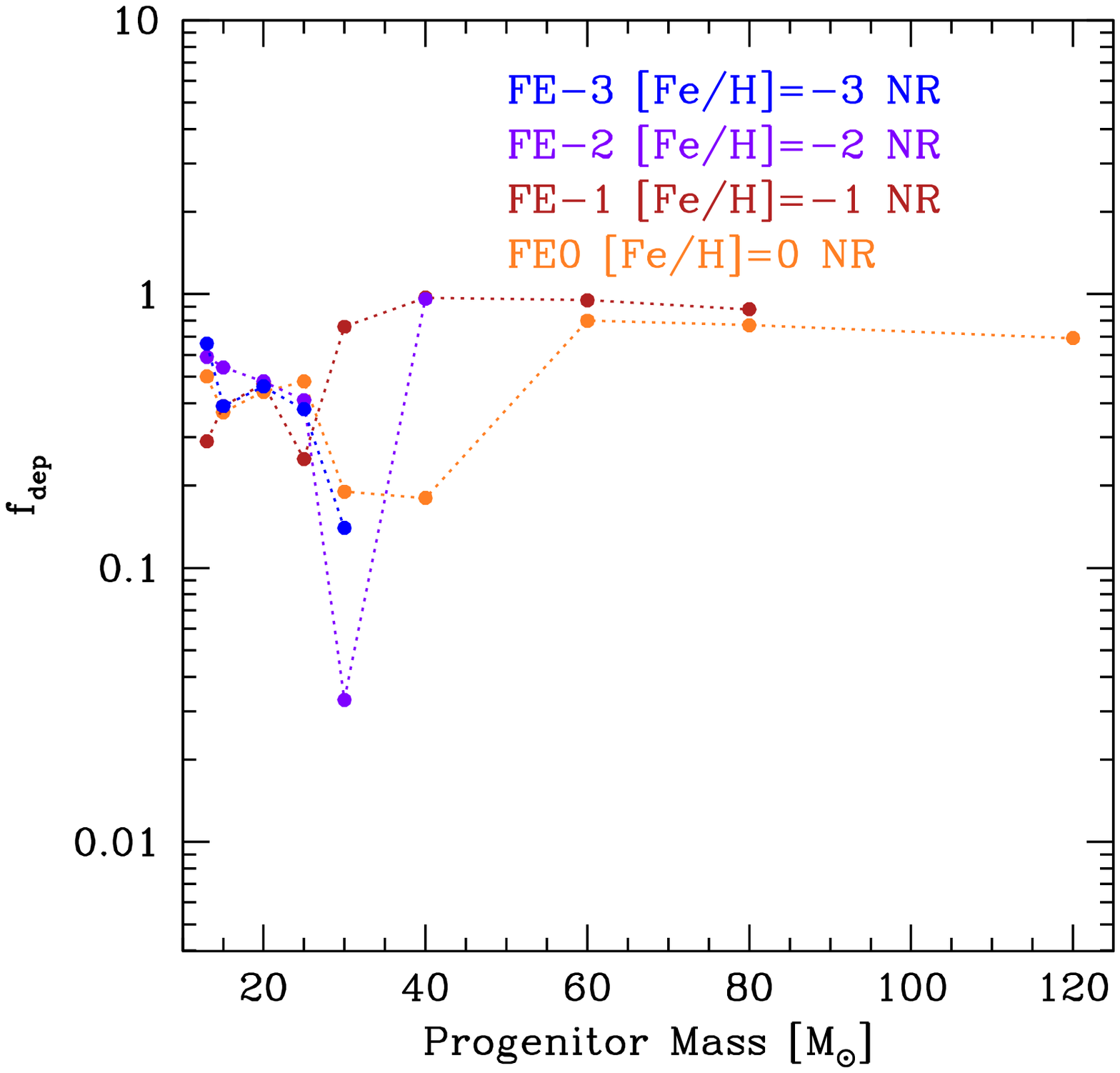}
\includegraphics[width=8.5cm]{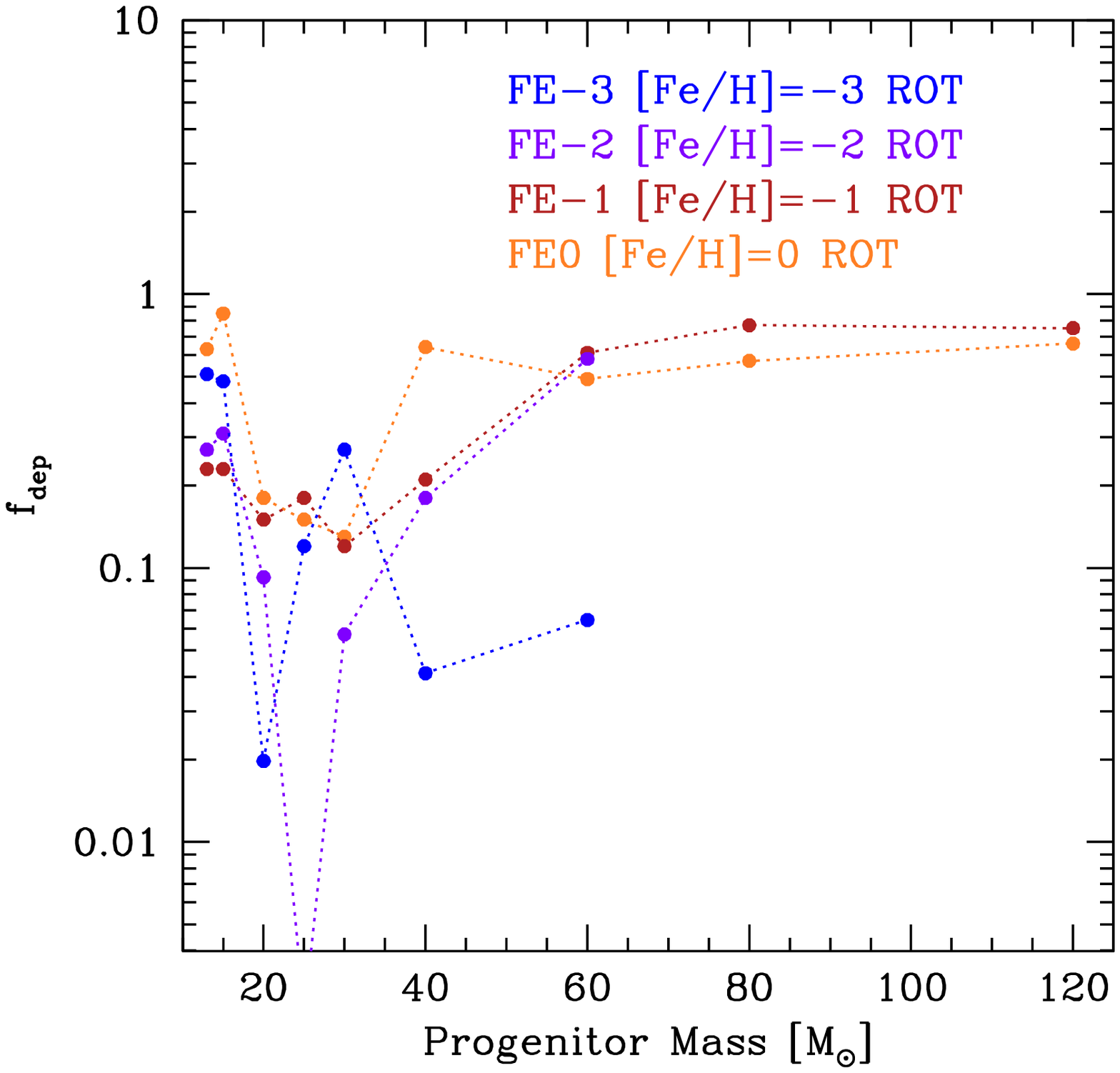}
\caption{The dust depletion factor as a function of the initial progenitor mass, 
for FE non rotating (left panel) and rotating (right panel) models with different metallicity (see the legenda).}
\label{FE_fdep}
\end{figure*}
\begin{figure*}
\vspace{\baselineskip}
\includegraphics[trim=0 2.5cm 0 0,width=8.5cm]{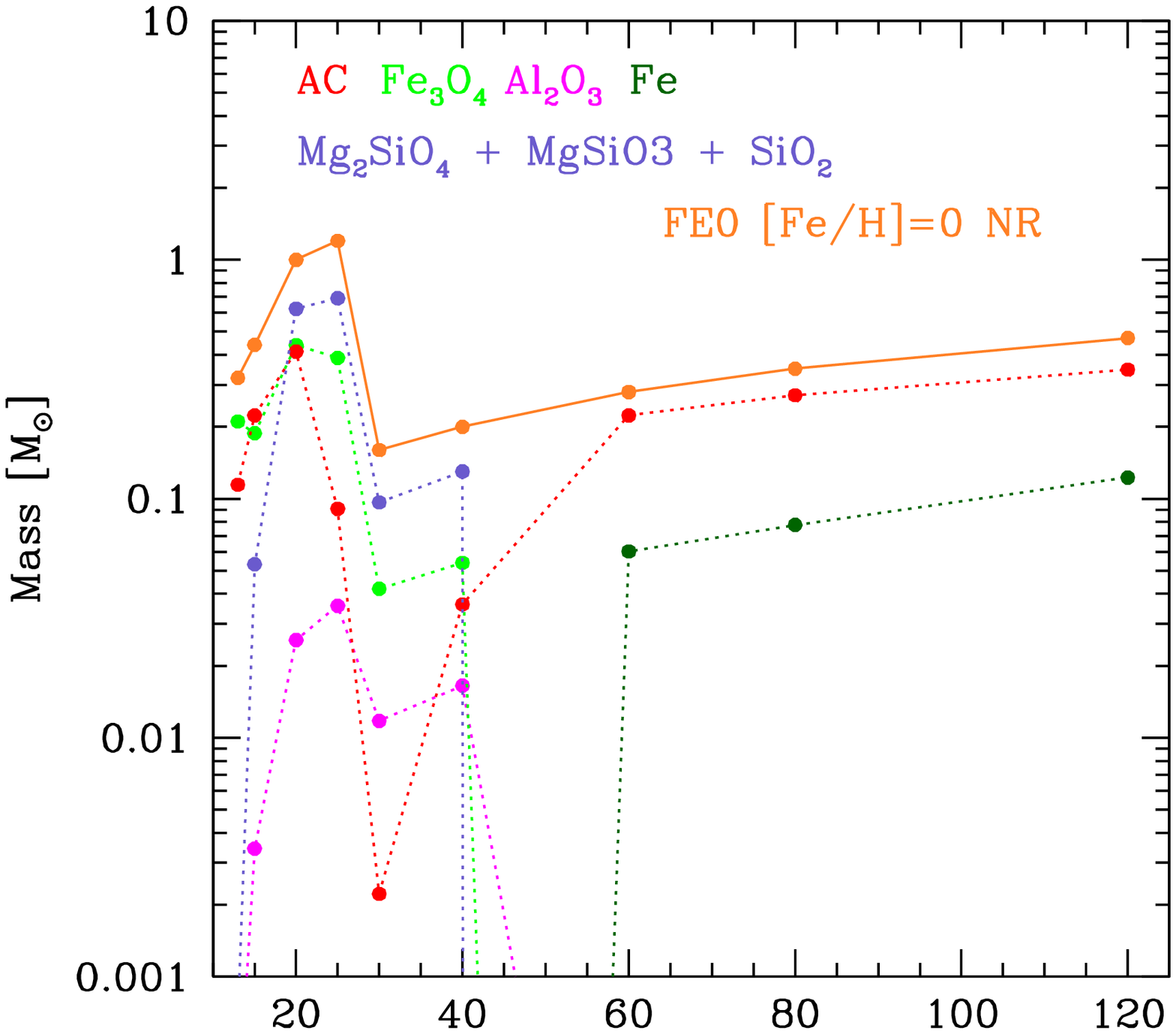}
\hspace{-2.5cm}
\includegraphics[trim=0 2.5cm 0 0,width=8.5cm]{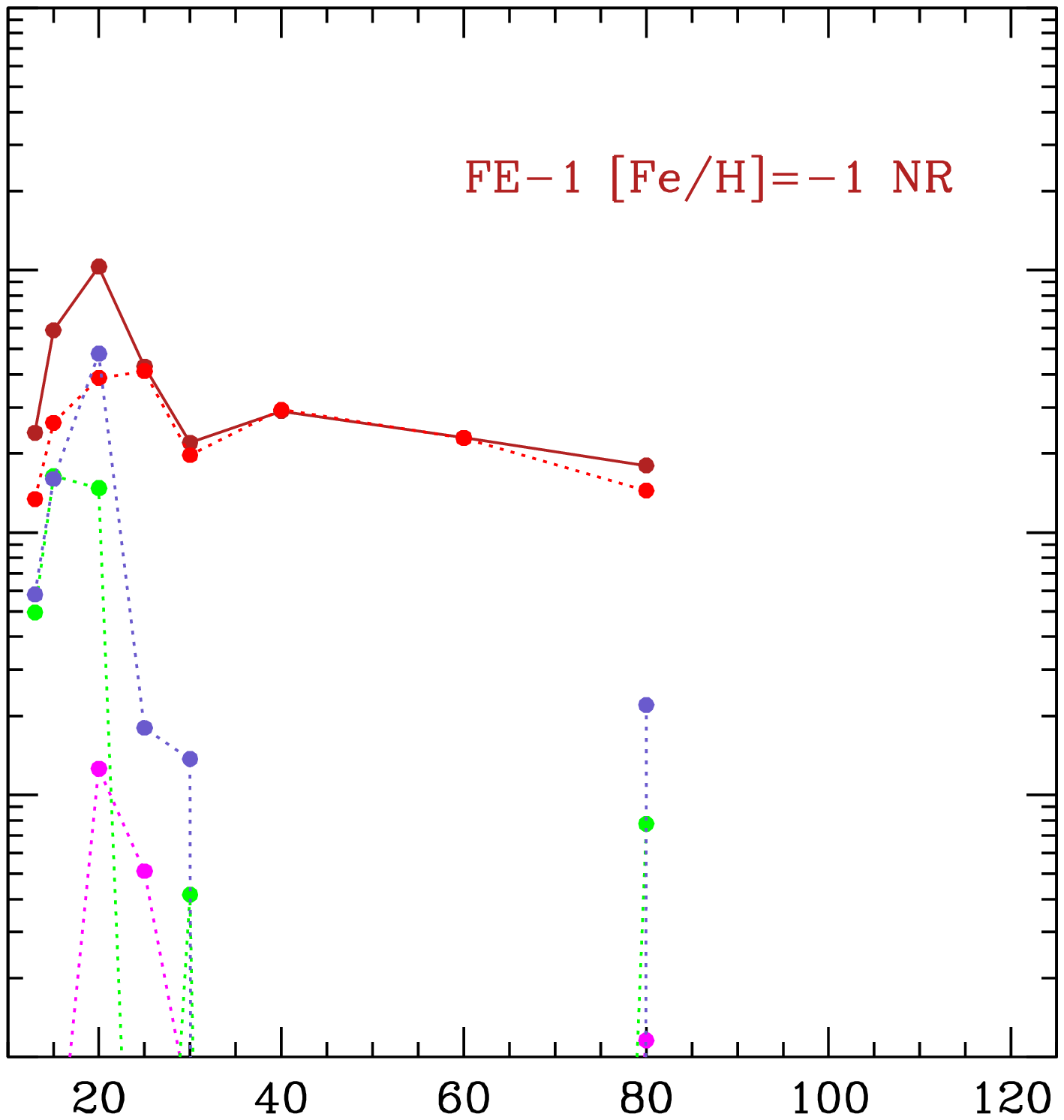}
\includegraphics[trim=0 0 0 2.5cm,width=8.5cm]{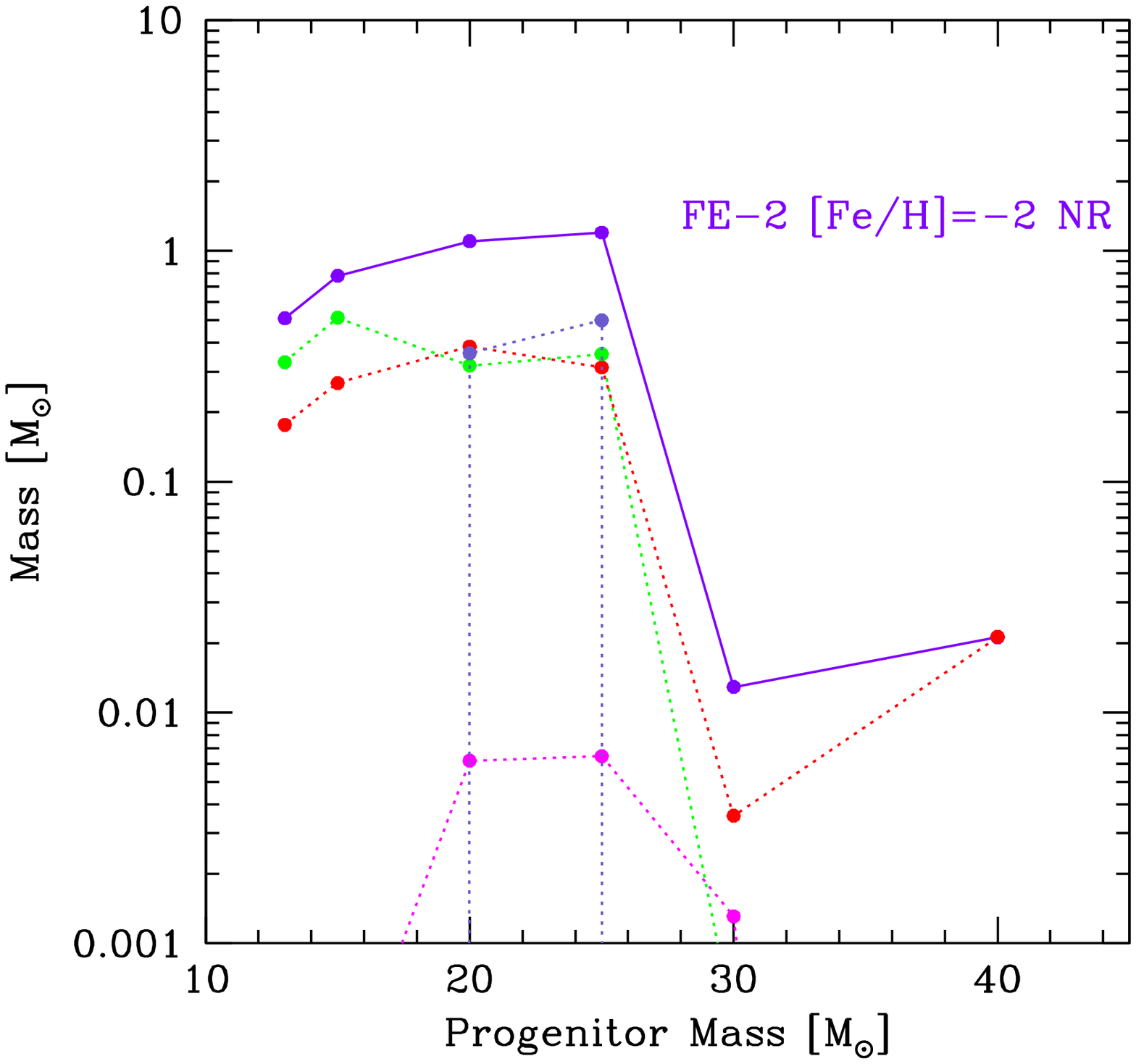}
\hspace{-2.5cm}
\includegraphics[trim=0 0 0 2.5cm,width=8.5cm]{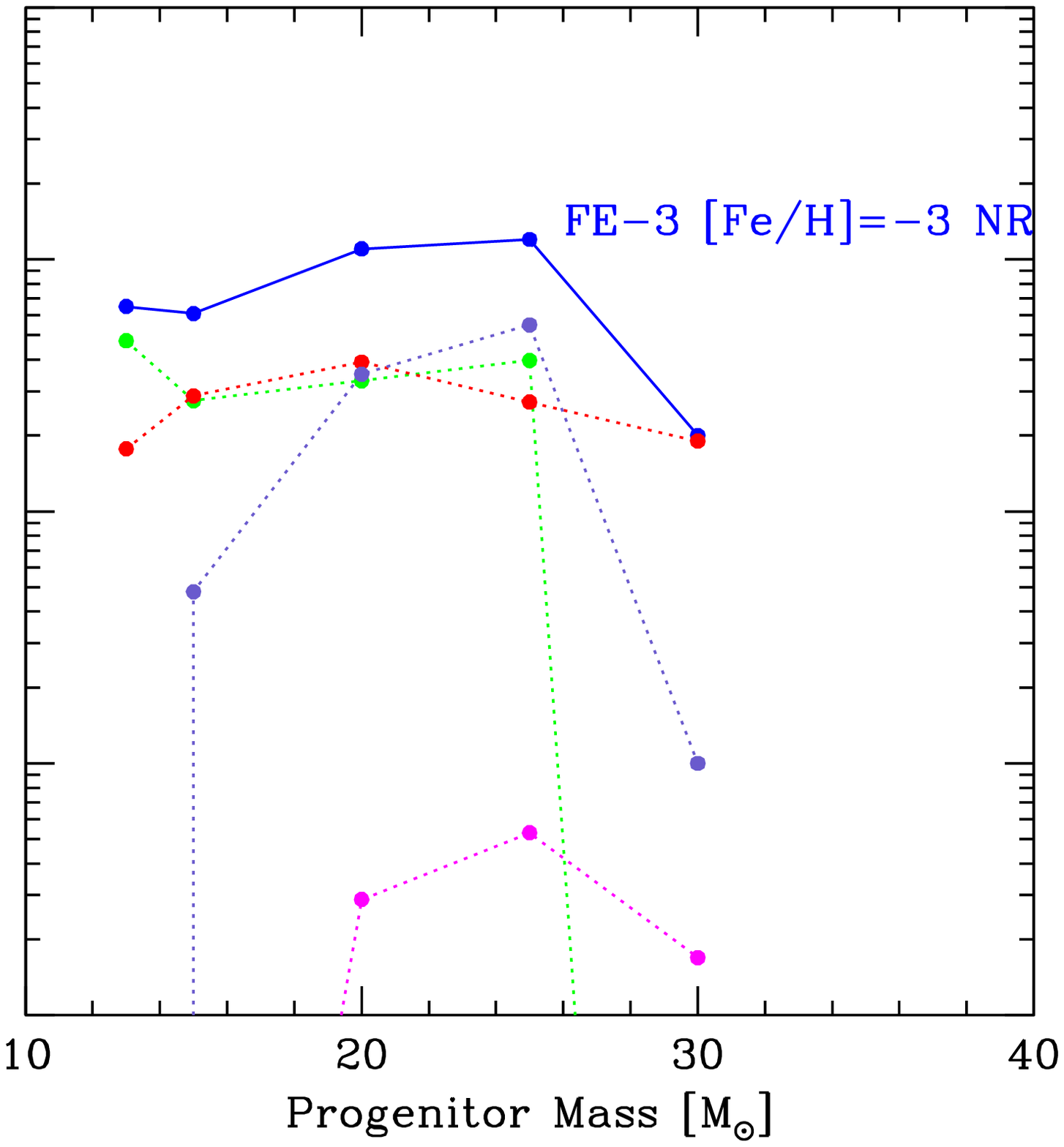}
\caption{Mass of dust in different grain species for FE non rotating models as 
a function of the progenitor mass. Each panel corresponds to a different metallicity.
In each panel, the upper solid line shows the total dust mass and the dotted lines 
represent the
contribution of AC grains (red), Al$_2$O$_3$ (magenta), Fe$_3$O$_4$
(green), solid iron (dark green) and silicates and quarz, MgSiO$_3$+Mg$_2$SiO$_4$+SiO$_2$
(light blue, see the legenda).}
\label{FE_grains_NR}
\end{figure*}
\begin{figure*}
\vspace{\baselineskip}
\includegraphics[trim=0 2.5cm 0 0,width=8.5cm]{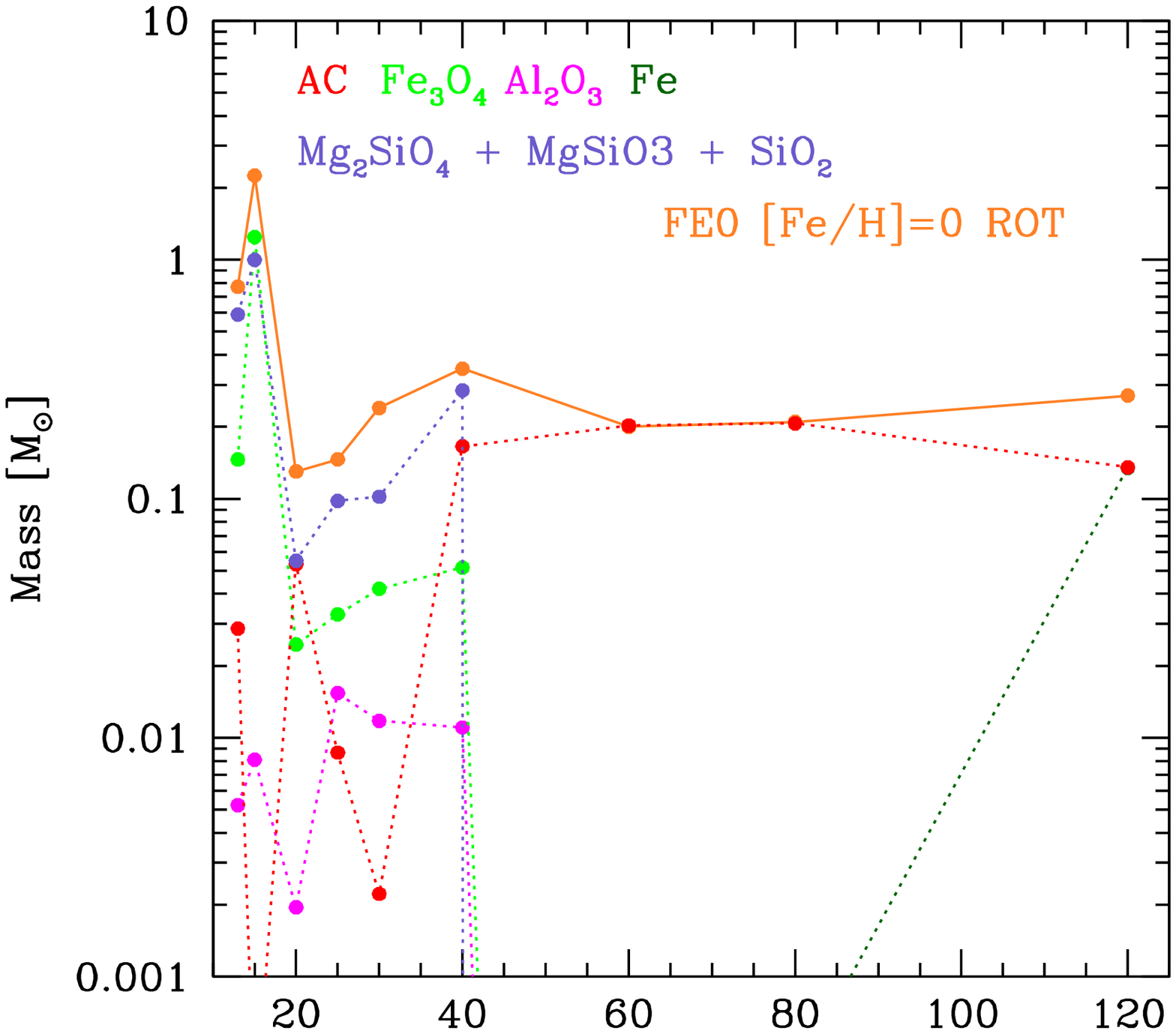}
\hspace{-2.5cm}
\includegraphics[trim=0 2.5cm 0 0,width=8.5cm]{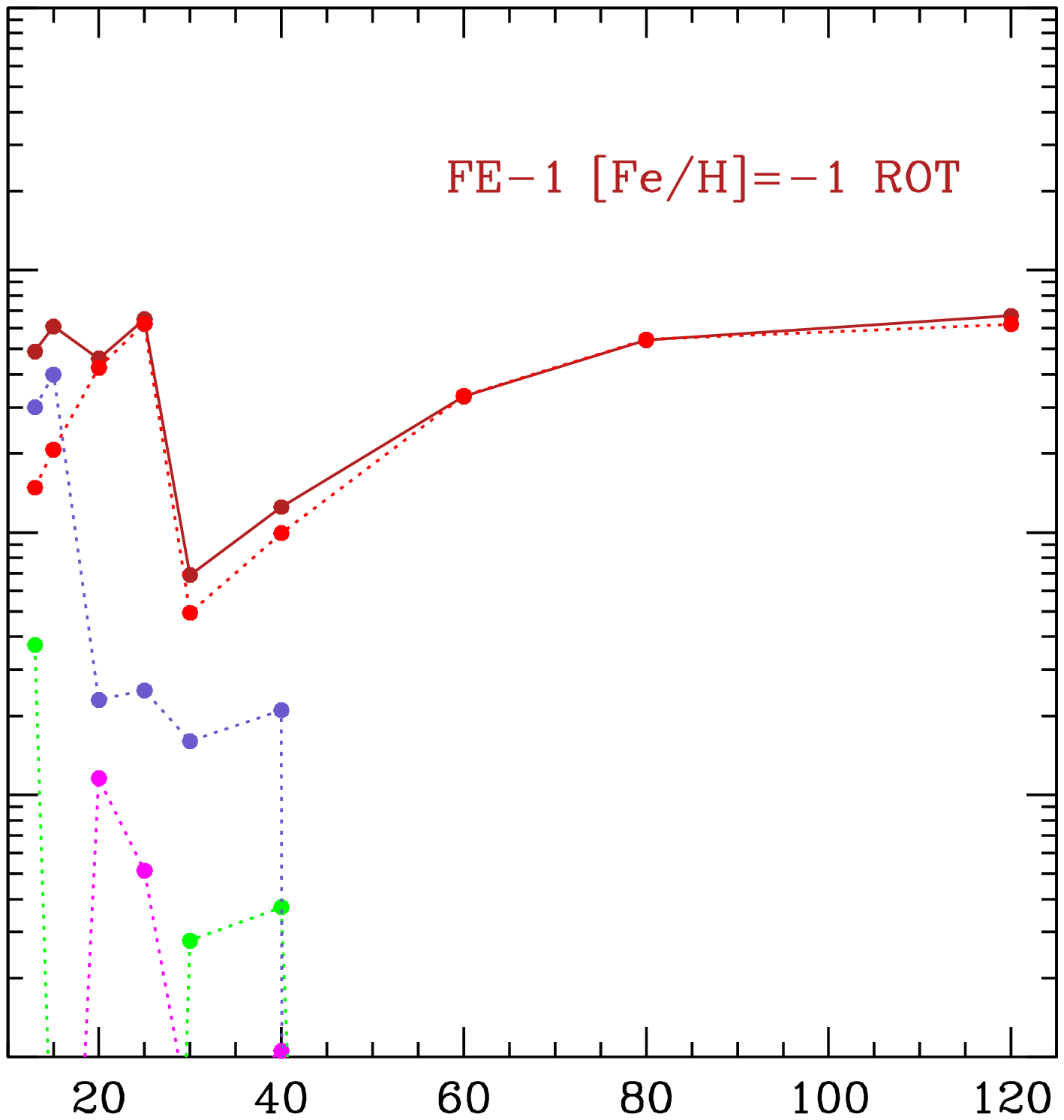}
\includegraphics[trim=0 0 0 2.5cm,width=8.5cm]{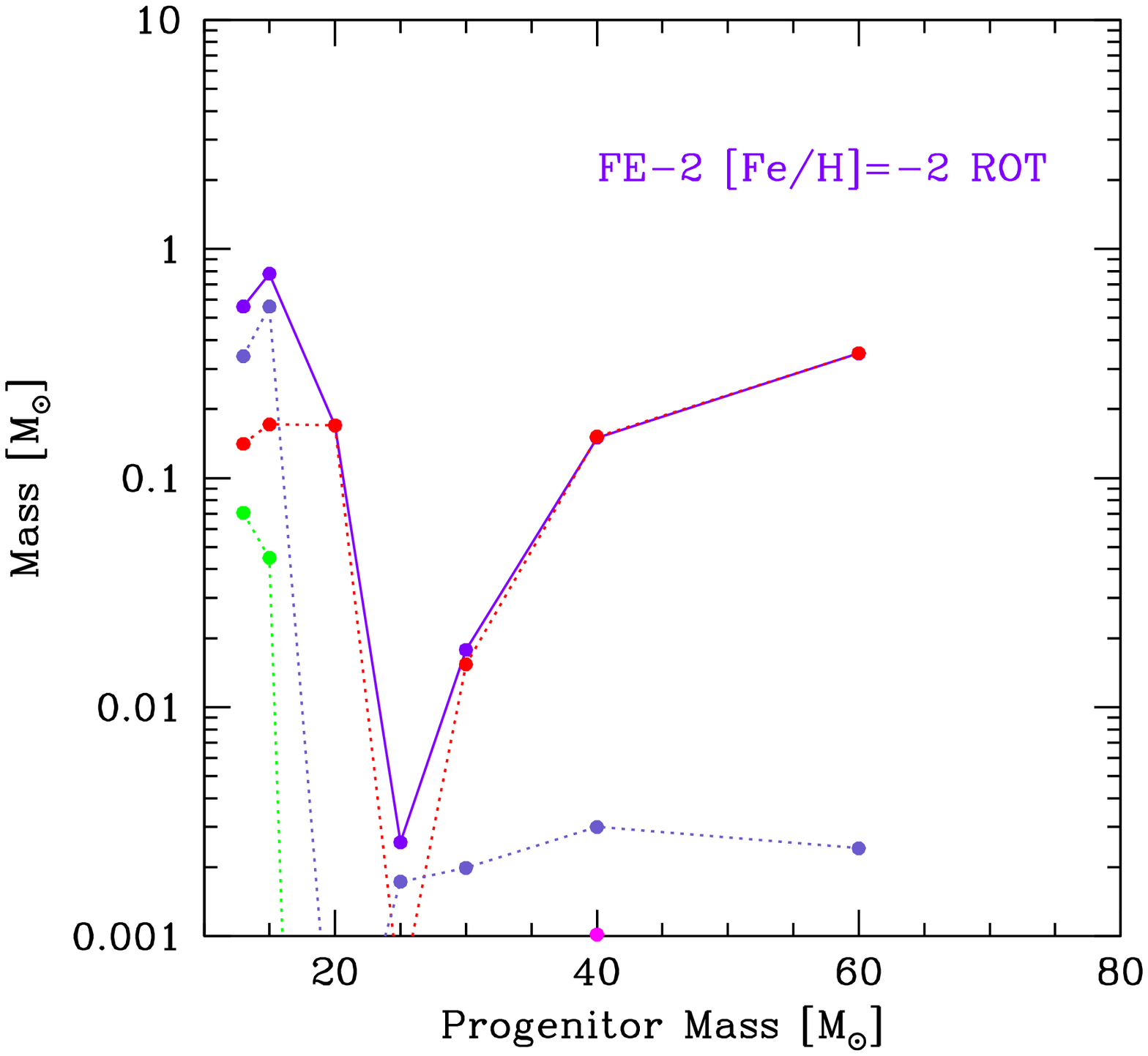}
\hspace{-2.5cm}
\includegraphics[trim=0 0 0 2.5cm,width=8.5cm]{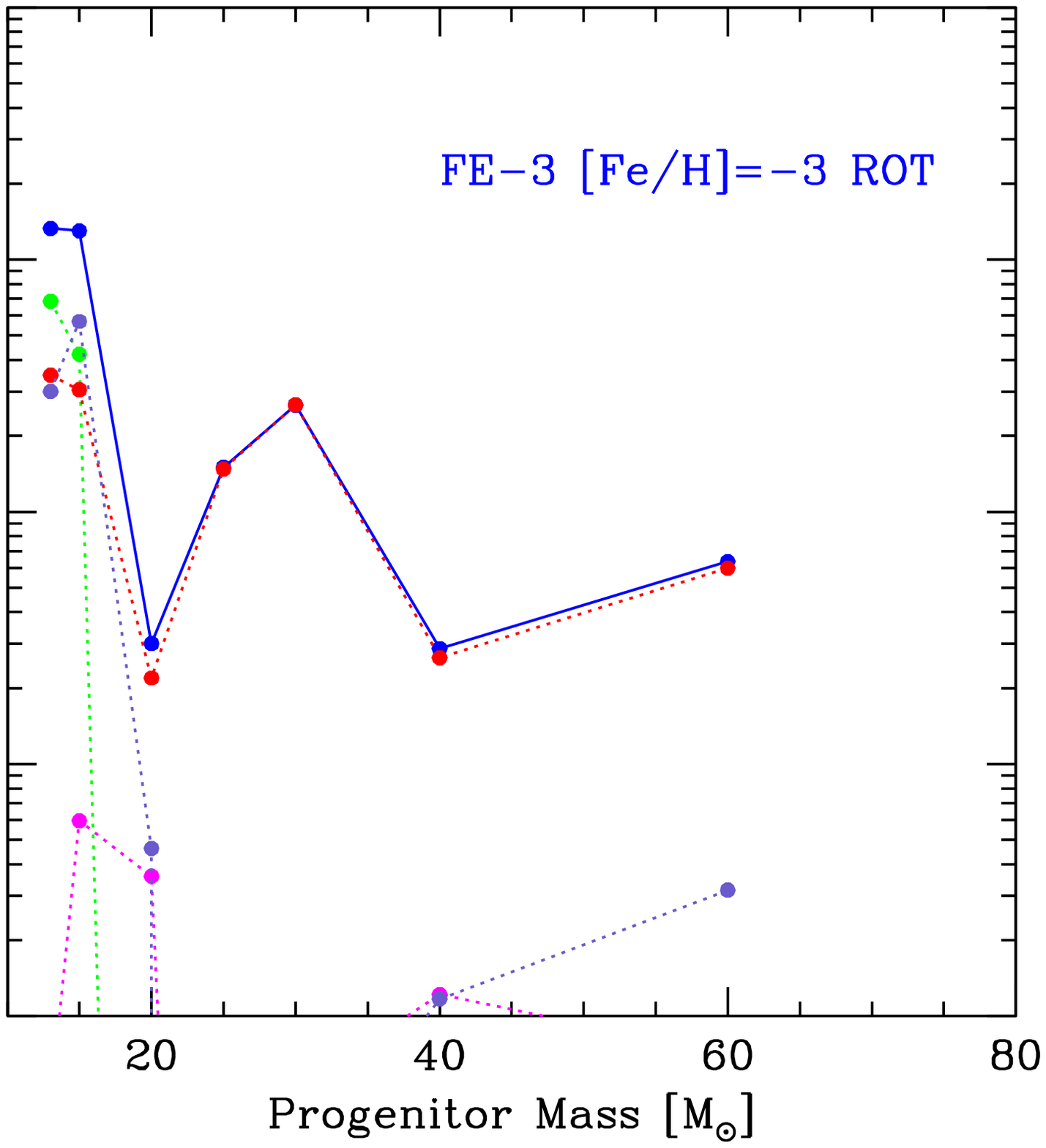}
\caption{Mass of dust in different grain species for FE rotating models as 
a function of the progenitor mass. Each panel corresponds to a different metallicity.
In each panel, the upper solid line shows the total dust mass and the dotted lines 
represent the contribution of AC grains (red), Al$_2$O$_3$ (magenta), Fe$_3$O$_4$
(green), solid iron (dark green) and silicates and quarz, MgSiO$_3$+Mg$_2$SiO$_4$+SiO$_2$
(light blue, see the legenda).}
\label{FE_grains_ROT}
\end{figure*}
\begin{figure*}
\vspace{\baselineskip}
\includegraphics[width=8.5cm]{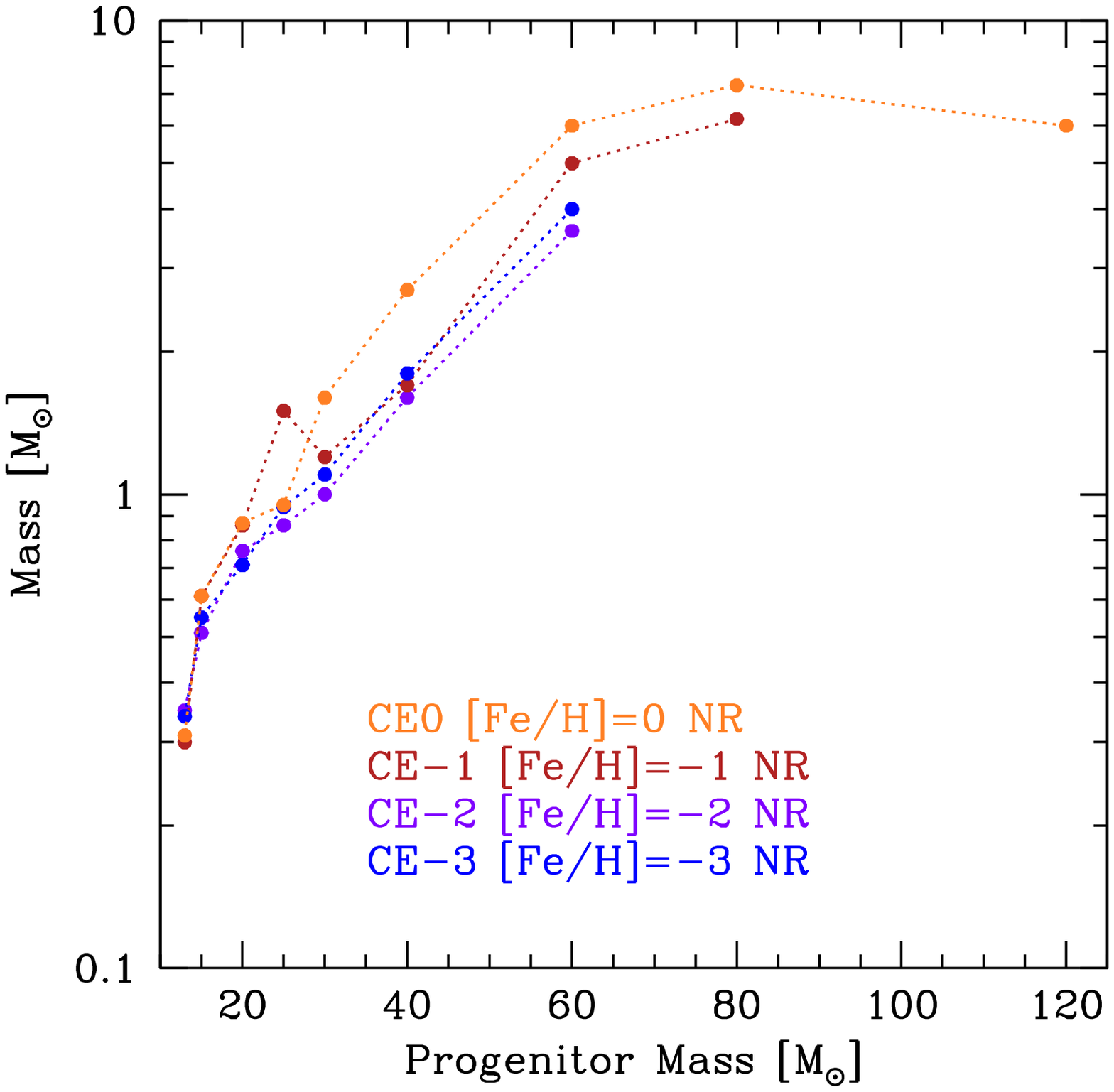}
\includegraphics[width=8.5cm]{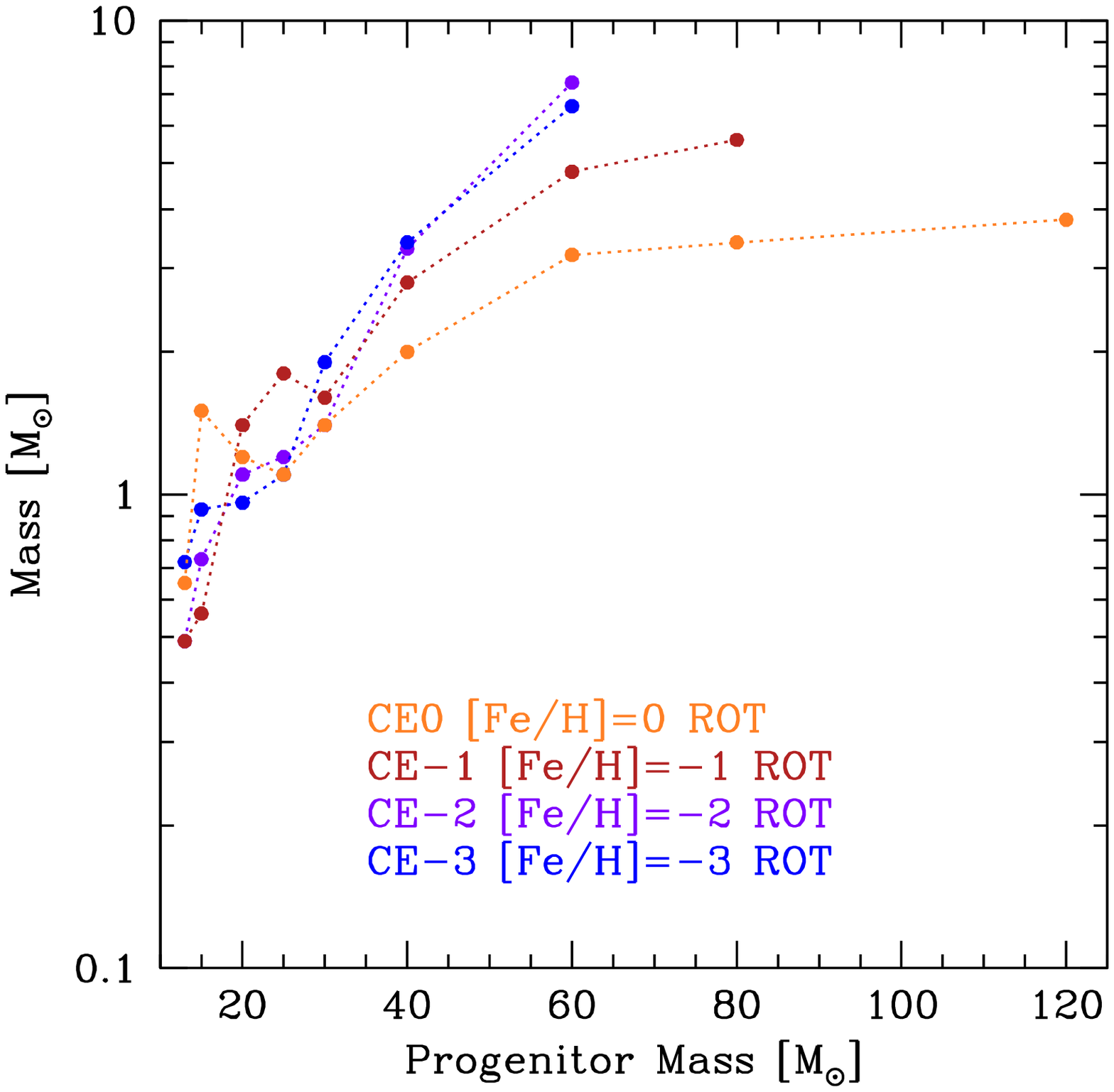}
\caption{Mass of dust formed in the ejecta of CE SN models as 
a function of the mass of the progenitor star, for different values of metallicity. 
Left and right panels compare the results for non rotating and rotating models.}
\label{CE_dust}
\end{figure*}
\begin{figure*}
\vspace{\baselineskip}
\includegraphics[width=8.5cm]{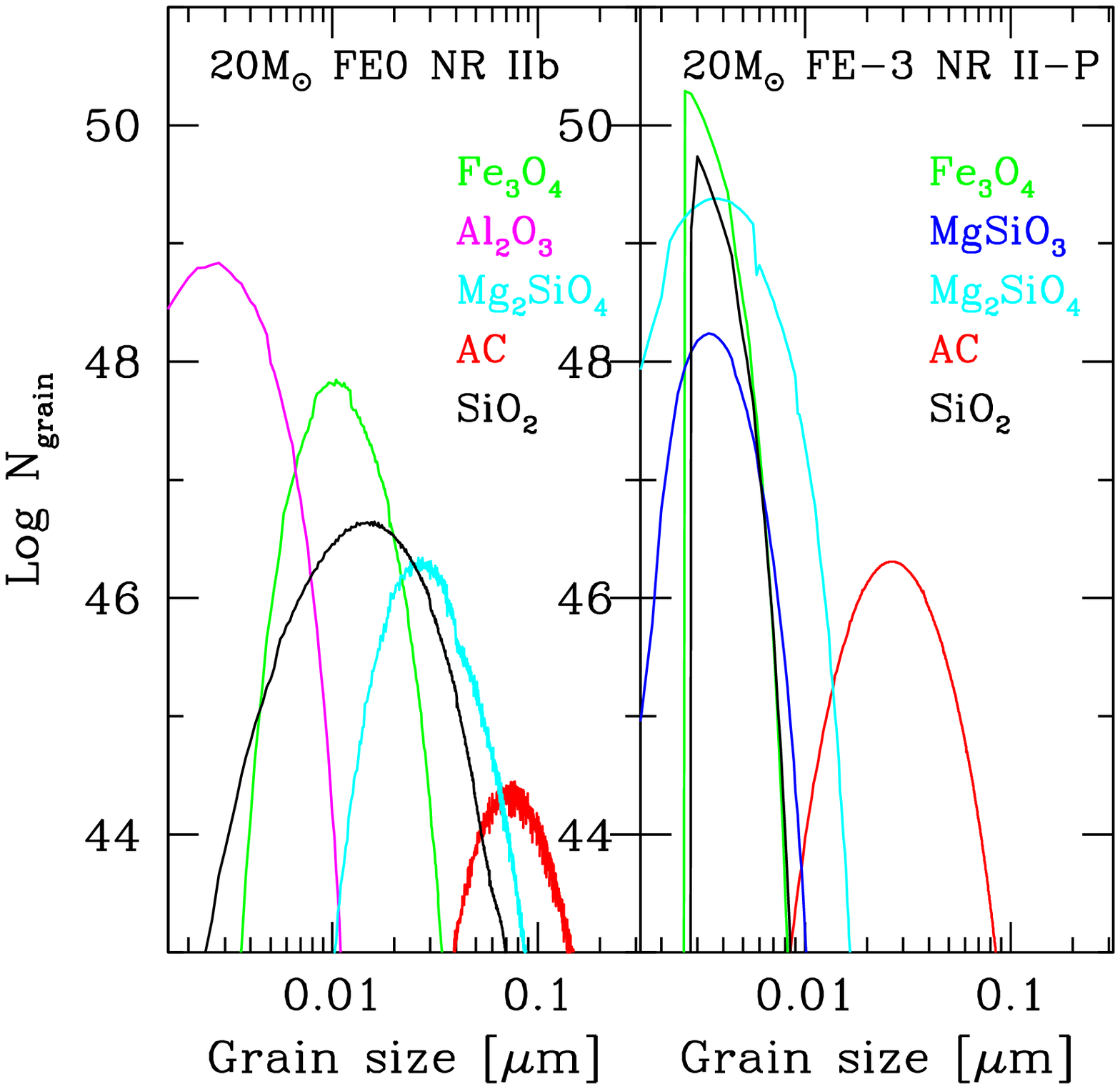} 
\includegraphics[width=8.5cm]{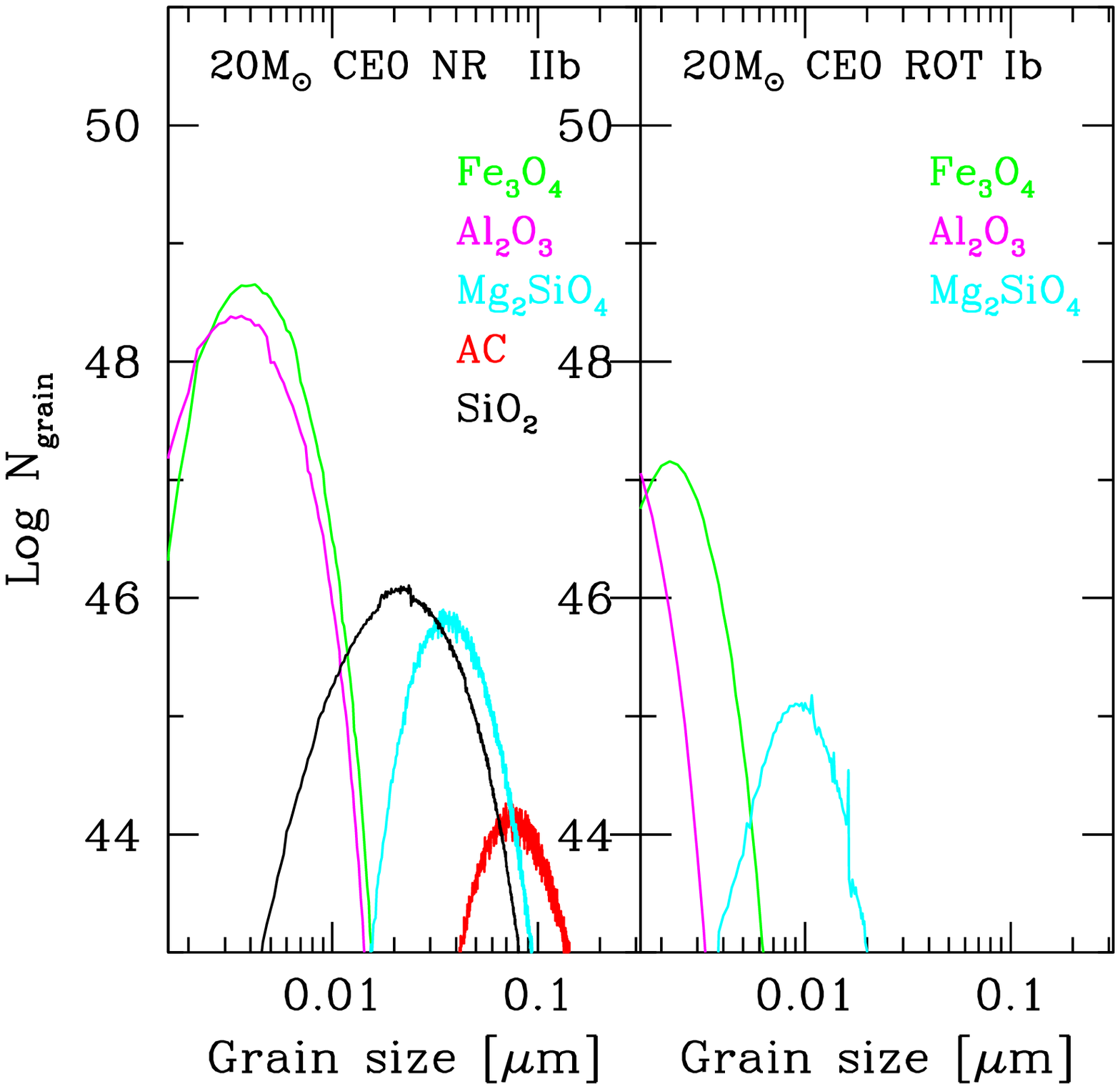} 
\caption{Grain size distribution function obtained for models
with the same initial SN progenitor mass ($20 M_\odot$), but
different metallicity, rotation rates and explosion
energies (see the legenda). Colour coding of the different grain size
distributions is the following: AC grains in red, Al$_2$O$_3$ in magenta, Fe$_3$O$_4$ in green,
SiO$_2$ in black, MgSiO$_3$ in cyan and Mg$_2$SiO$_4$ in blue.}
\label{size_grains}
\end{figure*}
\section{Dust grid: dependence on fallback, metallicity and rotation}
In this section we present the grid of SN dust yields for FE and CE SN models,
discussing the effects of fallback, metallicity and rotation.

We start by analyzing the total mass of dust formed in FE-SN models 
as a function of metallicity and rotation. The progenitor mass, the SN type and the total 
mass of dust condensed in rotating and non rotating models are reported at fixed
metallicity in Tables \ref{FE_NR} and \ref{FE_ROT}, respectively.
The same data is also shown in Fig.~\ref{FE_dust}.

For non-rotating models, we find the total mass of dust to be in the range $\rm [0.16 - 1.2]~M_{\sun}$ 
for set FE0, $\rm [0.18 - 0.59]~M_{\sun}$ for set FE-1, $\rm [1.3 \times 10^{-2}  - 1.2 ]~M_{\sun}$ 
for set FE-2, and $\rm [0.20 - 1.2]~M_{\sun}$ for set FE-3 (see Table \ref{FE_NR}). 
These results confirm previous theoretical predictions for type II-P SNe 
\citep{1989ApJ...344..325K,1991A&A...249..474K,2001MNRAS.325..726T,
2003ApJ...598..785N,2007MNRAS.378..973B,2010ApJ...713....1C}. 
We find that the effects of rotation depend on the mass and metallicity of the progenitor stars,
as we discuss below. The resulting dust masses in rotating FE SN models
are in the range $\rm [0.13 - 2.25]~M_{\sun}$ for set FE0, $\rm [6.9\times 10^{-2} - 0.65]~M_{\sun}$ 
for set FE-1, $\rm [2.6\times 10^{-3} - 0.78]~M_{\sun}$ for set FE-2, and $\rm [6.3 \times 10^{-2} - 1.3]~M_{\sun}$, 
for set FE-3 (see Table \ref{FE_ROT}).

Figure~\ref{FE_dust} shows that the dust mass does not monotonically increase with progenitor mass. Rather, it depends on the physical conditions of the ejecta, 
such as their temperature profile, their initial radius, and the gas-phase metal abundances that result from the formation and destruction of some key molecular species \citep{Marassi2015}. As shown in the left panel of Fig.~\ref{FE_dust}, for FE-NR models 
the dust mass increases with progenitor mass in the range $\rm [13 - 25]~M_{\sun}$ and then drops, due to fallback, particularly
when [Fe/H] $\leq -2$. The increasing trend in the range $\rm [13 - 25]~M_{\sun}$ is mainly 
due to efficient destruction of molecules by Compton electrons coming from the $\rm^{56}{Ni}$ decay 
in the ejecta, which favours dust condensation \citep{Marassi2015}.

In massive SN progenitors, the dust mass decreases due to the smaller amount of metals in the ejecta 
that survives the large fallback. In addition, the thermodynamical properties of the ejecta depend on the progenitor mass
and, at the onset of adiabatic expansion, more massive SN have small initial radii, 
$R_0$ (see Eq.~\ref{thermo}), and large ejecta density. The latter condition enables a more efficient formation 
of molecules, resulting in an efficient subtraction of gas-phase metals 
\citep{2014ApJ...794..100M,Marassi2015}.

In Fig. \ref{FE_fdep} we show the total dust depletion factor, $f_{\rm dep}$,
defined as the fraction of the initial ejecta metal mass (newly synthesized and pre-existing metals) 
that has condensed into dust, $f_{\rm dep} = M_{\rm dust}/M_{\rm met}$, for FE non rotating (left panel) and 
rotating (right panel) models with different metallicity. 
For FE-NR SN models with mass in the range $[13 - 25]~M_{\sun}$, $f_{\rm dep}$ varies 
between $\sim 0.3$ and $\sim 0.6$. 
For more massive SN models, the trend with progenitor 
mass changes depending on metallicity. 
The ejecta of FE0 and FE-1 models with $M > 40 M_\odot$ have $\rm C>O$ and, despite the increase of fallback with 
progenitor mass, AC grains can form, leading to $f_{\rm dep} \sim 0.7$ for both rotating and non rotating models. 
For most of the models in set FE-2 and FE-3, instead, $\rm C < O$ and the lower amount of metals in the ejecta leads to low
values of $f_{\rm dep}$. 

The dependence of the dust composition on the progenitor mass and metallicity for non rotating FE models is presented in Fig.~\ref{FE_grains_NR}.
This figure clearly shows that in models FE0 and FE-1 with progenitor masses $M < 30 - 40 M_\odot$ a variety of grain species form, as a result of the large abundance of both pre-existing and newly synthesized metals in the ejecta. For larger progenitor masses,
however, the most abundant dust species is AC. This is a consequence of the $\rm C > O$  composition and low abundances of heavier elements in the ejecta (see Fig.\ref{Input_metal_efixAB}). Interestingly, in FE0 models with massive progenitors 
($M \geq 60~M_{\sun}$) $\rm \sim (0.05 - 0.12)~M_{\sun}$ of iron grains form. The fomation of solid iron is favored by the efficient oxygen depletion onto CO molecules (which prevents the formation of Fe$_3$O$_4$ grains) and by the abundance of pre-exisiting iron in the ejecta. Finally, in set FE-2 and FE-3 silicates form only in the ejecta of $\rm [20 - 25]~M_{\sun}$ progenitors while in all the other models the dominant grain species are AC and \Magnetite.

In the right panel of Fig.~\ref{FE_dust} we show the dust mass as a function 
of the progenitor mass for FE rotating SN models (see also Table \ref{FE_ROT}). For all 
the metallicity data set, the most efficient dust factories are rotating models 
with masses in the range  $\rm [13 - 15]~M_{\sun}$. This is not unexpected as
metal abundances in the ejecta are larger than in non rotating models. 
In addition, in set FE0, FE-2, FE-3 these two 
progenitor masses have a non-zero 
$\rm ^{56}{Ni}$ mass in the ejecta (see Tables \ref{FE_A}, \ref{FE_B}, \ref{FE_C}, \ref{FE_D}), which favours dust formation. 
For FE-ROT SN models with $20~M_{\sun} < M < 40~M_{\sun}$, the resulting 
dust mass decreases with progenitor mass. This is primarily due to different 
thermodynamical conditions in the ejecta, which enable more efficient molecule formation subtracting 
gas-phase metals. Also, the depletion factor is more scattered as it varies between $\sim 10^{-3}$ and $0.9$. 
For $M  \geq 40~M_{\sun}$, the ejecta is mostly composed by carbon as a result of fallback, 
and the mass of dust is dominated by AC grains.

The dependence of the dust composition on progenitor mass and metallicity for rotating FE models is presented in  Fig~\ref{FE_grains_ROT}. The comparison with Fig.~\ref{FE_grains_NR} allows to assess the impact of rotation.
In general, rotating FE models are characterized by a dust composition that shows the same qualitative trends 
discussed for non rotating models: low progenitor masses are able to form a variety of dust species, while more
massive progenitors mostly form AC grains. However, independent of metallicity, rotation leads to a more efficient
formation of silicates and to a less efficient formation of AC grains in low mass progenitors. Indeed, the physical
conditions present in the ejecta of rotating models promote the formation of CO and SiO molecules that, in turn, 
affect the relative abundance of AC and silicates.

The dust mass produced by rotating and non-rotating CE models are shown in Fig.~\ref{CE_dust} for different
metallicity and progenitor masses. As expected, CE models lead to larger dust masses compared to FE models,
and the dust mass increases with progenitor mass.
In fact, the calibration of the explosion favours large ejecta and small remnant masses that, even for the 
more massive progenitors, never exceed $\sim 2 M_\odot$ (see the lower panels in Fig.~\ref{fig:mass_nr_rot_efix}). 
In non rotating models (left panels), the total dust 
mass is in the range $[0.31 - 6.0]~M_{\sun}$ for set CE0, $ [0.30 - 6.2]~M_{\sun}$ for set CE-1, 
$[0.35 - 5.0]~M_{\sun}$ for set CE-2, and $[0.34 - 5.1]~M_{\sun}$ for set CE-3. 

Rotation leads to more metal-rich ejecta, particularly by massive, low metallicity progenitors (as a consequence of the more efficient rotation-induced mixing), and 
this increases the mass of dust formed. The opposite is true for solar metallicity
massive progenitors, where a lower dust mass is produced in rotating models, because of the stronger mass loss suffered during the pre-SN phase. 

Overall, in rotating models we find that the total dust mass formed
is in the range $ [0.65 - 3.8]~M_{\sun}$ for set CE0, $ [0.49 - 5.6]~M_{\sun}$ for set CE-1, 
$ [0.49 - 7.4]~M_{\sun}$ for set CE-2, and $ [0.72 - 6.6]~M_{\sun}$ for set CE-3. 

For all CE models, we find that - independent of metallicity and rotation - the dust mass is dominated 
by silicates, as a consequence of the $\rm O > C$ composition and larger abundances
of heavier elements in the ejecta, compared to FE models (see Fig.~\ref{Input_metal_LC}).

Finally, in Fig.~\ref{size_grains} we show an example of how the grain size distribution
is affected by different properties of the SN models. We fix the mass of the SN progenitor
to be $\rm 20\,M_{\sun}$ and we consider a FE-NR model with $\rm [Fe/H] = 0$ and $\rm [Fe/H] = -3$ 
(leftmost panels) and a CE model with $\rm [Fe/H]=0$ and $v = 0, 300 \rm km/s$ (rightmost panels). 

The two SN models shown in the leftmost panels correspond to a type IIb SN at solar metallicity 
and to a type II-P SN at lower metallicity.
The grain species that form are different and - for the type II-P - the grain sizes are shifted to lower values. 
This shift is mainly due to the different ejecta density when adiabatic expansion starts: the ejecta of 
the type II-P SN has an intial radius $R_{0} \sim 1.5 \times 10^{15} \rm cm$, whereas 
the type IIb ejecta has a smaller $R_{0} \sim 3 \times 10^{14} \rm cm$ and, consequently,
a greater ejecta density. 

The FE-NR and CE-NR models shown in the first and third panels from the left
have the same metallicity (FE0, CE0) and the same progenitor, He-core and remnant masses (see Tables 
\ref{FE_A}, \ref{CE_A}), but a different amount of $\rm ^{56}Ni$ in the ejecta 
($0.33 \, M_{\sun}$ for FE-NR and $5.8 \times 10^{-2}~M_\odot$ for CE-NR), and slightly different initial 
conditions ($R_0=3.4\times 10^{14} \rm cm$ for FE-NR and $R_0=2.9\times 10^{14}\rm cm$ for FE-CE). 
The main difference is that the CE non-rotating model has a lower dust mass due to 
inefficient destruction of molecules. In addition, the distribution of \Magnetite~ grains
is shifted towards lower radii, since $\rm ^{56}Ni$ decay does not provide enough iron to 
grow large \Magnetite~ grains.
Finally in the rightmost panels of Fig.\ref{size_grains}, we compare the size distribution
functions predicted for the CE0-NR model (a Type IIb SN) with the CE0-ROT model 
(a Type Ib SN): rotation causes the grain size distribution to shift towards lower radii, 
mainly due to the lower density in the ejecta. 

\section{conclusions}

We have analized the dust mass produced in SN ejecta as a function 
of progenitor mass, rotation, metallicity and fallback. The analysis of
non rotating SN models that explode with a fixed energy of $1.2 \times 10^{51}$ erg, 
shows that fallback has a large impact on both the composition 
and the mass of dust grains that form in the ejecta. 
Fallback has a major impact on the efficiency of dust formation in more
massive SNe, particularly at low metallicity.
As a result, SNe with progenitors $\lesssim 20 - 25~M_\odot$ are generally more
efficient at producing dust. They form between $\sim 0.1~M_\odot$ and $\sim 1~M_\odot$ of dust with
a composition that is dominated by silicates, magnetite and carbon grains
for progenitors at solar metallicity, and by magnetite and carbon grains at lower metallicity.
SNe with more massive progenitors with [Fe/H]=-1,-2,-3 are able to form only carbon
grains (with a contribution of iron grains from progenitors with [Fe/H]=0). At lower metallicity fallback is too
strong to allow significant dust production.

The above conclusions are only slightly modified by the adopted initial
rotational velocity of the stars. 
When the same stars are assumed to be rotating with $\rm v = 300$~km/s, they reach the pre-supernova 
stage with larger cores and more metal-rich ejecta. This increases
the efficiency of dust production, particularly for more massive progenitors. In addition,
rotation favors more efficient silicate dust production by low mass progenitors at all metallicity values, because 
these models have a higher abundance of Si and Mg in the ejecta.

Overall, the analysis  of the SN sample with fixed explosion energy leads us to conclude that
type Ib SNe are less dusty than type II-P and IIb, but this difference is due mainly to fallback and rotation.     

It is interesting to note that massive stars that explode as SNe with a fixed explosion energy
leaving massive remnants with masses $\rm \gtrsim 5 M_\odot$ naturally lead to carbon
enhanced ejecta with [C/Fe] $> 0.7$, particularly if they are rotating. This agrees with previous
studies that suggest that carbon enhanced metal poor stars have formed in the ashes of 
faint SNe or spinstars \citep{2003Natur.422..834B, 2003ApJ...594L.123L,2005Sci...309..451I,2014ApJ...794..100M,
2006A&A...447..623M,2015A&A...580A..32M}. Our findings indicate an increase in the carbon 
enhancement with decreasing metallicity whose implications will be investigated in the context 
of stellar archaeology in a future work. 

The destructive effects of the reverse shock depend on the explosion energy, on the density of the circumstellar medium, 
on the clumpiness of the ejecta, and on the typical grain sizes and their distribution 
within the ejecta. A recent analysis of \citet{2016A&A...587A.157B} shows that models 
that are able to reproduce the dust mass inferred from observations of four nearby SN 
remnants predicts that only between 1 and 8\% of the currently observed mass will survive, 
resulting in an average SN effective dust yield of order $\rm 10^{-2} \, M_\odot$. Here we have 
focused on the first phase of dust production in SN ejecta and in the future we will address the
impact of the reverse shock destruction. Based on our current findings, we expect a 
smaller dust survival fraction for grains formed in type Ib SNe, that are characterized by
smaller sizes and hence are more easily destroyed.

Our findings are relevant to understand the role of SNe in dust enrichment at high
redshifts and in the local Universe. Due to their short evolutionary timescales, 
SNe can provide a fast and efficient dust formation channel at high redshift. 
The presence of dust grains in star forming regions provide a fundamental formation
pathway for the first low-mass and long-living stars 
\citep{2003Natur.422..869S,2005ApJ...626..627O,2012MNRAS.423L..60S,2012MNRAS.419.1566S,2014ApJ...794..100M,
2014MNRAS.439.3121C,2015MNRAS.446.2659C,2015MNRAS.451.2108D,2017MNRAS.465..926D} and 
can help to understand the early dust enrichment of the interstellar medium 
of $z > 6$ quasar host galaxies \citep{2009MNRAS.397.1661V,2011MNRAS.416.1916V,
2014MNRAS.444.2442V,2010A&A...522A..15M,2014MNRAS.438.2765C} and 
normal star forming systems (\citealt{2014MNRAS.443.1704H,2015A&A...582A..78M,2015MNRAS.451L..70M}, Graziani et al. in prep). 

Even in local galaxies, the short lifetimes of dust grains due to efficient destruction
by interstellar shocks \citep{2011A&A...530A..44J,2014A&A...570A..32B} require a fast replenishment by
stars, aided by grain growth in dense interstellar gas when the metallicity is $\rm Z \geq 0.1 Z_\odot$ 
\citep{2008A&A...479..453Z,2016ApJ...831..147Z,2013MNRAS.432..637A,2014MNRAS.442.1440S,
2016MNRAS.457.1842S,2018MNRAS.473.4538G,2017MNRAS.464..985G}.
The starting point to understand the complex cycling of dust in the interstellar medium is dust production by SNe, and 
our study aims to provide additional elements to assess how dust formation is affected by important physical properties 
of the progenitor stars and the explosions. 

\section*{Acknowledgments}
The research leading to these results has received 
funding from the European Research Council under 
the European Union's Seventh Framework Programme 
(FP/2007-2013) / ERC Grant Agreement n. 306476.
A.C. and M.L. acknowledge financial support from the PRIN INAF (2014):
Transient Universe: unveiling new types of stellar explosions with
PESSTO (P.I. A. Pastorello).
M.L. acknowledges the support by the ESO Visitor Program 2017-2018 and by the 
italian grants "Premiale 2015 MITiC" (P.I. B. Garilli) 
and "Premiale 2015 FIGARO" (P.I. G. Gemme).
\bibliographystyle{mnras}
\bibliography{bib_griglia}{}

\section{appendix}
In this appendix we report all the Tables. In particular, Tables \ref{FE_NR}, 
\ref{FE_ROT}, \ref{CE_NR}, \ref{CE_ROT} show - for non-rotating and rotating 
FE and CE models -  the progenitor mass, $M  [M_{\odot}]$, SN type, and the total mass of dust 
for each metallicity data set; in Tables \ref{FE_A}, \ref{FE_B}, \ref{FE_C}, \ref{FE_D}, 
\ref{CE_A}, \ref{CE_B}, \ref{CE_C}, \ref{CE_D} we report the physical 
properties of FE and CE non-rotating and rotating SN models: the progenitor mass 
$M  [M_{\odot}]$, $M_{\rm preSN} [M_{\odot}]$ (mass of the star at the time of explosion), 
SN type, $M_{\rm rem} [M_{\odot}]$, $M_{\rm He_{core}}[M_{\odot}]$, 
$M_{\rm eje} [M_{\odot}]$ and $\rm ^{56}Ni$ $[M_{\odot}]$. 

\begin{table*}
\caption{Properties of the non-rotating (NR) and rotating (ROT)  fixed-energy models, FE with [Fe/H] = 0 (FE0): 
progenitor mass ($M$), pre-SN mass ($M_{\rm preSN}$), SN-type, explosion energy ($E_{\rm exp}$), 
remnant-mass ($M_{\rm rem}$), helium core mass ($M_{\rm He_{core}}$), ejecta mass ($M_{\rm eje}$), and 
the nichel 56 mass ($\rm ^{56}Ni$).}
\label{FE_A}
\begin{tabular}{@{}lccccccc}
\hline
\multicolumn{8}{c}{FE0 NR}\\ 
\hline
 $M [M_{\sun}]$& $M_{\rm preSN} [M_{\sun}]$ &SN Type &$E_{\rm exp}$ & $M_{\rm rem} [M_{\sun}]$&$M_{\rm He_{core}}[M_{\sun}]$& $M_{\rm eje} [M_{\sun}] $ & $M({\rm^{56}Ni}) [M_{\sun}]$ \\
\hline
 13& 11.86 &II-P& 1.2&1.24 &4.0&10.62&0.15\\
\hline
 15& 13.23 &II-P& 1.2&1.25&4.88 &11.98 &0.14\\
\hline
 20& 7.54 &IIb &1.2&1.09&7.18& 6.45 &0.33\\
\hline
 25& 8.54 &IIb &1.2&1.27&8.54& 7.27 &0.24\\
\hline
 30& 10.83 &Ib &1.2&5.45&10.83&5.38 &0.00\\
\hline
 40& 14.14 &Ib &1.2&8.06&14.14&6.08&0.00\\
\hline
 60& 16.95 &Ib &1.2&12.62&16.95&4.33 &0.00\\
\hline 
 80& 22.71 &Ib &1.2& 18.93 &22.72&3.78 &0.00\\
\hline
 120& 27.87 &Ib &1.2&24.64&27.87&3.23 &0.00\\ 
\hline
\multicolumn{8}{c}{FE0 ROT}\\ 
\hline
 $M [M_{\sun}]$& $M_{\rm preSN} [M_{\sun}]$ &SN type &$E_{\rm exp}$ & $M_{\rm rem} [M_{\sun}]$&$M_{\rm He_{core}}[M_{\sun}]$& $M_{\rm eje} [M_{\sun}] $ & $M({\rm^{56}Ni}) [M_{\sun}]$ \\
\hline
 13& 5.35 &Ib &1.2&1.87&5.35& 3.48 &0.072\\
\hline
 15& 6.22 &Ib &1.2& 1.01&6.22& 5.21 &0.41\\
\hline
 20& 8.18 &Ib &1.2&4.61&8.18&3.57&0.00\\
\hline
 25& 9.48 &Ib &1.2&4.70 &9.48&4.78&0.00\\
\hline
 30&11.20 &Ib &1.2&4.52& 11.20&6.68&0.00\\
\hline
 40&13.81 &Ib &1.2&8.62&13.81&5.19&0.00\\
\hline
 60& 16.64 &Ib &1.2&12.53 &16.64& 4.11 &0.00\\
\hline
 80& 17.48 &Ib &1.2&13.35&17.48 & 4.13 &0.00\\
\hline
 120& 18.59 &Ib &1.2&14.69&18.58& 3.9 &0.00\\
\hline
\end{tabular}
\end{table*}
\begin{table*}
\caption{Properties of the non-rotating (NR) and rotating (ROT)  fixed-energy models, FE with [Fe/H] = -1 (FE-1): 
progenitor mass ($M$), pre-SN mass ($M_{\rm preSN}$), SN-type, explosion energy ($E_{\rm exp}$), 
remnant-mass ($M_{\rm rem}$), helium core mass ($M_{\rm He_{core}}$), ejecta mass ($M_{\rm eje}$), and 
the nichel 56 mass ($\rm ^{56}Ni$).}
\label{FE_B}
\begin{tabular}{@{}lccccccc}
\hline
\multicolumn{8}{c}{FE-1 NR}\\ 
\hline
 $M [M_{\sun}]$&$M_{\rm preSN} [M_{\sun}]$ & SN Type & $E_{\rm exp}$ & $M_{\rm rem} [M_{\sun}]$&$M_{\rm He_{core}}[M_{\sun}]$& $M_{\rm eje} [M_{\sun}] $ & $M({\rm^{56}Ni}) [M_{\sun}]$ \\
\hline
 13& 12.48 &II-P & 1.2&1.10&4.19&11.38 &0.00\\
\hline
 15& 14.17 &II-P &1.2&1.27&5.14&12.9 &0.00\\
\hline
 20& 18.35 &II-P &1.2&1.32&7.41&17.03 &0.00\\
\hline
 25& 20.57 &II-P &1.2&4.52&10.02&16.5 &0.00\\
\hline
 30& 28.27 &II-P &1.2&6.33&11.72&21.94 &0.00\\
\hline
 40& 28.71 &II-P &1.2&10.76&16.42&17.95 &0.00\\
\hline
 60& 41.96 &II-P &1.2&22.37&26.52&19.59 &0.00\\
\hline 
 80& 39.85 &Ib  &1.2&36.25&38.86&3.6 &0.00\\
\hline
\multicolumn{8}{c}{FE-1 ROT}\\ 
\hline
 $M [M_{\sun}]$&$M_{\rm preSN} [M_{\sun}]$ &SN type &$E_{\rm exp}$ & $M_{\rm rem} [M_{\sun}]$&$M_{\rm He_{core}}[M_{\sun}]$& $M_{\rm eje} [M_{\sun}] $ & $M({\rm^{56}Ni}) [M_{\sun}]$ \\
\hline
13& 10.67 &II-P &1.2&2.07 &5.46&8.6&0.00\\                     
\hline
15& 11.14 &II-P &1.2&2.24&5.92&9.0&0.00\\              
\hline
20& 17.09 &II-P &1.2&3.57&7.47&13.52&0.00\\                  
\hline
25& 18.38 &II-P &1.2&6.57&11.63&11.81&0.00\\                 
\hline
30& 15.95 &Ib &1.2&12.28&15.95&3.67 &0.00\\                
\hline
40& 20.68 &Ib &1.2&17.11&20.68& 3.57 &0.00\\               
\hline
60& 27.53 &Ib &1.2&24.13&27.53&3.4&0.00\\              
\hline
80&32.07 &Ib &1.2&28.83&32.07&3.24&0.00\\             
\hline
120&40.49 &Ib&1.2& 37.64&40.49&2.85&0.00\\ 
\hline
\end{tabular}
\end{table*}
\begin{table*}
\caption{Properties of the non-rotating (NR) and rotating (ROT)  fixed-energy models, FE with [Fe/H] = -2 (FE-2): 
progenitor mass ($M$), pre-SN mass ($M_{\rm preSN}$), SN-type, explosion energy ($E_{\rm exp}$), 
remnant-mass ($M_{\rm rem}$), helium core mass ($M_{\rm He_{core}}$), ejecta mass ($M_{\rm eje}$), and 
the nichel 56 mass ($\rm ^{56}Ni$). For failed SN models, we only show the pre-SN and final remnant masses.}
\label{FE_C}
\begin{tabular}{@{}lccccccc}
\hline
\multicolumn{8}{c}{FE-2 NR}\\ 
\hline
 $M [M_{\sun}]$ &$M_{\rm preSN} [M_{\sun}]$ &SN type &$E_{\rm exp}$ & $M_{\rm rem} [M_{\sun}]$&$M_{\rm He_{core}}[M_{\sun}]$& $M_{\rm eje} [M_{\sun}] $ & $M({\rm^{56}Ni}) [M_{\sun}]$ \\
\hline
 13& 12.96 &II-P &1.2&1.15&4.25&11.81  &0.24 \\
\hline
 15& 14.78 &II-P &1.2&1.10&5.07&13.68  &0.37  \\
\hline
 20& 19.72 &II-P &1.2&1.28&7.34& 18.44 &0.23 \\
\hline
 25& 24.65 &II-P &1.2&1.33&9.67& 23.32 &0.26  \\
\hline
 30& 29.86 &II-P &1.2&4.86&11.37& 25 &0.00\\
\hline
 40&39.74 &II-P &1.2&14.78&16.38&24.96&0.00\\
\hline
 60&59.61 & failed SN &1.2&28.87&- &- & - \\
\hline 
 80&78.59 &failed SN &1.2&42.34&- &- & - \\
\hline
\multicolumn{8}{c}{FE-2 ROT}\\ 
\hline
 $M [M_{\sun}]$ &$M_{\rm preSN} [M_{\sun}]$ &SN type &$E_{\rm exp}$ & $M_{\rm rem} [M_{\sun}]$&$M_{\rm He_{core}}[M_{\sun}]$& $M_{\rm eje} [M_{\sun}] $ & $M({\rm^{56}Ni}) [M_{\sun}]$ \\
\hline
 13&11.44 &II-P &1.2&1.97& 5.43& 9.47 &0.051\\
\hline
 15& 13.67 &II-P &1.2&2.31& 5.86& 11.36  &0.00\\
\hline
 20& 16.79 &II-P &1.2&6.39& 9.79& 10.4 &0.00\\
\hline
 25& 13.16 &Ib &1.2&8.49& 13.16&4.67 &0.00\\
\hline
 30& 15.06 &Ib &1.2&10.74& 15.06&4.32&0.00\\
\hline
 40&22.94 &Ib &1.2&19.43& 22.94&3.51&0.00\\
\hline
 60&37.41 &Ib &1.2&34.57& 37.41&2.84&0.00\\
\hline
\end{tabular}
\end{table*}
\begin{table*}
\caption{Properties of the non-rotating (NR) and rotating (ROT)  fixed-energy models, FE with [Fe/H] = -3 (FE-3): 
progenitor mass ($M$), pre-SN mass ($M_{\rm preSN}$), SN-type, explosion energy ($E_{\rm exp}$), 
remnant-mass ($ M_{\rm rem}$), helium core mass ($M_{\rm He_{core}}$), ejecta mass ($M_{\rm eje}$), and 
the nichel 56 mass ($\rm ^{56}Ni$). For failed SN models, we only show the pre-SN and final remnant masses.}
\label{FE_D}
\begin{tabular}{@{}lccccccc}
\hline
\multicolumn{8}{c}{FE-3 NR}\\ 
\hline
$M [M_{\sun}]$ &$M_{\rm preSN} [M_{\sun}]$ &SN type &$E_{\rm exp}$ & $M_{\rm rem} [M_{\sun}]$&$M_{\rm He_{core}}[M_{\sun}]$& $M_{\rm eje} [M_{\sun}] $ & $M({\rm^{56}Ni}) [M_{\sun}]$\\
\hline
 13& 12.97 &II-P &1.2&1.09&4.10& 11.88&0.34 \\
\hline
 15& 14.95 &II-P &1.2&1.31&5.07 & 13.64&0.20  \\
\hline
 20& 19.79 &II-P &1.2&1.29&7.14 & 18.5&0.24 \\
\hline
 25& 24.63 &II-P & 1.2&1.37&9.55 & 23.26&0.28  \\
\hline
 30& 29.97 &II-P &1.2&5.04&12 &24.93& 0.00 \\
\hline
 40& 39.96 & failed SN &1.2&17.55&- &- & -  \\
\hline
 60& 59.94 &failed SN &1.2&31.12&- &- & - \\
\hline 
 80& 79.90 &failed SN &1.2&45.62&- &- & - \\
\hline
\multicolumn{8}{c}{FE-3 ROT}\\ 
\hline
$M [M_{\sun}]$ &$M_{\rm preSN} [M_{\sun}]$ &SN type &$E_{\rm exp}$ & $M_{\rm rem} [M_{\sun}]$&$M_{\rm He_{core}}[M_{\sun}]$& $M_{\rm eje} [M_{\sun}] $ & $M({\rm^{56}Ni}) [M_{\sun}]$\\ 
\hline
 13& 12.77 &II-P &1.2&1.14& 4.87&11.63  &0.49\\
\hline 
 15& 13.78 &II-P &1.2&1.30& 6.06& 13.7 &0.30\\
\hline
 20& 19.96 &II-P &1.2&3.75& 7.85&16.21&0.00\\
\hline
 25& 13.29 &Ib &1.2&8.71& 12.91& 4.58 &0.00\\
\hline
 30& 17.06 &Ib &1.2&13.15& 16.80& 3.91 &0.00\\
\hline
 40& 24.48 &Ib &1.2&21.12& 24.48& 3.36 &0.00\\
\hline
 60& 38.14 & Ib&1.2&34.79&38.14&3.35&0.00\\
\hline
\end{tabular}
\end{table*}
\begin{table*}
\caption{Properties of the non-rotating (NR) and rotating (ROT)  calibrated-energy models, CE with [Fe/H] = 0 (CE0): 
progenitor mass ($M$), pre-SN mass ($M_{\rm preSN}$), SN-type, explosion energy ($E_{\rm exp}$), 
remnant-mass ($M_{\rm rem}$), helium core mass ($M_{\rm He_{core}}$), ejecta mass ($M_{\rm eje}$), and 
the nichel 56 mass ($\rm ^{56}Ni$).}
\label{CE_A}
\begin{tabular}{@{}lcccccccc}
\hline
\multicolumn{8}{c}{CE0 NR}\\ 
\hline
 $M [M_{\sun}]$&$M_{\rm preSN} [M_{\sun}]$ &SN Type &$E_{\rm exp}$ & $M_{\rm rem} [M_{\sun}]$&$M_{\rm He_{core}}[M_{\sun}]$& $M_{\rm eje} [M_{\sun}] $ & $M({\rm^{56}Ni}) [M_{\sun}]$ \\
\hline
 13&11.86 &II-P&0.56&1.50 &4.00&10.36&0.015\\ 
\hline
 15&13.23 &II-P&0.73&1.54 &4.88&11.69&0.024\\
\hline
 20& 7.54 &IIb&0.67&1.47 &7.18&6.07&0.058\\
\hline
 25& 8.54 &IIb&0.94&1.54 &8.54&7.0&0.11\\
\hline
 30& 10.83 &Ib&1.71&1.92 &10.83&8.91&0.19\\
\hline
 40& 14.14 &Ib&1.92&1.64&14.14&12.5&0.46\\
\hline
 60&16.95 &Ib&2.59&1.37&16.95&15.58&0.62\\
\hline 
 80&22.71 &Ib&3.49&1.35&22.71&21.36&0.77\\
\hline
 120&27.87 &Ib&5.94&1.51&27.87&26.36&0.97\\
\hline
\multicolumn{8}{c}{CE0 ROT}\\ 
\hline
 $M [M_{\sun}]$&$M_{\rm preSN} [M_{\sun}]$ &SN Type &$E_{\rm exp}$ & $M_{\rm rem} [M_{\sun}]$&$M_{\rm He_{core}}[M_{\sun}]$& $M_{\rm eje} [M_{\sun}] $ & $M({\rm^{56}Ni}) [M_{\sun}]$ \\
\hline
 13& 5.35 &Ib &1.23 &1.97 &5.35&3.38&0.015\\
\hline
 15& 6.22 &Ib &0.93&1.99&6.22&4.23&0.024\\
\hline
 20& 8.18 &Ib &2.03&2.27&8.18&5.91&0.058\\
\hline
 25& 9.48 &Ib &1.77&1.97 &9.48&7.51&0.11\\
\hline
 30&11.20 &Ib &1.78&1.80 &11.20&9.4&0.19\\
\hline
 40& 13.81 &Ib &2.49&1.83&13.81&11.98&0.46\\
\hline
 60& 16.64 &Ib &2.76&1.45&16.64&15.19&0.63\\
\hline
 80&17.48 &Ib &2.83&1.42&17.48&16.06&0.64\\
\hline
 120& 18.59 &Ib &3.01&1.44&18.59&17.15&0.68\\
\hline
\end{tabular}
\end{table*}
\begin{table*}
\caption{Properties of the non-rotating (NR) and rotating (ROT)  calibrated-energy models, CE with [Fe/H] = -1 (CE-1): 
progenitor mass ($M$), pre-SN mass ($M_{\rm preSN}$), SN-type, explosion energy ($E_{\rm exp}$), 
remnant-mass ($M_{\rm rem}$), helium core mass ($M_{\rm He_{core}}$), ejecta mass ($M_{\rm eje}$), and 
the nichel 56 mass ($\rm ^{56}Ni$).}
\label{CE_B}
\begin{tabular}{@{}lccccccc}
\hline
\multicolumn{8}{c}{CE-1 NR}\\ 
\hline
$M [M_{\sun}]$&$M_{\rm preSN} [M_{\sun}]$ &SN Type &$E_{\rm exp}$ & $M_{\rm rem} [M_{\sun}]$&$M_{\rm He_{core}}[M_{\sun}]$& $M_{\rm eje} [M_{\sun}] $ & $M({\rm^{56}Ni}) [M_{\sun}]$ \\
\hline
 13& 12.48 &II-P &0.50&1.52&4.19&10.96&0.015\\
\hline
 15&14.17 &II-P &0.82&1.60&5.14&12.57&0.024\\
\hline
 20&18.35 &II-P &0.88&1.61&7.41&16.74&0.058\\
\hline
 25& 20.57 &II-P &1.49&1.97&10.00&18.06&0.11\\
\hline
 30&28.27 &II-P &1.58 &1.88&11.72&26.39&0.19\\
\hline
 40& 28.71 &II-P &1.87&1.64&16.46&27.07&0.46\\
\hline
 60& 41.96 &II-P &3.56&1.39&26.58& 40.57 &0.85\\
\hline 
 80&39.85 &Ib  &5.68&1.46&38.86& 38.39 &1.23\\
\hline
\multicolumn{8}{c}{CE-1 ROT}\\ 
\hline
$M [M_{\sun}]$&$M_{\rm preSN} [M_{\sun}]$ &SN Type &$E_{\rm exp}$ & $M_{\rm rem} [M_{\sun}]$&$M_{\rm He_{core}}[M_{\sun}]$& $M_{\rm eje} [M_{\sun}] $ & $M({\rm^{56}Ni}) [M_{\sun}]$ \\
\hline
13&10.67 &II-P &1.30&2.07 &5.46&8.6& 0.015 \\
\hline
15& 11.14 &II-P &1.44& 2.10&6.02&9.04& 0.024\\
\hline
20& 17.09 &II-P &1.92&2.21&7.54&14.88& 0.058\\
\hline
25& 18.38 &II-P &2.22&2.24&11.83&16.14&0.11\\
\hline
30& 15.95 &Ib &3.08&2.50&15.95&13.45&0.19\\
\hline
40&20.68 &Ib &3.93&2.36&20.68&18.32&0.46\\
\hline
60& 27.53 &Ib &5.11&2.01&27.53& 25.52 &0.76\\
\hline
80& 32.07 &Ib &5.90&1.88&32.07& 30.19 &0.79\\
\hline
\end{tabular}
\end{table*}
\begin{table*}
\caption{Properties of the non-rotating (NR) and rotating (ROT)  calibrated-energy models, CE with [Fe/H] = -2 (CE-2): 
progenitor mass ($M$), pre-SN mass ($M_{\rm preSN}$), SN-type, explosion energy ($E_{\rm exp}$), 
remnant-mass ($M_{\rm rem}$), helium core mass ($M_{\rm He_{core}}$), ejecta mass ($M_{\rm eje}$), and 
the nichel 56 mass ($\rm ^{56}Ni$).}
\label{CE_C}
\begin{tabular}{@{}lccccccc}
\hline
\multicolumn{8}{c}{CE-2 NR}\\ 
\hline
$M [M_{\sun}]$&$M_{\rm preSN} [M_{\sun}]$ &SN Type &$E_{\rm exp}$ & $M_{\rm rem} [M_{\sun}]$&$M_{\rm He_{core}}[M_{\sun}]$& $M_{\rm eje} [M_{\sun}] $ & $M({\rm^{56}Ni}) [M_{\sun}]$ \\
\hline
 13& 12.96 &II-P &0.64&1.58 &4.24& 11.38 &0.015\\
\hline
 15& 14.78 &II-P &0.64&1.54&5.10&13.24&0.024\\
\hline
 20&19.72 &II-P &0.88&1.63 &7.34& 18.09 &0.058\\
\hline
 25& 24.65 &II-P &1.04&1.71 &9.67& 22.94&0.11\\
\hline
 30& 29.86 &II-P &1.30&1.74 &11.37& 28.12 &0.19\\
\hline
 40& 39.74 &II-P &1.92&1.69&16.38& 38.05 &0.46\\
\hline
 60&59.61 &II-P &3.68&1.43&26.85& 58.18 &0.88\\
\hline 
 80&78.59 &II-P &5.35&1.43&37.84&77.16 &1.15\\
\hline
\multicolumn{8}{c}{CE-2 ROT}\\ 
\hline
$M [M_{\sun}]$&$M_{\rm preSN} [M_{\sun}]$ &SN Type &$E_{\rm exp}$ & $M_{\rm rem} [M_{\sun}]$&$M_{\rm He_{core}}[M_{\sun}]$& $M_{\rm eje} [M_{\sun}] $ & $M({\rm^{56}Ni}) [M_{\sun}]$ \\
\hline
 13&11.44 &II-P &1.23&2.02&5.43&9.42&0.015\\
\hline
 15&13.67 &II-P &1.51&2.15&5.86&11.52&0.024\\
\hline
 20&11.44 &II-P &2.38&2.65&9.79&8.79&0.058\\
\hline
 25&13.16 &Ib &2.51&2.37&13.00 &10.79&0.11\\
\hline
 30&15.06 &Ib &2.70&2.33&15.06&12.73&0.19\\
\hline
 40&22.94 &Ib &4.26&2.44&22.94&20.5&0.46\\
\hline
 60&37.41 &Ib &4.83&1.95&37.41&35.46 &1.49\\
\hline
\end{tabular}
\end{table*}
\begin{table*}
\caption{Properties of the non-rotating (NR) and rotating (ROT)  calibrated-energy models, CE with [Fe/H] = -3 (CE-3): 
progenitor mass ($M$), pre-SN mass ($M_{\rm preSN}$), SN-type, explosion energy ($E_{\rm exp}$), 
remnant-mass ($M_{\rm rem}$), helium core mass ($M_{\rm He_{core}}$), ejecta mass ($M_{\rm eje}$), and 
the nichel 56 mass ($\rm ^{56}Ni$).}
\label{CE_D}
\begin{tabular}{@{}lccccccc}
\hline
\multicolumn{8}{c}{CE-3 NR}\\ 
\hline
$M [M_{\sun}]$&$M_{\rm preSN} [M_{\sun}]$ &SN Type &$E_{\rm exp}$ & $M_{\rm rem} [M_{\sun}]$&$M_{\rm He_{core}}[M_{\sun}]$& $M_{\rm eje} [M_{\sun}] $ & $M({\rm^{56}Ni}) [M_{\sun}]$ \\
\hline
 13&12.97 &II-P &0.54&1.55 &4.12& 11.42 & 0.015 \\
\hline
 15&14.95 &II-P &0.88&1.68 &5.07&13.27& 0.024\\
\hline
 20&19.79 &II-P &0.89&1.65 &7.18&18.14& 0.058\\
\hline
 25&24.63 &II-P &1.14&1.75 &9.55&22.88& 0.11\\
\hline
 30&29.97 &II-P &1.38&1.72 &11.99& 28.25 & 0.19\\
\hline
 40&39.96 &II-P &2.26&1.74&16.98&38.22 & 0.46\\
\hline
 60&59.94 &II-P &4.03&1.39 &27.01&58.55&0.95\\
\hline 
 80&79.90 &II-P &5.68 &1.44 &37.93& 78.46 & 1.16\\
\hline
\multicolumn{8}{c}{CE-3 ROT}\\ 
\hline
$M [M_{\sun}]$&$M_{\rm preSN} [M_{\sun}]$ &SN Type &$E_{\rm exp}$ & $M_{\rm rem} [M_{\sun}]$&$M_{\rm He_{core}}[M_{\sun}]$& $M_{\rm eje} [M_{\sun}] $ & $M({\rm^{56}Ni}) [M_{\sun}]$ \\
\hline
 13&12.77 &II-P &0.78& 1.80 &4.57&10.97&0.015\\
\hline 
 15&13.78 &II-P &1.10&2.09&5.52&11.69&0.024\\
\hline
 20&19.96 &II-P &1.86&2.18&7.85&17.78&0.058\\
\hline
 25&13.29 &Ib &2.52&2.38&12.79&10.91& 0.11\\
\hline
 30&17.06 &Ib &2.93&2.34&16.78&14.72&0.19\\
\hline
 40&24.48 &Ib &4.59&2.52&24.48 &21.96& 0.46 \\
\hline
 60&38.14 &Ib &7.07&2.13&38.14&36.01  & 0.96 \\
\hline
\end{tabular}
\end{table*}
\begin{table*}
\caption{Dust masses produced by the non-rotating (NR) fixed-energy models (FE) of different metallicity: 
FE0 ([Fe/H] = 0), FE-1 ([Fe/H] = -1), FE-2 ([Fe/H] = -2), FE-3 ([Fe/H] = -3). 
Progenitor mass ($M$), SN type and dust mass ($M_{\rm dust}$). We also indicate
the failed SN or pair unstable models for which dust formation has not been computed.}
\label{FE_NR}
\begin{tabular}{@{}lcccccccc}
\hline
\multicolumn{9}{c}{FE NR} \\
\hline
& \multicolumn{2}{c}{FE0}&\multicolumn{2}{c}{FE-1}&\multicolumn{2}{c}{FE-2}&\multicolumn{2}{c}{FE-3}\\
\hline
 $M [M_{\odot}]$& SN type &$M_{\rm dust} [M_{\sun}]$&SN type &$M_{\rm dust} [M_{\sun}]$&SN type &$M_{\rm dust} [M_{\sun}]$&SN type &$M_{\rm dust} [M_{\sun}]$ \\
\hline
 13& II-P &0.32&II-P&0.24&II-P&0.51&II-P&0.65\\
\hline
 15& II-P &0.44&II-P&0.59&II-P&0.78&II-P &0.61 \\
\hline
 20& IIb &1.0&II-P&1.03&II-P&1.1&II-P &1.1\\
\hline
 25& IIb &1.2&II-P&0.43&II-P&1.2&II-P &1.2\\
\hline
 30& Ib  &0.16&II-P&0.22&II-P& $1.3\times 10^{-2}$&II-P&0.20\\
\hline
 40& Ib  &0.24&II-P&0.29&II-P& $2.1\times 10^{-2}$ &failed SN &- \\
\hline
 60& Ib &0.28&II-P&0.23 &failed SN &- &failed SN& - \\
\hline 
 80& Ib  &0.35&Ib&0.18  &failed SN & -  &failed SN &-\\
\hline
120 &Ib & 0.47& pair-unstable &-& pair-unstable &-& pair-unstable &-\\

\hline
\end{tabular}
\end{table*}
\begin{table*}
\caption{Dust masses produced by the rotating (ROT) fixed-energy models (FE) of different metallicity: 
FE0 ([Fe/H] = 0), FE-1 ([Fe/H] = -1), FE-2 ([Fe/H] = -2), FE-3 ([Fe/H] = -3). 
Progenitor mass ($M$), SN type and dust mass ($M_{\rm dust}$). We also indicate
the pair unstable models for which dust formation has not been computed.}
\label{FE_ROT}
\begin{tabular}{@{}lcccccccc}
\hline
\multicolumn{9}{c}{FE ROT} \\
\hline
& \multicolumn{2}{c}{FE0}&\multicolumn{2}{c}{FE-1}&\multicolumn{2}{c}{FE-2}&\multicolumn{2}{c}{FE-3}\\
\hline
$M [M_{\odot}]$& SN type &$M_{\rm dust} [M_{\sun}]$&SN type &$M_{\rm dust} [M_{\sun}]$&SN type &$M_{\rm dust} [M_{\sun}]$&SN type &$M_{\rm dust} [M_{\sun}]$ \\
\hline
 13& Ib &0.77&II-P&0.49&II-P&0.56&II-P&1.3\\
\hline
 15& Ib &2.25 &II-P&0.61&II-P&0.78&II-P&1.3\\
\hline
 20& Ib &0.13 &II-P&0.46&II-P&0.17&II-P&$3.0\times 10^{-2}$\\
\hline
 25& Ib &0.15 &II-P&0.65&Ib& $2.6\times 10^{-3}$&Ib& 0.15\\
\hline
 30& Ib &0.24 &Ib &$6.9\times 10^{-2}$&Ib&$1.8\times 10^{-2}$&Ib&0.27\\
\hline
 40& Ib & 0.35 &Ib &0.12&Ib&0.15&Ib&$2.9\times 10^{-2}$\\
\hline
 60& Ib & 0.20 &Ib&0.33&Ib &0.35&Ib&$6.3\times 10^{-2}$\\
\hline 
 80& Ib & 0.21  &Ib&0.54& pair-unstable &-&pair-unstable&-\\
\hline
 120 &Ib& 0.27  &Ib&0.62&pair-unstable&-&pair-unstable&-\\
\hline
\end{tabular}
\end{table*}
%
\begin{table*}
\caption{Dust masses produced by non-rotating (NR) calibrated-energy models (CE) of different metallicity: 
CE0 ([Fe/H] = 0), CE-1 ([Fe/H] = -1), CE-2 ([Fe/H] = -2), CE-3 ([Fe/H] = -3). 
Progenitor mass ($M$), SN type and dust mass ($M_{\rm dust}$). We also indicate
the pair unstable models for which dust formation has not been computed.}
\label{CE_NR}
\begin{tabular}{@{}lcccccccc}
\hline
\multicolumn{9}{c}{CE NR} \\
\hline
& \multicolumn{2}{c}{CE0}&\multicolumn{2}{c}{CE-1}&\multicolumn{2}{c}{CE-2}&\multicolumn{2}{c}{CE-3}\\
\hline
$M [M_{\odot}]$& SN type &$M_{\rm dust} [M_{\sun}]$&SN type &$M_{\rm dust} [M_{\sun}]$&SN type &$M_{\rm dust} [M_{\sun}]$&SN type &$M_{\rm dust} [M_{\sun}]$ \\
\hline
 13& II-P &0.31 & II-P &0.30 &II-P &0.35 &II-P&0.34\\
\hline
 15& II-P &0.61 & II-P &0.61  &II-P&0.51 &II-P&0.55\\
\hline
 20& IIb &0.87 & II-P &0.86 &II-P&0.76 &II-P&0.71 \\
\hline
 25& IIb &0.95 & II-P &1.50 &Ib&0.86 &Ib&0.94 \\
\hline
 30& Ib  &1.6  & Ib &1.2 &Ib&1.0 &Ib&1.1 \\
\hline
 40& Ib  &2.7  & Ib &1.7 &Ib&1.6 &Ib&1.8 \\
\hline
 60& Ib & 6.0  &Ib &5.0 & Ib&3.6 &Ib&4.0 \\
\hline 
 80& Ib & 7.3 &Ib &6.2 &pair-unstable&-&pair-unstable&-\\
\hline
 120 &Ib &6.0  &pair-unstable&-&pair-unstable&-&pair-unstable&-\\
\hline 
\end{tabular}
\end{table*}
\begin{table*}
\caption{Dust masses produced by rotating (ROT) calibrated-energy models (CE) of different metallicity: 
CE0 ([Fe/H] = 0), CE-1 ([Fe/H] = -1), CE-2 ([Fe/H] = -2), CE-3 ([Fe/H] = -3). 
Progenitor mass ($M$), SN type and dust mass ($M_{\rm dust}$). We also indicate
the pair unstable models for which dust formation has not been computed.}
\label{CE_ROT}
\begin{tabular}{@{}lcccccccc}
\hline
\multicolumn{9}{c}{CE ROT} \\
\hline
& \multicolumn{2}{c}{CE0}&\multicolumn{2}{c}{CE-1}&\multicolumn{2}{c}{CE-2}&\multicolumn{2}{c}{CE-3}\\
\hline
$M [M_{\odot}]$& SN type &$M_{\rm dust} [M_{\sun}]$&SN type &$M_{\rm dust} [M_{\sun}]$&SN type &$M_{\rm dust} [M_{\sun}]$&SN type &$M_{\rm dust} [M_{\sun}]$ \\
\hline
 13& Ib &0.65&II-P&0.49 &II-P&0.49 &II-P&0.72  \\
\hline
 15& Ib &1.5  &II-P&0.56 &II-P&0.73 &II-P&0.93 \\
\hline
 20& Ib &1.2  &II-P&1.4 &II-P&1.1 &II-P&0.96 \\
\hline
 25& Ib &1.1  &II-P&1.8 &Ib&1.2&Ib&1.1\\
\hline
 30& Ib &1.4  &Ib&1.6  & Ib&1.4 &Ib&1.9 \\
\hline
 40& Ib &2.0  &Ib&2.8  &Ib&3.3 &Ib&3.4 \\
\hline
 60& Ib &3.2  &Ib&4.8  &Ib&7.4 &Ib&6.6 \\
\hline 
 80& Ib &3.4  &Ib&5.6 &pair-unstable&-&pair-unstable&- \\
\hline
 120& Ib &3.8 &pair-unstable&-&pair-unstable&-&pair-unstable&-\\
\hline
\end{tabular}
\end{table*}
\label{lastpage}
\end{document}